\documentclass[final]{elsarticle}
\usepackage[letterpaper, margin=1.0in]{geometry}
\usepackage{color}
\usepackage{amsmath}
\usepackage{amssymb}
\usepackage{wrapfig}
\usepackage{sidecap}
\usepackage{amsfonts}
\usepackage{mathrsfs}
\usepackage{subfigure}
\usepackage{booktabs}
\usepackage{caption}

\usepackage{lineno,hyperref}

\journal{Journal of \LaTeX\ Templates}

\biboptions{authoryear}

\begin{document}

\begin{frontmatter}

\title{Exploring the Venus global super-rotation using a\\ comprehensive General Circulation Model}

\author[mymainaddress,mysecondaryaddress]{J. M. Mendon\c ca\corref{cor1}}
\ead{joao.mendonca@csh.unibe.ch}
\author[mysecondaryaddress]{P. L. Read}

\cortext[cor1]{Corresponding author}
\address[mymainaddress]{Center for Space and Habitability, University of Bern, Siddlerstrasse 5, Bern 3012, Switzerland}
\address[mysecondaryaddress]{Department of Physics, University of Oxford, Clarendon Laboratory, Parks Road, Oxford, U.K.}

\begin{abstract}
The atmospheric circulation in Venus is well known to exhibit strong super-rotation. However, the atmospheric mechanisms responsible for the formation of this super-rotation are still not fully understood. In this work, we developed a new Venus general circulation model to study the most likely mechanisms driving the atmosphere to the current observed circulation. Our model includes a new radiative transfer, convection and suitably adapted boundary layer schemes and a dynamical core that takes into account the dependence of the heat capacity at constant pressure with temperature. 

The new Venus model is able to simulate a super-rotation phenomenon in the cloud region quantitatively similar to the one observed. The mechanisms maintaining the strong winds in the cloud region were found in the model results to be a combination of zonal mean circulation, thermal tides and transient waves. In this process, the semi-diurnal tide excited in the upper clouds has a key contribution in transporting axial angular momentum mainly from the upper atmosphere towards the cloud region.  The magnitude of the super-rotation in the cloud region is sensitive to various radiative parameters such as the amount of solar radiative energy absorbed by the surface, which controls the static stability near the surface.  In this work, we also discuss the main difficulties in representing the flow below the cloud base in Venus atmospheric models.

Our new radiative scheme is more suitable for 3D Venus climate models than those used in previous work due to its easy adaptability to different atmospheric conditions. This flexibility of the model was crucial to explore the uncertainties in the lower atmospheric conditions and may also be used in the future to explore, for example, dynamical-radiative-microphysical feedbacks. 

\end{abstract}

\begin{keyword}
\texttt Venus \sep Planetary atmospheres
\end{keyword}

\end{frontmatter}


\section{Introduction}
\label{sec:intro}
The Venus atmospheric circulation has several characteristics which remain poorly understood, such as the mechanism of formation and maintenance of the atmospheric super-rotation below 100 km. The super-rotation is characterized by a much faster rotation of the atmosphere compared to the rotation rate of the solid planet. This phenomenon quantifies the excess of axial angular momentum that an atmosphere possesses when compared with an atmosphere co-rotating with the underlying planet. From \cite{1986Read} this global phenomenon is quantified using the following equation,
\begin{equation}
\label{eqn:S}
S = \frac{M_t}{M_0}-1,
\end{equation}
where $S$ is defined as the global super-rotation index, $M_t$ is the total axial angular momentum of the atmosphere and $M_0$ is the axial angular momentum of the atmosphere with zero zonal wind velocities relative to its underlying planet. The total angular momentum of the atmosphere per unit mass ($M_t$) is defined by:
\begin{equation}
\label{eqn:M}
M_t = \int \int \int \frac{m a^2 \cos \phi}{g} d\phi d\lambda dp
\end{equation}
where $a$ is taken to be the radius of the planet, $\phi$ is the latitude, $\lambda$ is the longitude, $p$ is pressure, $g$ is the gravitational acceleration and $m$  is the angular momentum per unit mass ($m = a \cos \phi (\Omega a \cos \phi + u)$), $\Omega$ is the rotation rate of the planet and $u$ is the zonal component of the wind velocity).  Using equations \ref{eqn:S} and \ref{eqn:M} and the observational vertical profiles of zonal winds and their uncertainties from \cite{1985Kerzhanovich}, we can estimate $S = 7.7_{-3.6}^{+4.2}$ (S$ \sim 1.5\times10^{-2}$ for the Earth, \citealt{1986Read}). The vertical profiles from \cite{1985Kerzhanovich} are associated with low latitudes. Three different profiles corresponding to the lowest, mean and highest wind values observed at each altitude, were used to build three global wind fields to compute three different values of $S$ (the mean value and its uncertainties). The three global wind fields were defined using the observational vertical wind profiles at the equator and were extrapolated to the pole region assuming that the atmospheric circulation follows a solid body rotation profile at each altitude. These three dimensional wind fields are axisymmetric. In general, the solid body rotation assumption slightly underestimates the values of the zonal winds at high latitudes, but is a very good approximation for the altitudes where the winds have the largest contribution to angular momentum density (at around 20 km, \citealt{1983Schubert}).

In general, dynamical motions in the atmosphere of Venus are driven by a differential insolation in latitude, which might be expected to induce atmospheric circulation in the form of cells with rising atmospheric flow at low latitudes and descending at high latitudes.  The presence of middle or high latitude local super-rotation, which refers to the typically barotropically unstable jet, is a consequence of the existence of those cells that transport angular momentum from low toward high latitudes. The presence of a large equator-to-pole Hadley circulation in each hemisphere (e.g., \citealt{1980Schubert}) is due to the slow planetary rotation, which weakens the Coriolis acceleration, increasing the efficiency of the latitudinal heat transport of the atmosphere. The total axial angular momentum of the atmosphere is mainly controlled by the mechanical surface-atmosphere interaction.  During the ``spin-up'' of the atmosphere this mechanical interaction pumps axial angular momentum into the atmosphere. More difficult to explain is the presence of the observed equatorial super-rotation. The strong winds in the equatorial region are not produced or maintained by the influence of zonal mean mechanisms (\citealt{1969Hide}). Such super-rotation requires the presence of non-axisymmetric eddy motions, unless super-rotation was its initial condition.

Using the equations of motion we can learn more about possible atmospheric mechanisms for generating and maintaining the strong winds in the equatorial region (\citealt{1997Gierasch}). The zonal momentum equation can be written as a conservation equation for axial angular momentum in the form:
\begin{equation}
\underbrace{\frac{\partial}{\partial t}(\rho m) }_{[A]}+ \underbrace{\frac{}{}\vec{\nabla}\cdot (\rho \textbf{v}m)}_{[B]} + \underbrace{\frac{\partial p}{\partial \lambda}}_{[C]} = \underbrace{\frac{}{}\vec{\nabla}\cdot(\tau \cdot \hat{\textbf{z}}\times\textbf{r})}_{[D]}
\end{equation}
 where $\hat{\textbf{z}}$ is the unit vector in the direction of the planetary angular velocity ($\Omega$) and $\tau$ is the viscous stress tensor. $m$ in this equation is the axial angular momentum per unit mass as mentioned previously. Zonally averaging this equation and assuming the friction is negligible, the terms [C] and [D] drop out. This simplification makes it easier to interpret the conservation of angular momentum in a circulating atmospheric cell. Unless there is convergence of the angular momentum flux, $\textbf{F}_m = \rho \textbf{v} m$, towards a location of maximum angular momentum per unit of mass, it is not possible to produce or sustain, for example, the observed strong prograde winds at low latitudes in the Venus atmosphere. Note that the term ``prograde'' in this work refers to winds in the direction of the planet's rotation. It was demonstrated in \cite{1969Hide} that global or local super-rotation cannot be obtained in a purely inviscid, axisymmetric system that evolved from rest. This result is frequently called ``Hide's first theorem'' (\citealt{1986Read}), and implies that the excess of angular momentum, $S>0$, can only be obtained from non-axisymmetric motions. This non-axisymmetric phenomenon can be represented by the known Reynolds' stress terms (associated with zonal pressure torques) that are defined by:
\begin{equation}
\label{eqn:fm}
[F_m] = \rho a \cos \phi ([u^{\star}v^{\star}],[u^{\star}w^{\star}]) 
\end{equation}
where $\rho$ is the atmospheric density, $a$ is the planet radius, the square brackets denote a zonal average, $v$ is the meridional component of the wind velocity and $w$ is the wind speed in the vertical direction. The stars on each variable mean that they are disturbances in relation to their respective zonal average. The first and second terms are the meridional and vertical components of the eddy fluxes. From Eq. (\ref{eqn:fm}), the weight of each component of the Reynold's stress (horizontal and vertical) is in general related to different possible mechanisms that contribute to the formation and$/$or maintenance of the super-rotation. To be able to complete the puzzle as to the real nature of the general super-rotation, there is a need to identify the atmospheric processes involved in these two terms and quantify their contribution to the phenomenon.

The Venus atmosphere has been explored by several space missions in the past: notably the Venera orbiters and entry probes, Pioneer Venus and Magellan, and more recently the European Venus Express mission. These missions made atmospheric data available which increased the interest in the development of global circulation models capable of interpreting these data and guiding their analysis. Numerical modelling of the global Venus atmospheric circulation started more than forty years ago, with the complexity and accuracy improving along the years (e.g. \citealt{1975Kalnay}, \citealt{2003Yamamotoa}, \citealt{2007Lee3} and \citealt{2010Lebonnois}). Recently, typical Venus numerical models that use very simplified representations of radiation and boundary layer processes (e.g., \citealt{2003Yamamotoa}; \citealt{2007Lee3}; \citealt{2013Lebonnois}), have suggested that the global atmospheric super rotation is at least partially maintained by the equatorward momentum transport via synoptic eddies from high latitude  barotropically unstable jets. This mechanism is commonly known as the GRW mechanism (\citealt{1975Gierasch}; \citealt{1979RossowWilliams}). Further studies using simplified GCMs have also highlighted other possible mechanisms involving interactions between mid-latitude Rossby waves and equatorial Kelvin waves to form unstable modes that can lead to zonal acceleration in the tropics (\citealt{2010Mitchell}; \citealt{2014Potter}). Evolving towards more complex physically-based models we find the work by \cite{2007Ikeda} and \cite{2010Lebonnois}, who included for instance a self-consistent computation of temperature using a radiative transfer formulation. In these cases the diurnal cycle is not neglected, which revealed to be an important factor in the atmospheric dynamics produced. The diurnal cycle excites migrating thermal tides especially in the Venus cloud region due to the large extinction of solar energy there. The thermal tides play an important role maintaining the super rotation, since they transport prograde momentum vertically and predominantly at low latitudes, from above the cloud region towards the upper cloud deck (e.g., \citealt{1992Newman} and \citealt{2010Lebonnois}). These three momentum transport mechanisms: high latitude barotropic eddies, tropical Rossby-Kelvin instabilities or thermal tides, are thought to be the main possible mechanisms for the formation and maintenance of strong zonal winds at low latitudes: via the $[u^{\star}v^{\star}]$ and $[u^{\star}w^{\star}]$ terms respectively. 

In this study we have developed a new Venus General Circulation Model called the Oxford Planetary Unified (Model) System for Venus (\textbf{OPUS-Vr}) that includes a new radiation scheme (\citealt{2015Mendonca}). We use it to simulate the Venus atmosphere in a physically consistent manner. The main advantage of our radiation code against previous works is the explicit calculation of the solar fluxes and the easy adaptability to different optical structures. The OPUS-Vr is aimed at studying the atmospheric mechanisms that transport momentum, and explore the range of atmospheric conditions favourable to the formation of an atmospheric circulation similar to the one observed.

In the next section we describe the numerical model used in this work. In section \ref{sec:Dbaseline} the results from the reference simulation are explored, and the main momentum transport mechanisms and waves are identified. In section \ref{sec:Dbaseline} we also compared the model results obtained with available observational data. In section \ref{sec:SurfAlb} a sensitivity test to the surface albedo is explored. Finally in sections \ref{sec:dics} and \ref{sec:conclu} a discussion on possible super-rotation mechanisms working within the lower Venus atmosphere is presented followed by the general conclusions.

\section{OPUS-Vr}
\label{sec:opus-vr}
The new version of the Oxford Planetary Unified (Model) System for Venus (OPUS-Vr) that includes a relatively complete, physically-based radiative transfer formulation is an expansion and improvement of an earlier simplified version (here designated OPUS-Vs, \citealt{2006Lee2}; \citealt{2007Lee3}). New parameterisations more suitable to simulate the extreme Venus climate were developed and implemented, such as: a new radiation scheme for the solar and thermal spectral regions, a new convection scheme, an adapted diffusive soil model and a physically-based boundary layer scheme, together with a modified dynamical core which takes into account the temperature dependence of the atmospheric heat capacity at constant pressure.

The model's core is based on version 4.5.1 of the UK Meteorological Office Portable Unified Model (UM). This Earth GCM is not the most recent version of the UK Met Office GCM at the time of writing but it was used in the past with success in numerical Earth weather prediction and for Earth climate research (e.g., \citealt{2000Stott}; \citealt{2008Reichler}), and it is still being used in the climateprediction.net project (e.g., \citealt{2013Yamazaki}). The structure of the model is divided into two main sections: a dynamical core and a library of physical parameterisations.  The main planetary parameters such as: the rotation rate, planetary radius, acceleration due to gravity, planetary obliquity and surface pressure, are the same as the ones used in \cite{2006Lee2} and \cite{2007Lee3}.

\subsection{Dynamical Core}
The dynamical core utilises a numerical formulation that solves the nonlinear primitive equations, which describes the resolved global atmospheric flow. Based on a finite-difference formulation, the dynamical equations used in this work were derived from \cite{1995White}, who solved the Navier-Stokes equations in a rotating spherical configuration using a less approximated form than is common for the meteorological primitive equations (``traditional approximation''). The model does not assume a shallow atmosphere approximation but retains a more complete representation of the full Coriolis acceleration and the planetary radius is replaced by a ``pseudo-radius'' which varies with the geopotential height.  The main adaptation to the original dynamical core is done in the `forward' step of the  `forward-backward' explicit routine used. The equations used in this scheme to update the zonal wind velocity and the geopotential height were changed from the original structure to take into account the variation of $c_p$ with temperature (these two equations depended explicitly on a constant $c_p$; see \citealt{2013Mendonca}). The observational results summarised in the Venus International Reference Atmosphere (VIRA, \citealt{1980Seiff}) indicate that the specific heat capacity at constant pressure (c$_p$) of the atmosphere varies from 1181 J~kg$^{-1}$K$^{-1}$ near the planet's surface to around half of this value for temperatures typically obtained near the top of our model domain (around 100 km). An analytic expression that evaluates efficiently the approximate dependence of c$_p$ with temperature was suggested by \cite{2010Lebonnois},
\begin{equation}
\label{eq:cp}
c_{p}= c_{p0}\times \Big(\frac{T}{T_{0}}\Big)^\nu
\end{equation}
where c$_{p0}$ =  1000 J~kg$^{-1}$K$^{-1}$, T$_{0}$ = 460 K and $\nu$ = 0.35. Some fundamental thermodynamic variables need to be redefined from their typical form if the dependence of c$_p$ with temperature is taken into account (for Earth studies, constant c$_{p}$ is a good approximation). For example, in \cite{2010Lebonnois} the potential temperature is derived using Equation (\ref{eq:cp}), giving the following result
\begin{equation}
\label{eq:pt}
\theta^{\nu} = T^{\nu} + \nu T_0^{\nu}\ln\Big(\frac{p_{ref}}{p}\Big)^{k_0}
\end{equation}
where the constants $\nu$, $k_0 = \frac{R}{C_{p0}}$ and $T_0$ are the same as in Equation (\ref{eq:cp}). The Exner function which converts absolute temperature into potential temperature in several parts of the GCM code was replaced by Eq. \ref{eq:pt}.

\begin{figure}
\centering
\includegraphics[width=0.55\textwidth]{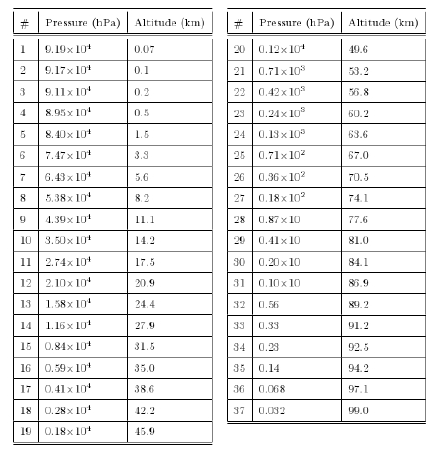}
\captionsetup{labelformat=empty}
\caption{Table 1: Vertical discretisation of the OPUS-Vr. The full level pressure values were obtained using the surface pressure equal to 92 bar (1 bar = $10^5$ Pa) and VIRA temperature profile to compute the altitudes.}
\label{tableopusvr}
\end{figure}

The typical horizontal and vertical resolutions are $5^\circ\times5^\circ$ for the horizontal and 37 layers for the vertical, which covers the atmosphere from the surface up to an altitude of roughly 100 km with a maximum vertical spacing of 3.5 km (see table \ref{tableopusvr} and \citealt{2013Mendonca}).

To ensure stability the model has a Fourier filter implemented to avoid the use of very short time steps constrained by the CFL condition at high latitudes, and horizontal numerical diffusion which controls the numerical noise and qualitatively represents the physical phenomena of turbulence and eddy viscosity on the sub-grid scale. The horizontal numerical diffusion applied was a sixth-order operator with a coefficient that controls the strength of the diffusion equals to $1.0\times10^9$m$^2$s$^{-\frac{1}{2}}$ (the same setting as in \citealt{2007Lee3}). Note that vertical diffusion is not used in our work. The time-step for dynamics in OPUS-Vr was typically five minutes. More details concerning the integration schemes at work in the dynamical core can be found in \cite{1992Cullen} and \cite{2013Mendonca}.  

\subsection{Physical parameterizations}
The new set of physical parameterization schemes is more complex and complete than the OPUS-Vs version (\citealt{2006Lee2}, \citealt{2007Lee3}, \citealt{2013Mendonca}).  Only the sponge layer parameterisation that acts at the top of the model's domain was maintained from the old version (e.g., \citealt{2006Lee2}). The physical time-step used in the simulations explored in this work is ten minutes, which is the same as the one used for the OPUS-Vs simulations presented in \cite{2006Lee2}. The prognostic variables in the dynamical core are updated two times per each physical time-step. In the next sub-sections we describe the new physical parameterisations developed for OPUS-Vr.

\subsubsection{Soil model}
OPUS-Vr includes a simplified parameterisation to simulate the evolution of the temperatures in the subsurface. The surface temperature is also affected by this formulation due to the exchanges of heat via thermal conduction in the soil. The conduction equation that is used to update the temperatures is
\begin{equation}
  \frac{\partial T}{\partial t} = -\frac{\lambda}{C}\frac{\partial^2 T}{\partial z^2},
\end{equation}
where $C$ is the specific heat per unit volume and $\lambda$ is the soil conductivity. The soil thermal inertia (I = $\sqrt{\lambda C}$) used is equal to 2200 Jm$^{-2}$K$^{-1}$s$^{-\frac{1}{2}}$ which corresponds to a basalt soil (\citealt{1999Rees}).

The parameterisation also uses a multilayer soil formulation (with ten layers) that was developed by \cite{1986Warrilow}, and produces accurate results for responses to variations of the temperature forcing at the surface. The code for this routine was adapted from the Oxford Mars GCM code but it was first developed and tested for the LMD Mars GCM (\citealt{1993Hourdin}).

\subsubsection{Boundary Layer}
The Rayleigh friction parameterisation used in OPUS-Vs to represent the mechanical interaction between the surface and the atmosphere is replaced in this new version by a more physically-based scheme. The code used in OPUS-Vr is an adaptation of one of the schemes available in the HadCM3 for the Earth (\citealt{1993Smitha} and \citealt{1993Smithb}). This boundary layer scheme is based on a bulk transport turbulent mixing parameterisation (\citealt{2005Jacobson}). This code was already explored under Venus conditions by \cite{2006Lee2}, but here the parameterisation was revised and adapted to take into account the variations of the atmospheric $c_p$ with temperature (e.g., in the computation of the layer depths). Further improvements are related to the direct interaction of this code with the radiative transfer model and the use of the surface thermodynamic properties from the simplified soil model described above. More details concerning the boundary layer model and scientific description can be found in \cite{1993Smitha}, \cite{1993Smithb} and \cite{2013Mendonca}.

At the three top layers of the model's domain an identical sponge parameterisation to the one used in the OPUS-Vs is used to damp the eddy component of the wind field only to zero (\citealt{2006Lee2}).

\subsubsection{Radiative transfer model}
The radiative transfer code implemented in the OPUS-Vr is the same as the one presented by \cite{2015Mendonca}. This formulation computes the radiative heating and cooling rates in the atmosphere and is very flexible to adapt to different atmospheric conditions. This code, suitable for climate studies, is divided into two parts that work in two distinct spectral ranges: solar radiation (0.1-5.5$\mu$m) and thermal radiation (1.7-260$\mu$m).

The solar radiation code is a two-stream-type scheme based on an adding layer method and a $\delta$-Eddington approximation from \cite{1992Briegleb}. These two methods represent the multiple scattering phenomena. The $\delta$-Eddington approximation scheme calculates the transmissivity and reflectivity of each layer in a vertical column, that are later combined using the adding layer method. See more details about these two methods in \cite{1992Briegleb} or in \cite{2015Mendonca}. The spectral range analysed is divided into fifty-five bands with 0.1 $\mu$m resolution each. The physical properties in each layer are assumed to be horizontally homogenous. Several components of the atmosphere that interact with the solar radiation were taken into account: three gases (CO$_2$, H$_2$O and SO$_2$), three size distribution modes of cloud droplets assumed to be 75$\%$ sulphuric acid and $25\%$ water, a UV absorber component and the Rayleigh scattering effect due to the presence of CO$_{2}$ and N$_{2}$ molecules. The volume mixing ratios assumed for the gases were taken from the VIRA model (\textbf{V}enus \textbf{I}nternational \textbf{R}eference \textbf{A}tmosphere, \citealt{1985Kliore}) and the cloud amount from \cite{1986Crisp}. 

The amount of incoming solar radiation at the top of the vertical column for a specific band and the length of the optical paths are controlled by the solar zenith angle. This angle was corrected to include the effect of the atmospheric spherical curvature. The equation from \cite{1967Rodgers} gives a good estimate of the real zenith angle, taking into account the effective solar path length. This formulation has been used with some success in Earth and Mars GCMs (e.g., \citealt{1967Rodgers} and \citealt{1993Hourdin}), and it is also a good approximation for the Venus case because of the similar planet dimensions,
\begin{equation}
\frac{1}{\mu}=\frac{35}{\sqrt{1224\times\mu_0 +1}}
\end{equation}
where $\mu$ is the effective zenith angle cosine corrected and $\mu_0$ is the original value (not corrected).

The final total solar upward and downward spectral fluxes at the interface of each atmospheric layer are combined in each layer to compute the solar heating rates, which depends on $c_p$ varying with temperature as well. The solar heating rates are updated typically every 1.5 Earth days during the OPUS-Vr simulations. 

The thermal radiation code is based on an absorptivity/emissivity formulation (neglecting scattering). This method combines all the thermal emission from isolated layers of a vertical column to compute total upward and downward thermal fluxes at each spectral band. The angular integration is done in this parameterisation using a diffusivity factor that depends on the layer optical depth. The spectral range analysed is divided into two hundred and seventy three bands with variable resolution. 

The physical properties in each layer are  assumed to be horizontally homogenous as in the solar radiation scheme. The optical properties of three gases (CO$_2$, H$_2$O and SO$_2$), three size distribution modes of cloud droplets assumed to be 75$\%$ sulphuric acid and $25\%$ water and the continuum absorption due to CO$_{2}$-CO$_{2}$  and H$_2$O-H$_2$O collisions, were taken into account in this scheme. The details on the opacity data used are explained in \cite{2015Mendonca}. In this work there was no need to increase empirically the magnitude of the collision-induced absorption coefficients of CO$_2$ as suggested in \cite{2009Eymet} to increase the surface temperature. The cloud structure used in our work was adopted from \cite{1986Crisp} (for more information you can also read \citealt{2015Mendonca}). The cloud deck is divided in three parts: lower (1000 - 1400 hPa), middle (380-1000 hPa) and upper (36-380 hPa) clouds. The optical properties of the three cloud modes including the empirical UV absorber were also calculated from \cite{1986Crisp}. In this step, we interpolate the scattering and extinction efficiencies from  \cite{1986Crisp} to our spectral resolution as we explain in \cite{2015Mendonca}. The UV absorber is distribuited in the upper cloud region and the empirical absorption computed following the method suggested in \cite{1986Crisp} to match the solar flux radiometer aboard the Pioneer Venus probe (see also \citealt{2015Mendonca}).

As in the solar radiation scheme the thermal cooling rates are calculated in this parameterisation from the total flux differences across each layer. These rates are updated every 1.5 Earth hours during the OPUS-Vr simulations. This code has a routine implemented which reduces the number of spectral bands from a binning average process. In the simulations studied in this work, a factor of four in the binning average process was chosen. This procedure saved a significant fraction of the GCM's integration time and did not compromise the accuracy of the calculated radiative cooling rates (\citealt{2013Mendonca} and \citealt{2015Mendonca}).

This model thus includes a complete radiative transfer formulation, which makes it very flexible to adapt to different atmospheric conditions and to calculate more realistically the radiative cooling and heating rates in the atmosphere and surface. The flexibility of our model is an improvement from recent Venus GCMs such as the one from the LMD Venus GCM (\citealt{2010Lebonnois}), which is restricted to using precomputed solar tables, and an off-line IR matrix with fixed opacity distribution.

\subsubsection{Convection}
The convection parameterisation implemented in the OPUS-Vr is the same as the convective adjustment routine used in \cite{2015Mendonca}. This new formulation for a dry atmosphere was developed taking into account that, during the mixing of the potential temperature in unstable columns ($\delta \theta/\delta p > 0$), the total specific entropy of the system is conserved. In this parameterisation the dependence of $c_p$ with temperature is also included.

\section{Reference Simulation}
\label{sec:Dbaseline}

We initialized the reference simulation with the atmospheric composition and surface described in previous section. The surface short-wave albedo is assumed to be heuristically high for the reference simulation, 0.95, and the reason for the high value is justified in section \ref{sec:SurfAlb}. The planet surface is flat and the atmosphere started from a rest state with every vertical temperature profile close to the VIRA profile (\citealt{1980Seiff}), and with a surface pressure at 92 bars (1 mb = 1 hPa). The results presented in this work were very similar to the ones with a real topography. In order to simplify the analysis of the results and focus on the main results presented here, we have decided not to include the results with topography. The distribution of the clouds is assumed to be constant in the longitude and latitude directions. 

The available potential energy stored in the atmosphere is converted to kinetic energy during the initial model integration, which is forced by radiation processes, convection and by instabilities induced by the rotation of the planet. This conversion starts the flow moving in the stably stratified atmosphere at rest. The term ``spin-up phase'' is applied to GCMs to describe the time period required by these complex models to reach a statistical equilibrium state (defined in the Venus case as when the global kinetic energy and the total axial angular momentum converges to a statistically steady value as a function of time). The period of integration chosen, 215.5 Venus days (25000 Earth days), has been shown by other works to be enough for representing the spin-up phase typically required by the Venus GCMs (e.g., \citealt{2006Lee2} and \citealt{2010Lebonnois}). This long period is associated with the large time-scale for thermal adjustment of the deep atmosphere. In our work a Venus day refers to a Venus solar day with 116.7 Earth days.

The global super-rotation index ($S$, Eq. \ref{eqn:S}) in the last 42 Venus days of the long integration tended to stabilise at 0.18 magnitude, which indicates that a statistical steady state was reached. The variable $S$ depends on the total angular momentum of the atmosphere, which is dynamically redistributed in the atmosphere by atmospheric advection, modified by the action of waves, and is created or destroyed by the mechanical interaction of the atmosphere with the surface.

\begin{figure}
\centering
\subfigure[Global mean temperature.]{\label{fig:ux-tempg}\includegraphics[width=0.45\textwidth]{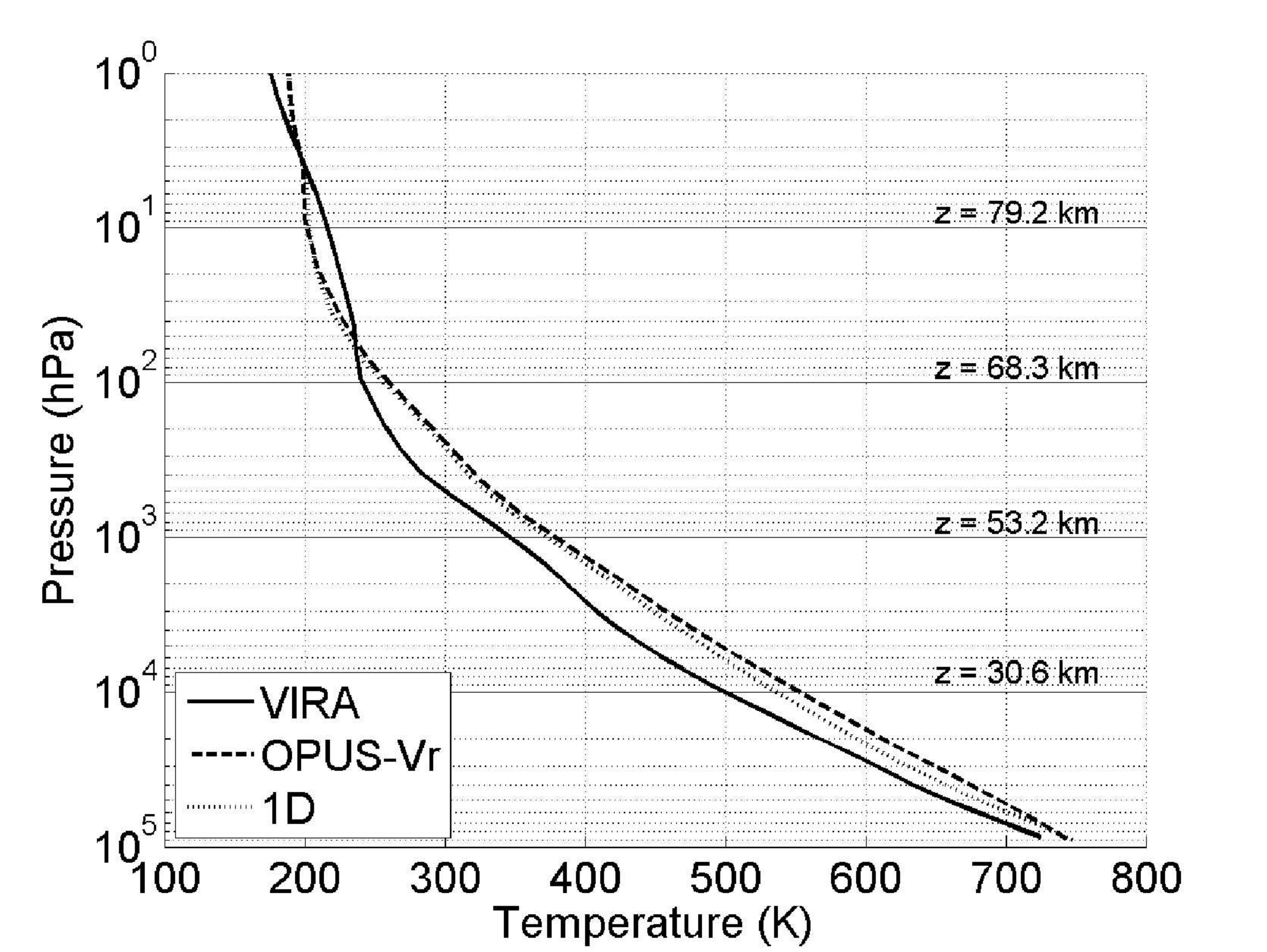}}
\subfigure[Stability.]{\label{fig:ux-sta}\includegraphics[width=0.45\textwidth]{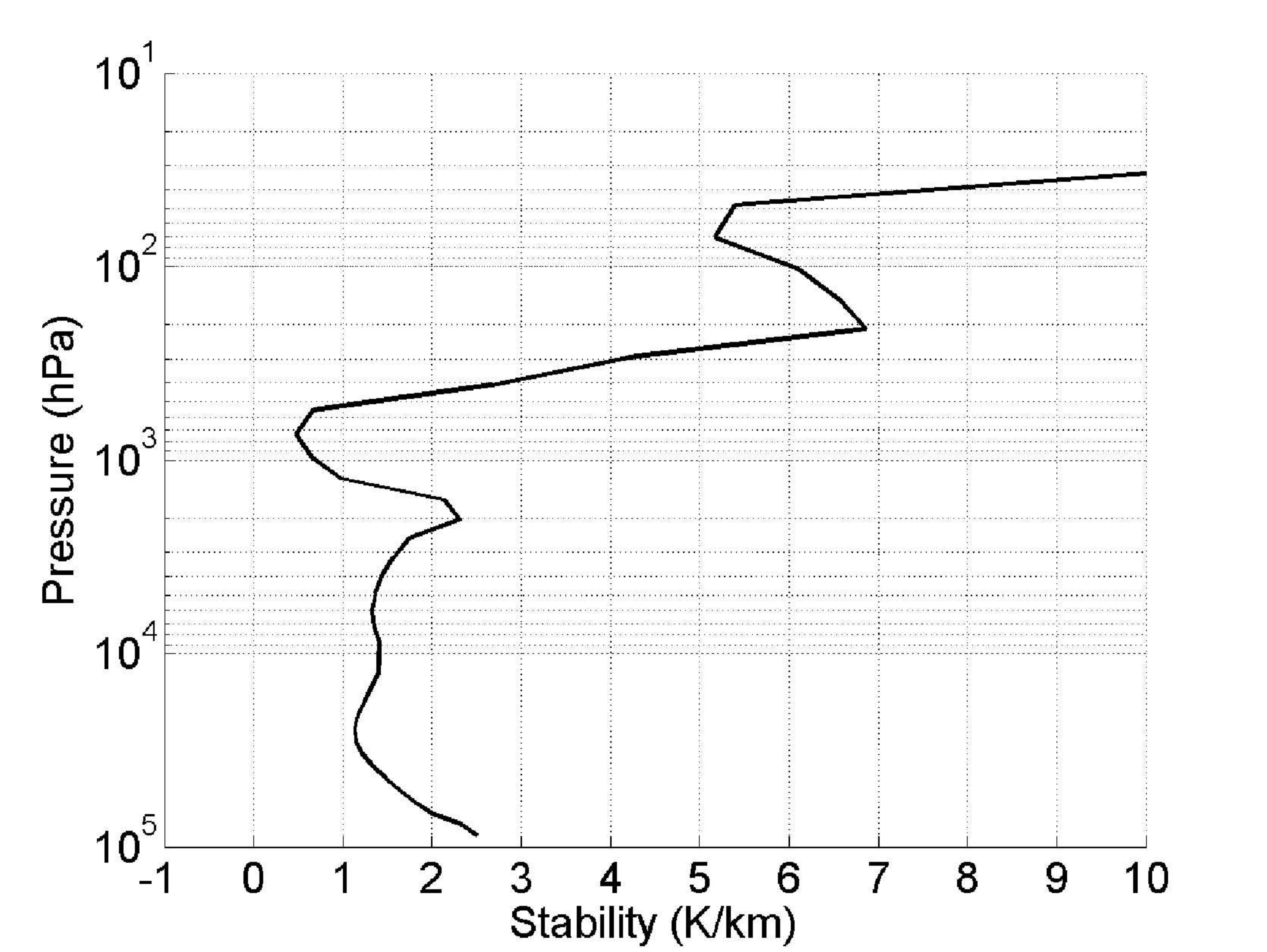}}
\subfigure[Mean zonal winds.]{\label{fig:ux-u}\includegraphics[width=0.45\textwidth]{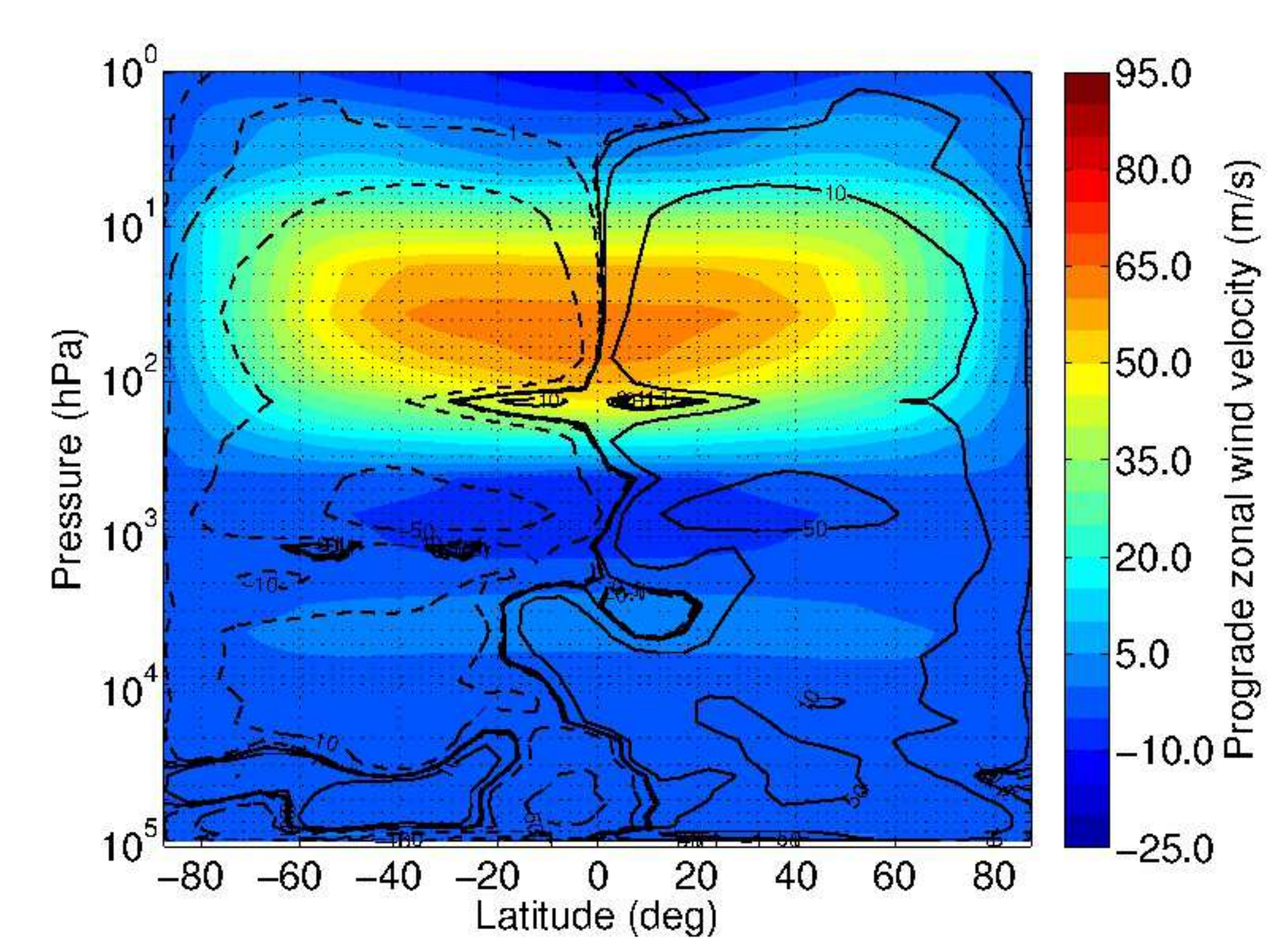}}
\subfigure[Temperature anomaly.]{\label{fig:ux-tempa}\includegraphics[width=0.45\textwidth]{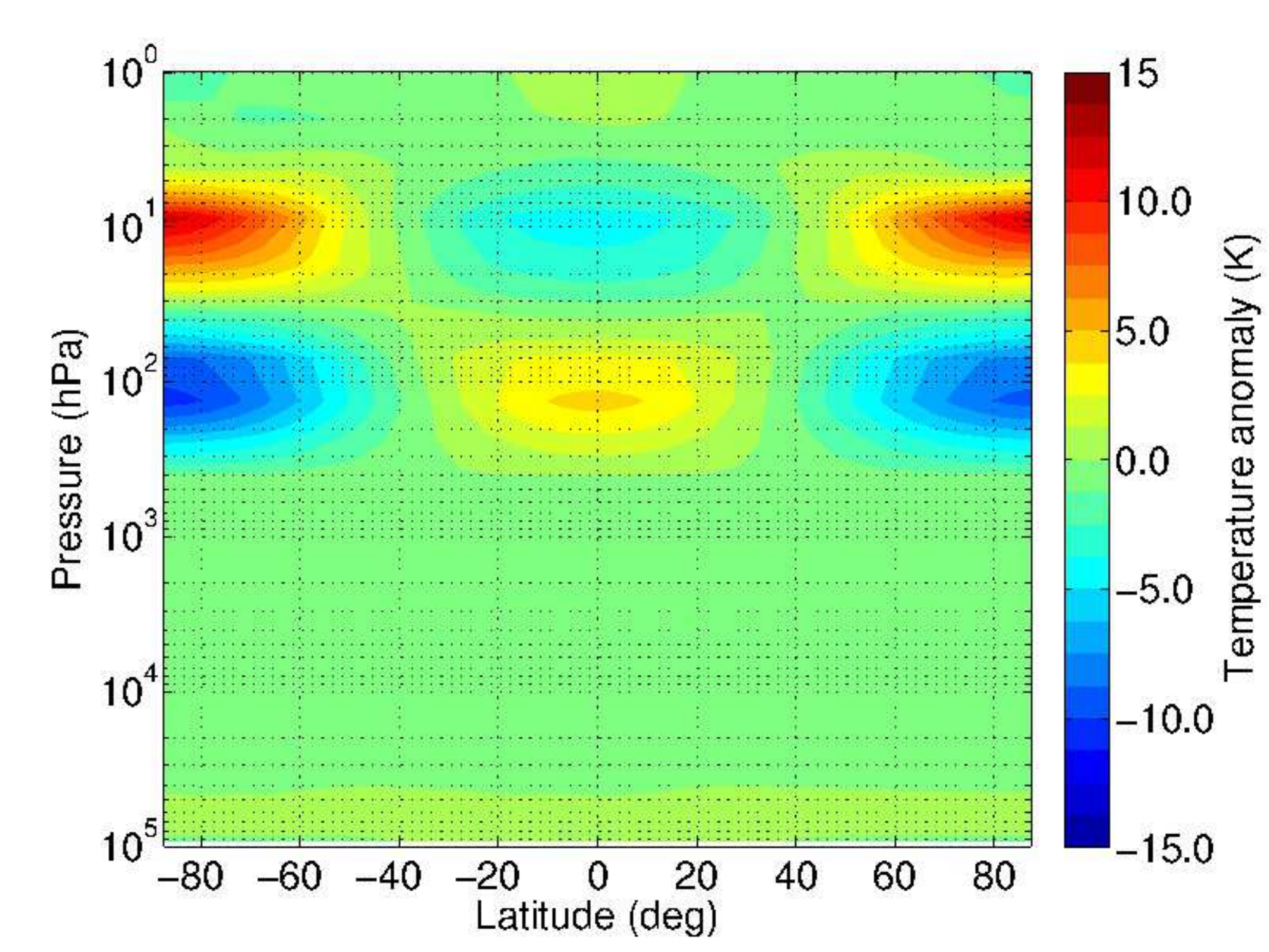}}
\caption[Reference results.]{ Reference results - Global mean profiles of stability and temperature and mean zonal winds and temperature anomaly maps. \textbf{(a)} and \textbf{(b)} are horizontal and time averages (over five Venus days) of absolute temperature and atmospheric static stability ($d\theta/dz$, where $\theta$ is the potential temperature). \textbf{(c)} and \textbf{(d)} are the zonal and time  averaged (over five Venus days) zonal winds and temperature anomaly maps obtained around day 214. The contours in \textbf{(c)} represent the averaged mass stream function (in units of $10^9$ kg/s).}
\label{ux-12}
\end{figure}

The mean vertical temperature profile shown in Fig. \ref{ux-12}(a) is very similar to the one obtained by the 1D radiative-convective model presented in \cite{2015Mendonca}. This 1D model includes the same radiation and convective adjustment schemes presented above, and it also uses the same atmospheric structure. The discrepancies between the modelled temperature profiles and the VIRA temperature profile are related to uncertainties in the optical properties of the gas and clouds as demonstrated e.g., in \cite{2015Lebonnois}. Fig. \ref{ux-12}(d) shows the mean temperature anomaly map. The temperature values ($T_a$) in this map are deviations from the area-weighted horizontal mean on pressure levels. The latitudinal variations in temperature shown in this map are largely influenced by the atmospheric circulation produced. The very small 1-2 K variations in the atmosphere at deeper levels than the $10^3$ hPa pressure level are consistent with the results obtained by the LMD Venus GCM (\citealt{2010Lebonnois}) but much smaller in comparison with the observations, which can evidently reach equator-to-pole differences of 30K (\citealt{2009Tellmann}). The zonal winds produced in this region and shown in  Fig. \ref{ux-12}(c) are also small, with the zonal winds reaching maxima of 1-2 m/s. The atmosphere in this region is characterized by a weak static stability (Fig. \ref{ux-12}b) with two relative maxima of stability, one near the surface and another one at roughly $2\times10^3$ hPa ($\sim 45$ km). See Fig. 2 of \cite{2007Zasova} for a comparison with the measured stability profile. The atmospheric circulation near the surface is not symmetric about the equator in Fig. \ref{ux-12}(c), but is dominated by the turbulent influence of the interchanges between the surface and atmosphere.

A zonal wind equatorial maximum ($\sim$3 m s$^{-1}$) is obtained at $5\times10^3$ hPa, which is close to a relative minimum in the atmospheric stability profile. Despite the existence of this equatorial maximum, the magnitude of the winds are much weaker than the values observed. Possible explanations for the poor representation of the atmospheric circulation in the lower atmosphere are discussed in section \ref{sec:dics}.

At $10^3$ hPa there is a region of weak stability, which is associated with the location of the lower cloud deck. In this region most of the upwelling infrared radiative energy from the deepest regions of the atmosphere is blocked by the clouds. The location of the weakly stable atmospheric region is in agreement with recent observational results of the VeRa experiment (\citealt{2009Tellmann}) and also with the LMD Venus GCM (\citealt{2010Lebonnois}). This region of weak stability is a region where upward convection motions are likely to occur. 

Going up in altitude, we find just above 10$^3$ hPa a region with weak retrograde winds. The formation of these retrograde winds is associated with the momentum transport by waves mainly excited in the upper cloud region and instabilities generated from the convection formed at the cloud base ( Fig. \ref{fig:fil_eli}). We will come back to this subject later on, when we discuss the transport of axial angular momentum and analyze the Eliassen-Palm flux maps (section \ref{sec:MT}).

Above the pressure level of 500 hPa that is located just over the convective layer in lower/middle cloud region, the zonal winds start to increase their magnitude with altitude, reaching a maximum of roughly 60 m/s in the equatorial region at 70 hPa.  This zonal wind pattern is consistent with a cyclostrophic regime. From the thermal wind equation we expect that in the region where there is a negative poleward gradient in temperature, the zonal winds increase with height (70-500 hPa). The gradient of temperature is reversed above 70 hPa and the zonal wind then decreases in magnitude with increasing height. The hotter polar region above 70 hPa is consistent with the model results of \cite{2010Lebonnois} and observational data (e.g., \citealt{1980Taylor}; \citealt{2007Zasova}; \citealt{2010Grassi}), and is likely a product of adiabatic heating from air compression into the poleward branch of the equator-to-pole Hadley cell (associated with a maximum temperature difference of 20 K between high and low latitudes at the same pressure level). From the mass streamfunction contours it is also possible to infer upward flow at low latitudes, which becomes stronger largely due to the significant absorption of incoming solar radition in the upper cloud region. 

\begin{figure}
\centering
\subfigure[Day = 173]{\label{fig:ux-alb-1}\includegraphics[width=0.4\textwidth]{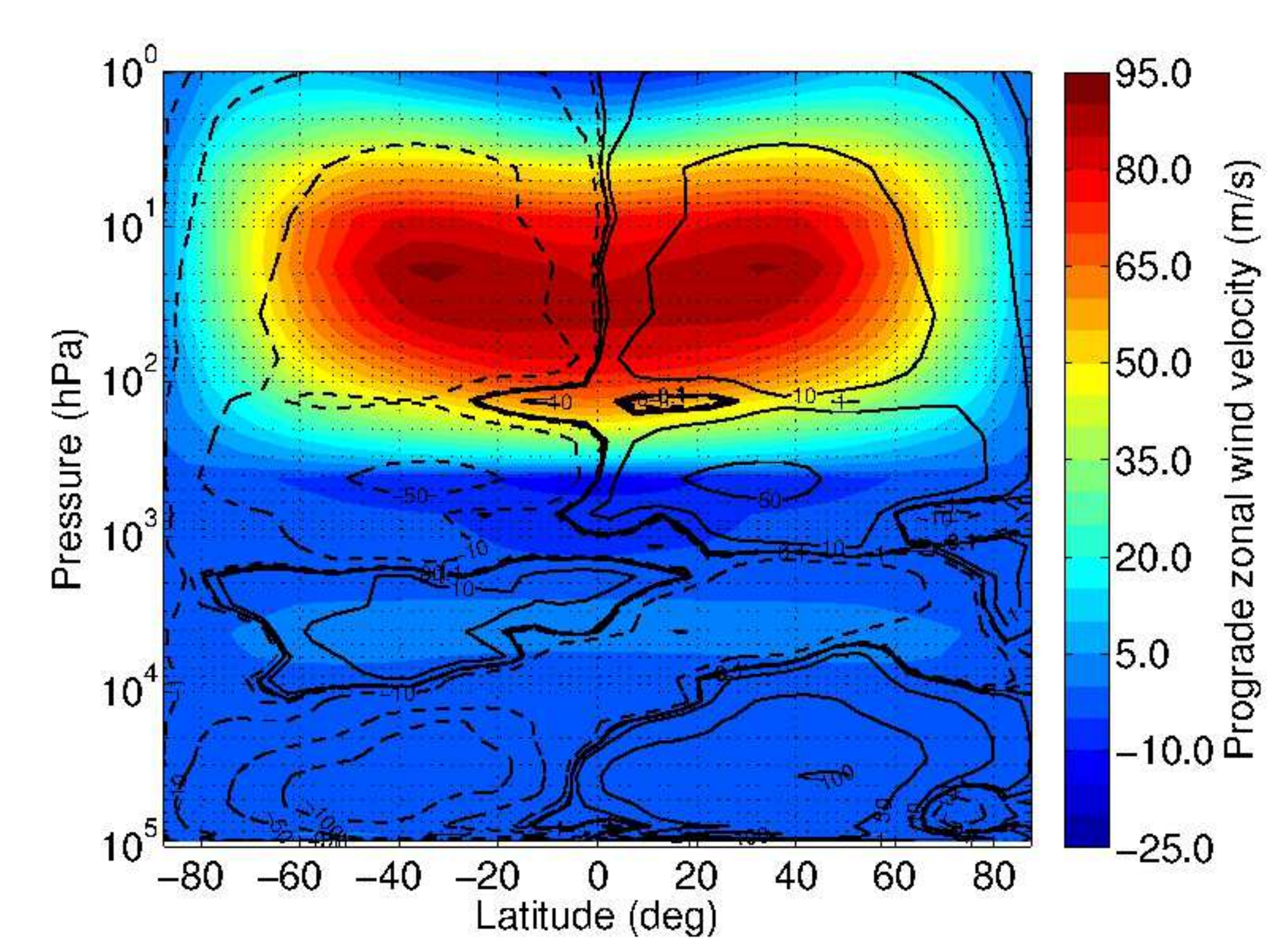}}
\subfigure[Day = 214]{\label{fig:ux-alb-2}\includegraphics[width=0.4\textwidth]{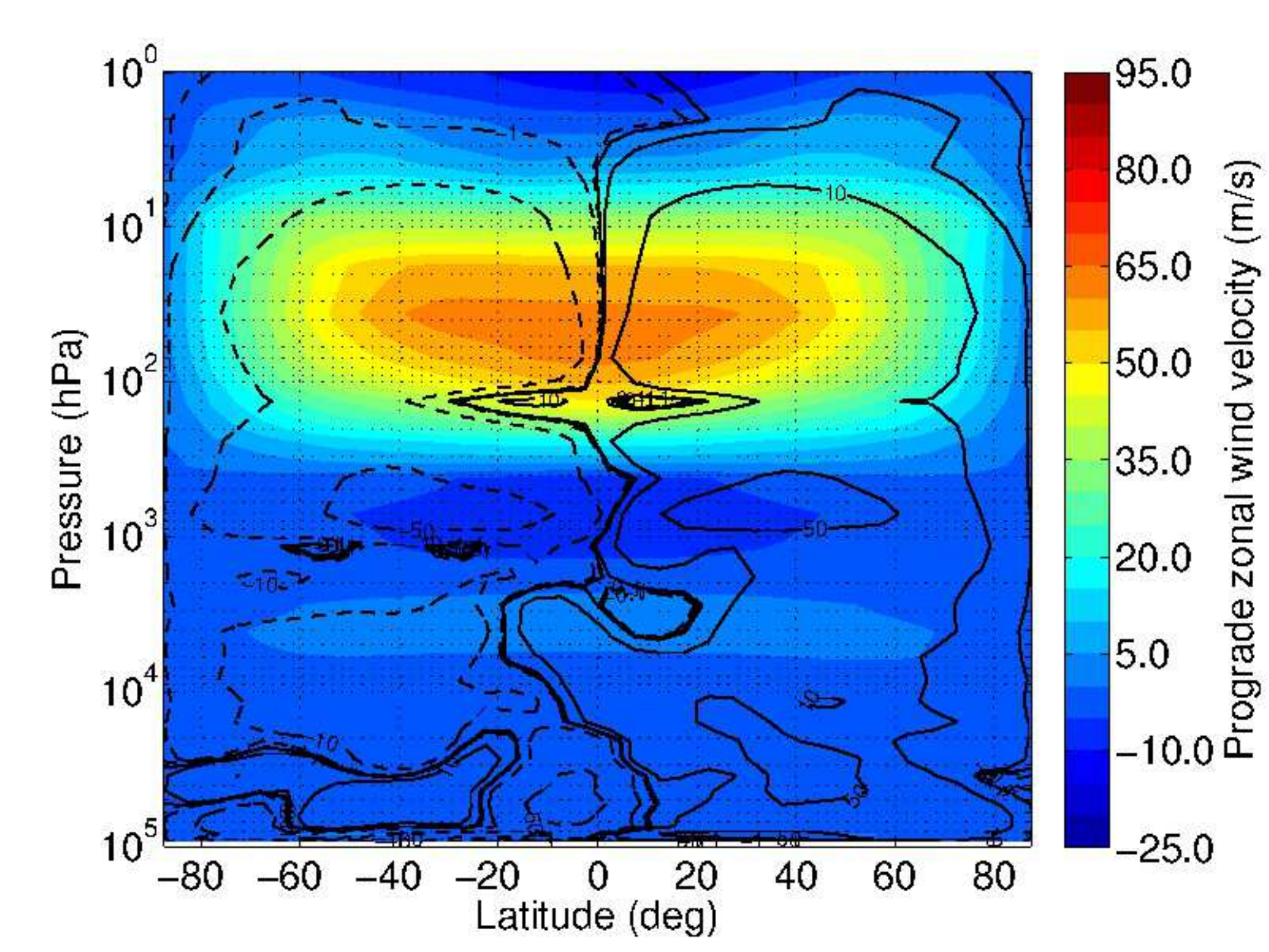}}
\subfigure[Day = 223]{\label{fig:ux-alb-3}\includegraphics[width=0.4\textwidth]{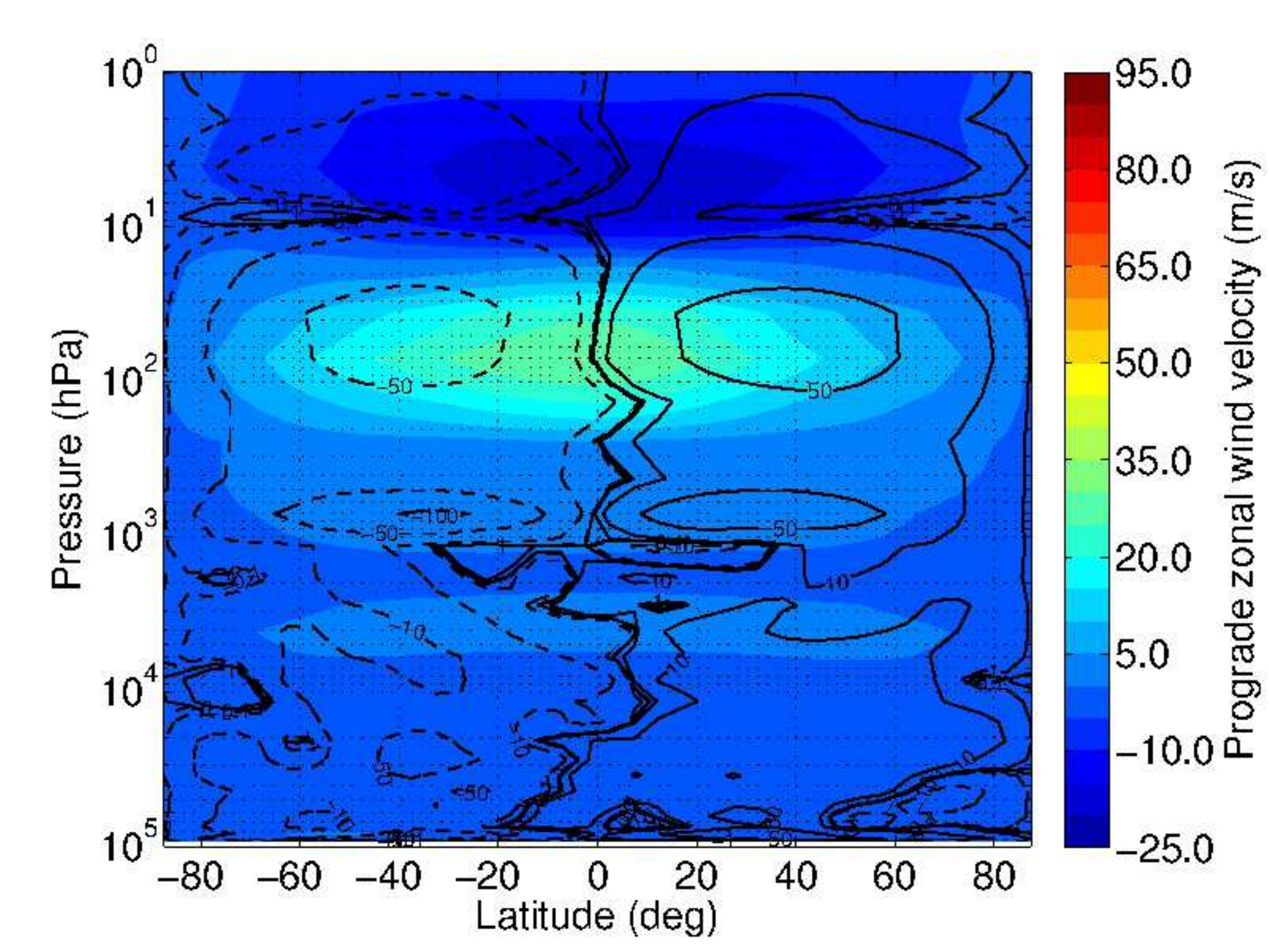}}
\subfigure[Day = 244]{\label{fig:ux-alb-4}\includegraphics[width=0.4\textwidth]{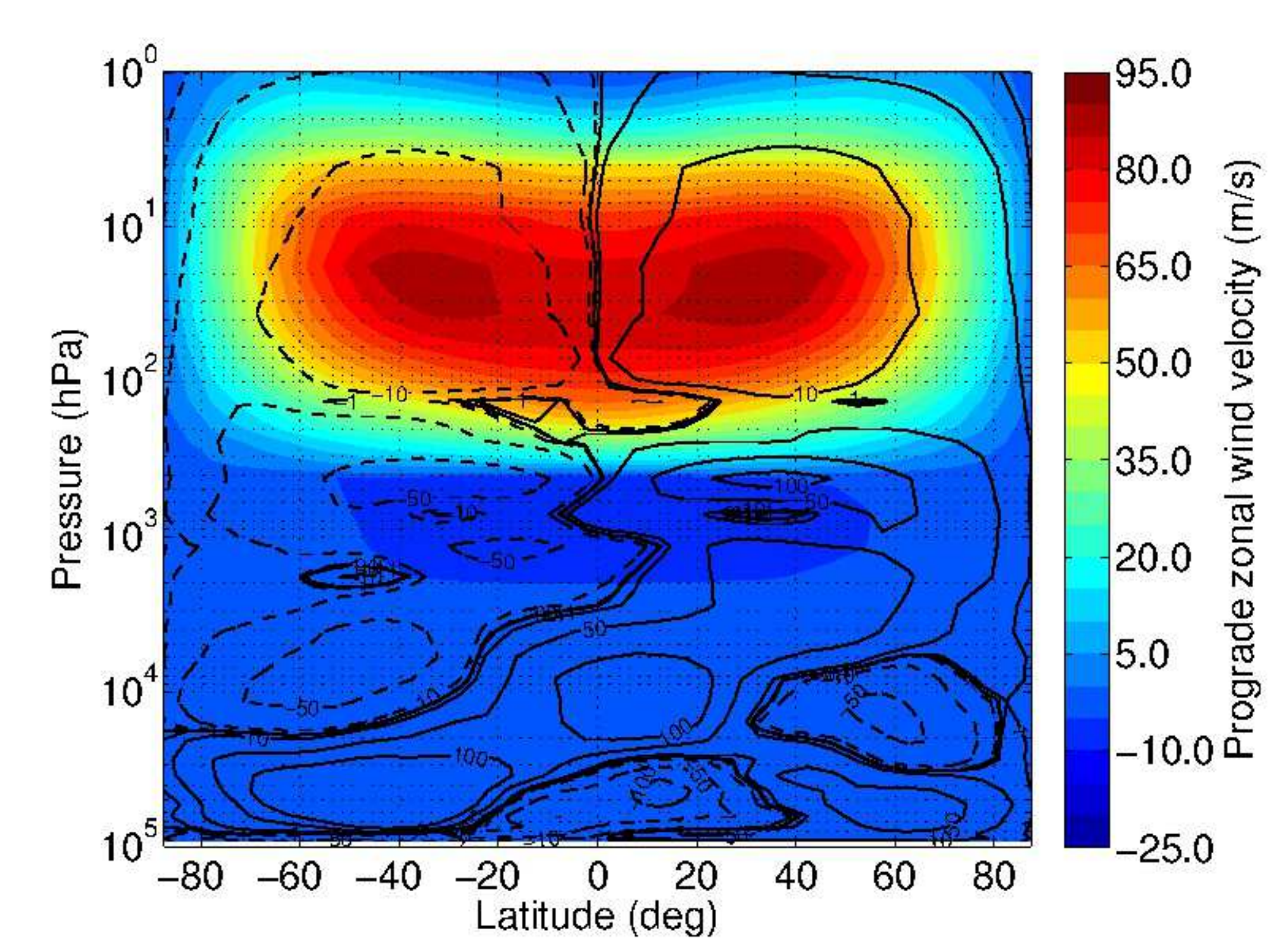}}
\caption[Zonal winds and mass stream function of the reference simulation at four different days.]{Zonally and time averaged (over 5 Venus days) zonal wind (m/s) and mass stream function (in units of $10^9$kg/s) of the reference simulation at four different days: (\textbf{a}) 173, (\textbf{b}) 214, (\textbf{c}) 223 and (\textbf{d}) 244.}
\label{fig:ux-3}
\end{figure}

Although our time sampling is not extensive, we see an indication of a cyclic behaviour as the distribution and strength of the winds in the cloud region occurring in day 173 is repeated later at day 244, as can be seen from the four maps of Fig. \ref{fig:ux-3}. In this figure the reference simulation was extended by 30 Earth days to have a more detailed analysis of the complete cycle. The wind profile shown at day 214 is very similar to that shown in \cite{2010Lebonnois}. The weakening/strengthening of the prograde winds in the upper cloud region is followed by a strengthening/weakening of the upward mean circulation in the same region. This long term variability of the zonal winds in the cloud region is related to a large-scale variation in the atmospheric circulation pattern in the lower atmosphere. Clear signs in these maps of the variation in the lower atmosphere are evident in the anti-symmetric patterns of the mass stream function below $10^3$ hPa between the days 173 and 244, and in the formation of a weak prograde equatorial jet at 5$\times 10^3$ hPa in the intermediate days. 

\subsection{Comparing with observational wind data} 
\label{sec:obs}
\begin{figure}
\centering
\subfigure[Vertical profile.]{\label{fig:ux-comp-v2}\includegraphics[width=0.49\textwidth]{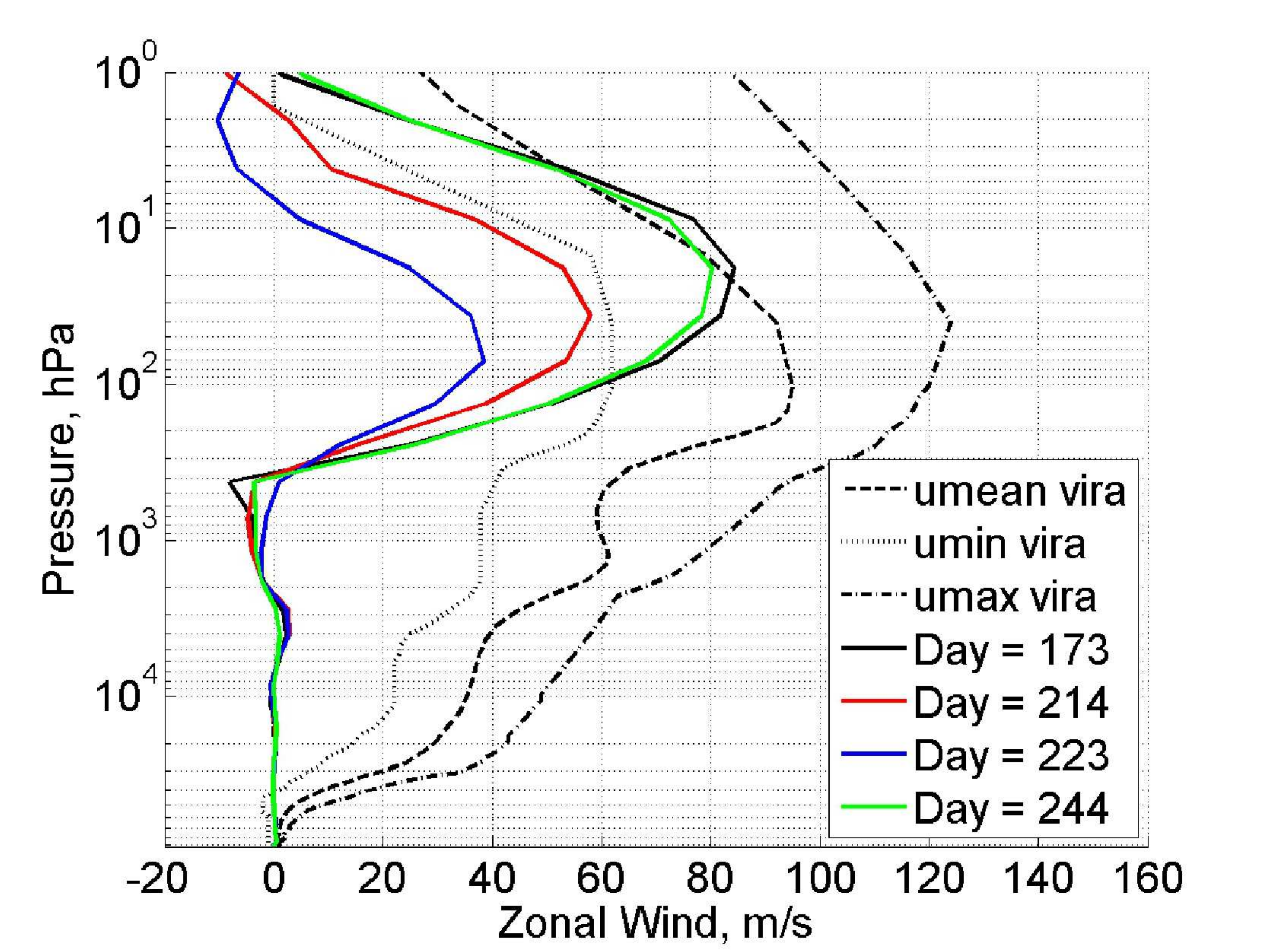}}
\subfigure[Latitudinal profile.]{\label{fig:ux-comp-h2}\includegraphics[width=0.49\textwidth]{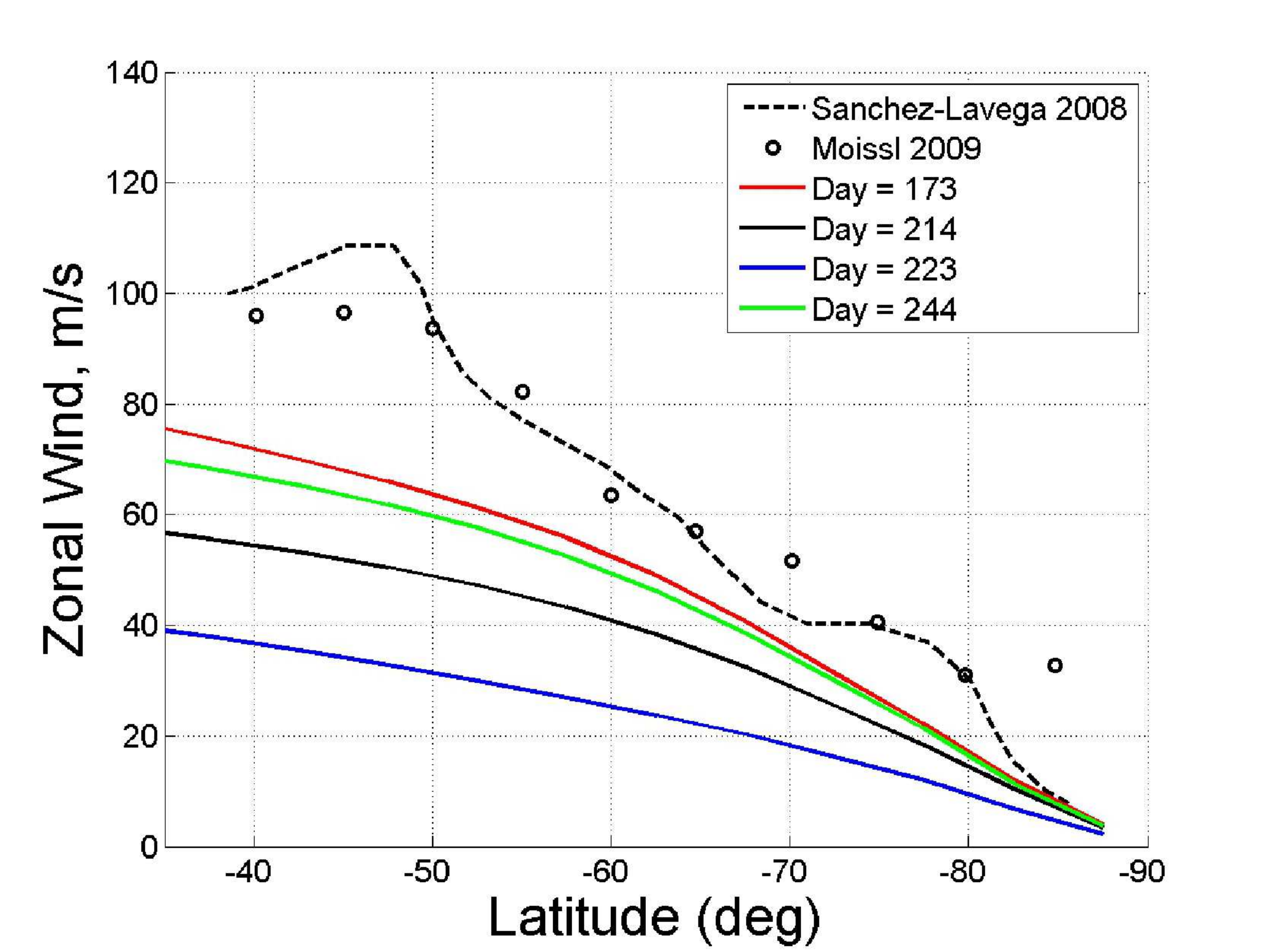}}
\caption[Vertical and latitudinal profiles of zonal winds, comparing the different days results obtained in the reference simulation with the observations.]{Vertical and latitudinal profiles of zonal winds, comparing the results for different days obtained in the reference simulation with the observations. The observational data from the plot \textbf{(a)} was available in \cite{1985Kerzhanovich} and the mean winds from cloud tracking results at roughly 70 km shown in \textbf{(b)} were from \cite{2008Lavega} and \cite{2009Moissl}. In both figures the model's mean zonal winds were zonally and time averaged for five Venus days. In the vertical profile the simulated winds were just averaged between 40$^\circ$S and 40$^\circ$N to be consistent with the observational data.}
\label{ux-comp-u2}
\end{figure}

In Fig. \ref{ux-comp-u2}(a) we used available observational data from a reference model of the Venus atmospheric circulation (\citealt{1985Kerzhanovich}) to do a direct comparison with the reference results obtained with OPUS-Vr at different simulation days. The OPUS-Vr values compared refer to vertical profiles of zonally and time averaged (over five Venus days) zonal winds between the surface and 1 hPa. The observational data correspond to the latitudinal region between 40$^\circ$S and 40$^\circ$N, and our model values were also averaged in that region. The winds below 100 hPa are consistently weaker than the observations. On days 173 and 244, however, the zonal wind magnitudes above pressure level 100 hPa are in good agreement with the observations. These two days are in the phase of the long term oscillation where stronger zonal winds are obtained. There is also a good correspondence between the relative maxima altitudes in the observations and the values modelled.

Oscillations with periods similar to the one found here (tens of Venus days or longer) in the Venus atmospheric state are not new in numerical or observational studies. \cite{2011Parish} used a GCM with physical routines similar to the ones from OPUS-Vs (\citealt{2006Lee2}) and high horizontal spatial resolution ($1^\circ\times1^\circ$), and found large scale oscillations in the zonal wind distribution with a period of ten Earth years. In those results, the largest variations were located below the cloud base, where the zonal winds varied with a period of ten Earth years between super-rotation and sub-rotation.  From the observational side, indications of long term oscillations were found (on periods of tens of Venus days) in the zonal flow, temperature and composition, in the cloud region and above it: e.g., \cite{1990Delgenio}, \cite{2008Belyaev} and \cite{2012Clancy}.

Using more recent observational data from Venus Express, we compared results from cloud tracking methods (\citealt{2008Lavega} and \citealt{2009Moissl}) that retrieved zonal winds at the cloud top ($\sim$ 70 km), with the OPUS-Vr latitudinal profile of the zonal winds at roughly the same altitude. As would be expected from Fig. \ref{ux-comp-u2}(a), Fig. \ref{ux-comp-u2}(b) shows a better agreement for the days 173 and 244 in comparison to values retrieved from cloud motions at $\sim70$ km.  The values modelled have a latitudinal gradient of the zonal momentum very similar to solid body rotation. These results suggest the presence of atmospheric wave activity such as barotropic eddies, which tend to mix the atmosphere towards a state of approximately uniform absolute vorticity  (\citealt{1999Schubert}; \citealt{2012Mendonca}), with consequent implications for the distribution of axial angular momentum and other variables.

\begin{figure}
\centering
\subfigure[Zonal wind Vs Time.]{\label{fig:ux-var1}\includegraphics[width=0.49\textwidth]{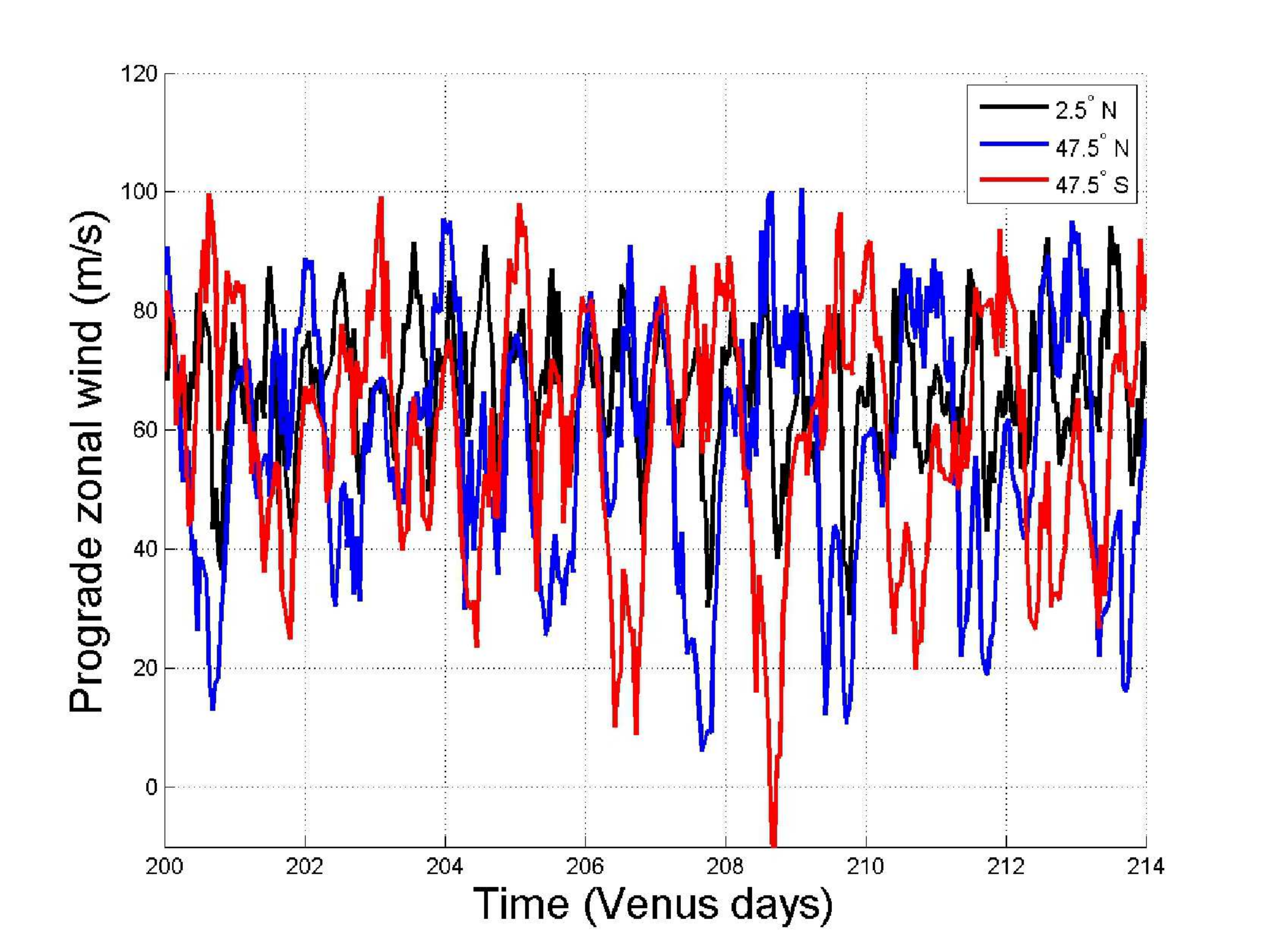}}
\subfigure[Spectrum.]{\label{fig:ux-var2}\includegraphics[width=0.49\textwidth]{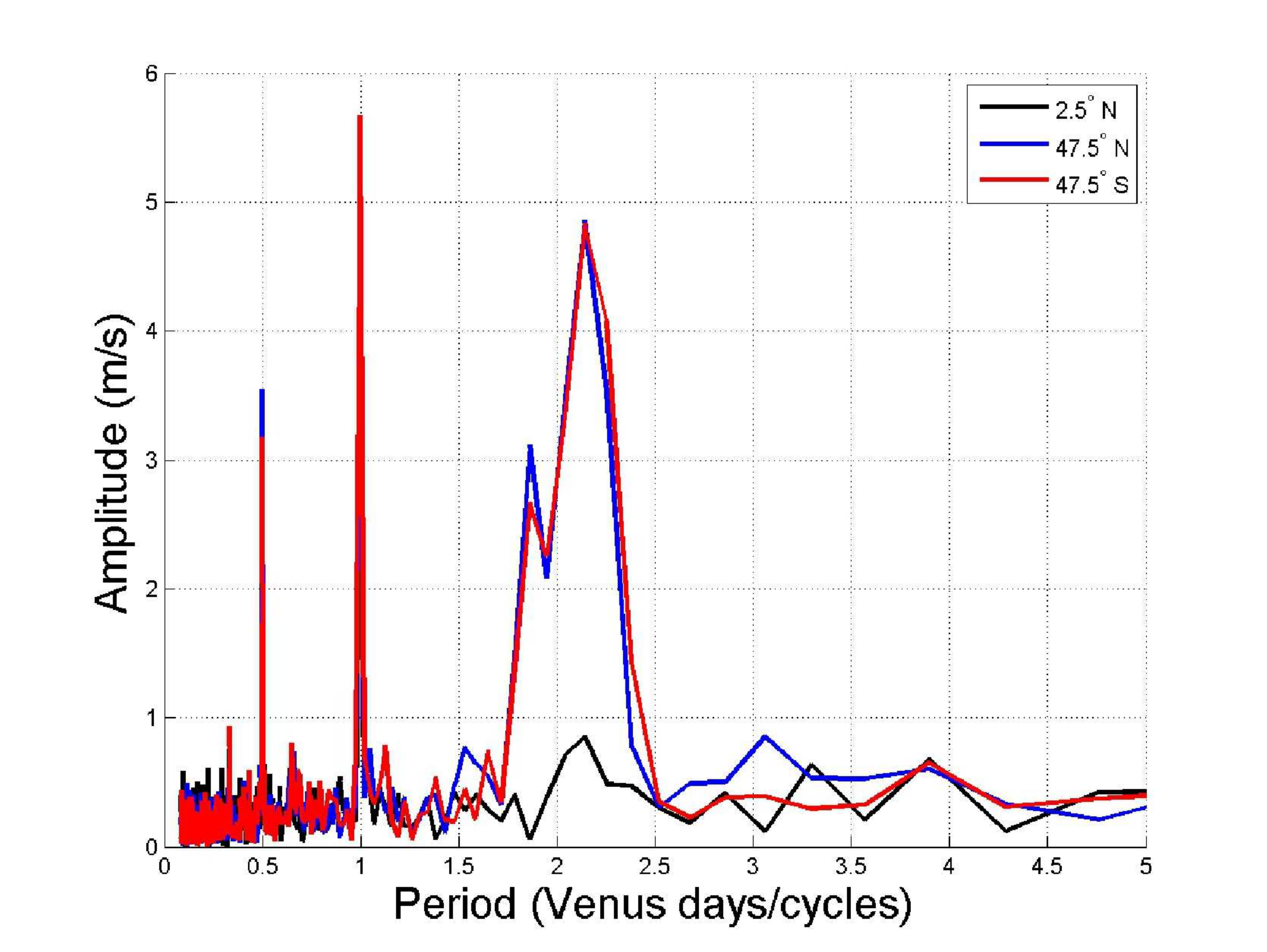}}
\caption[Zonal wind time series and spectrum from the reference simulation.]{ Zonal wind time series and spectrum from the reference simulation. The time series were retrieved from three locations at 36 hPa and 0$^{\circ}$ longitude: two mid-latitude points (47.5$^{\circ}$S and 47.5$^{\circ}$N) and one near the equator (2.5$^{\circ}$S). These results correspond to data obtained in the last fourteen Venus days of the reference simulation.}
\label{fig:ux-var}
\end{figure}

In Fig. \ref{fig:ux-var}(a), we show three time series of the winds in the zonal direction in three different geographical locations. These points are at the same pressure level of 36 hPa (cloud top in the model) and longitude of 0$^{\circ}$, but different latitudes: two mid-latitude points (47.5$^{\circ}$S and 47.5$^{\circ}$N) and one near the equator (2.5$^{\circ}$S). We can see a periodically varying wind, superimposed on a nearly constant background wind field. In this sample, which corresponds to the last fourteen Venus days of the simulation, the phase difference between the maxima of the two mid latitude points seems to be roughly $\pi$ most of the time. Upon studying this variability in more detail we compute a Fourier-transform analysis of those winds and the results are shown in Fig. \ref{fig:ux-var}(b). For the three locations the influence of the thermal tides is large, with the amplitude of the diurnal tide (with a period of one Venus day) being roughly 5.5 m s$^{-1}$ and the semidiurnal tide roughly 3.5 m s$^{-1}$. Two broad peaks show up at mid latitude locations (mid latitude jets) with periods of 1.86 and 2.14 Venus days. The causes of these long term oscillations are explored in a later wave analysis section. In a recent work from \cite{2013Khatuntsev}, long-term wind variations and periodicities were analysed using a cloud-tracking method covering a period of ten Venus years of data from the Venus Monitoring Camera (VMC) instrument aboard the Venus Express probe. The data revealed that, from a Fourier analysis of the wind field, there are periodicities related to the thermal tides (diurnal and semidiurnal tides), faster oscillations of the jets with periods of roughly 4.8 Earth days, and three long-term periodicities of 22.1, 2.05 and 1.90 Venus days. The period of 22.1 Venus days detected was not discussed in \cite{2013Khatuntsev} and its existence is arguable since the period is longer than the Venus Express mission. The two other maxima in the spectrum (2.05 and 1.90 Venus days), despite appearing to be reproduced in the results of the OPUS-Vr, were claimed in \cite{2013Khatuntsev} to be the result of the sampling periodicities in the VMC observations, which makes it difficult to compare the observations and simulation results on this topic.

\subsection{Momentum transport} 
\label{sec:MT}
To gain a better understanding of the main momentum transport mechanisms responsible for the simulated  atmospheric circulation, we calculated the zonally averaged transport of angular momentum separated into three different contributions: mean circulation [A], stationary waves [B] and transient waves [C]. Note that in this analysis we are not taking into account the mechanical transport due to the interaction of the surface with the atmosphere. A representation of momentum mixing during convection is not included in this reference simulation, which was shown in \cite{2013Mendonca} to have a low impact on the atmospheric circulation compared to the other mechanisms. To analyse the total meridional transport of angular momentum ($M$) in the reference results we used the following equation (\citealt{1992Peixoto}):
\begin{equation}
[\overline{vM}] = \underbrace{[\overline{v}][\overline{M}]}_\text{[A]} + \underbrace{[\overline{v^{\star}}\overline{M^{\star}}]}_\text{[B]} + \underbrace{[\overline{v'M'}]}_\text{[C]}
\label{eq:mom_trans}
\end{equation}
where $v$ is the meridional component of the wind velocity. The bars over the variables indicate temporal averages and the square brackets a zonal mean. The disturbances in relation to these two averages are represented by: $M' = M - \overline{M}$ and $M^{\star} = M - [M]$.  To analyse the vertical transport, the variable $v$ is replaced by $w$ in m s$^{-1}$.

For each GCM grid cell, the respective relative angular momentum $M$ is calculated using
\begin{equation}
M = m \times u \times a \cos \phi
\label{eq:mom}
\end{equation}
where $m$ is the atmospheric mass of a grid cell, $u$ the zonal wind, $\phi$ the latitude and $a$ the planetary radius. 

\subsubsection{Solid Planet Frame}
\begin{figure}
\centering
\includegraphics[width=1.0\textwidth]{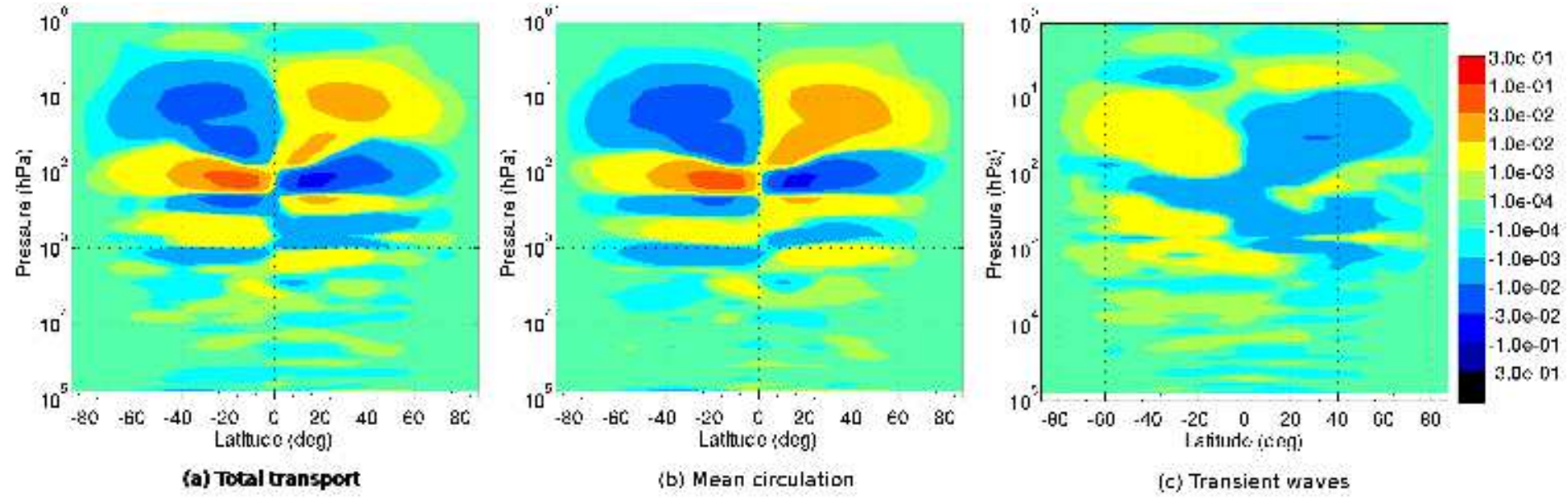}
\caption[Meridional transport of angular momentum of the reference results.]{Meridional transport of angular momentum of the reference results. The data used to produce these results correspond to the last five days  of the $\sim$214 Venus day-long simulation. $[\overline{vM}]$ is the net transport, $[\overline{v}][\overline{M}]$ is the mean circulation and $[\overline{v'M'}]$ is the transient waves. The units of the colour bars are in $10^{24}\times$(kg m$^3$ s$^{-2}$). }
\label{fig:ux-ht}
\end{figure}

In Fig. \ref{fig:ux-ht}, two components of the meridional transport of angular momentum and the net contribution are shown. The map of the stationary wave contribution is not included, because in this case its magnitude is much weaker than the other two means of transport. Fig. \ref{fig:ux-ht}(b) shows the calculated transport by the mean meridional circulation, and, as expected, it is in good agreement with the mass stream function contour lines in Fig. \ref{ux-12}(c). The transport by the mean circulation is the dominant mechanism to transport angular momentum in the atmosphere in the meridional direction, as can be seen from the similarities between Fig. \ref{fig:ux-ht}(b) (mean meridional circulation) and Fig \ref{fig:ux-ht}(a) (net transport). In general, prograde angular momentum is transported from low latitudes to high latitudes by the large thermally direct meridionally overturning cells above the cloud base region. At roughly 100 hPa two eddy-driven indirect meridional circulation cells, confined at low latitudes (one in each hemisphere),  reinforce the transport of axial angular momentum towards the equatorial region. The formation of these eddy-driven cells are related with the larger eddy momentum convergence at low latitudes over the atmospheric forcing related to the latitudinal gradient of the radiative heating. The main feature in the horizontal transient wave representation is the equatorward transport of angular momentum above the 100 hPa pressure level.

\begin{figure}
\centering
\includegraphics[width=1.0\textwidth]{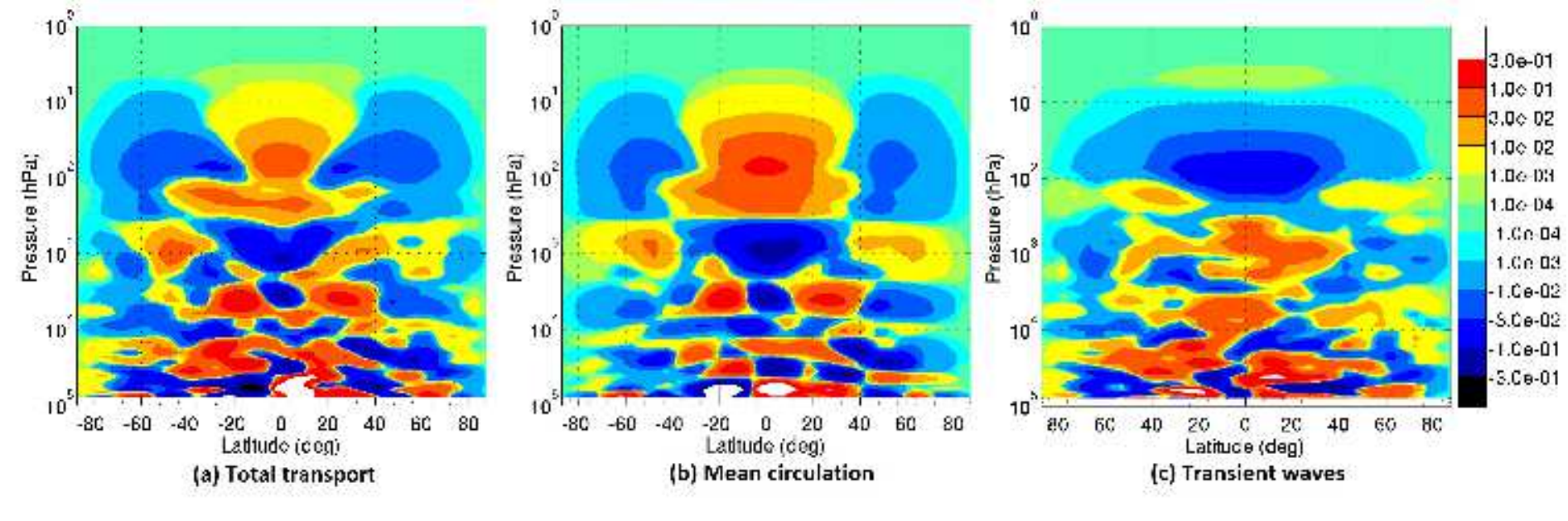}
\caption[Vertical transport of angular momentum of the reference results.]{Vertical transport of angular momentum of the reference results. The data used to produce these results correspond to the last five days  of the $\sim$214 Venus day-long simulation. $\bar{wM}$ is the net transport, $[\overline{w}][\overline{M}]$ is the mean circulation and $[[\overline{w^{\star}}][[\overline{M^{\star}}]$ is the transient waves. The units of the colour bars are in $10^{24}\times$(kg m$^3$ s$^{-2}$).}
\label{fig:ux-vt}
\end{figure}

The vertical transport of angular momentum is largely influenced by the mean circulation (Fig. \ref{fig:ux-vt}). The mean vertical circulation transports momentum upward at low latitudes and downward at high latitudes in the cloud region. This large circulation is confined between the cloud base at 1400 hPa and cloud top at 36 hPa. The negative values at the cloud base are due to the presence of a retrograde equatorial jet. In the upper clouds, the positive values of vertical transport mean that the atmospheric mean circulation is transporting prograde momentum to higher altitudes. Comparing the vertical with the horizontal transport by transient waves we see clearly that the vertical transport by transient waves has a more relevant role accelerating the atmosphere in the cloud top region. In this case prograde angular momentum is brought from higher levels in the atmosphere to the upper cloud region. The prograde relative maximum just above 10$^4$ hPa is located in a region of a long-term disturbance that is breaking the symmetry between the two hemispheres. Below this region, the analysis of vertical momentum transport becomes more complex due to the turbulent flow in the deepest regions of the atmosphere. The stationary waves have again a much weaker impact than that due to transient waves (which includes thermal tides) or the mean meridional circulation. 

\subsubsection{Sun-Synchronized Frame}
\begin{figure}
\centering
\subfigure[Stationary waves.]{\label{fig:ux2-ht-sinc}\includegraphics[width=0.49\textwidth]{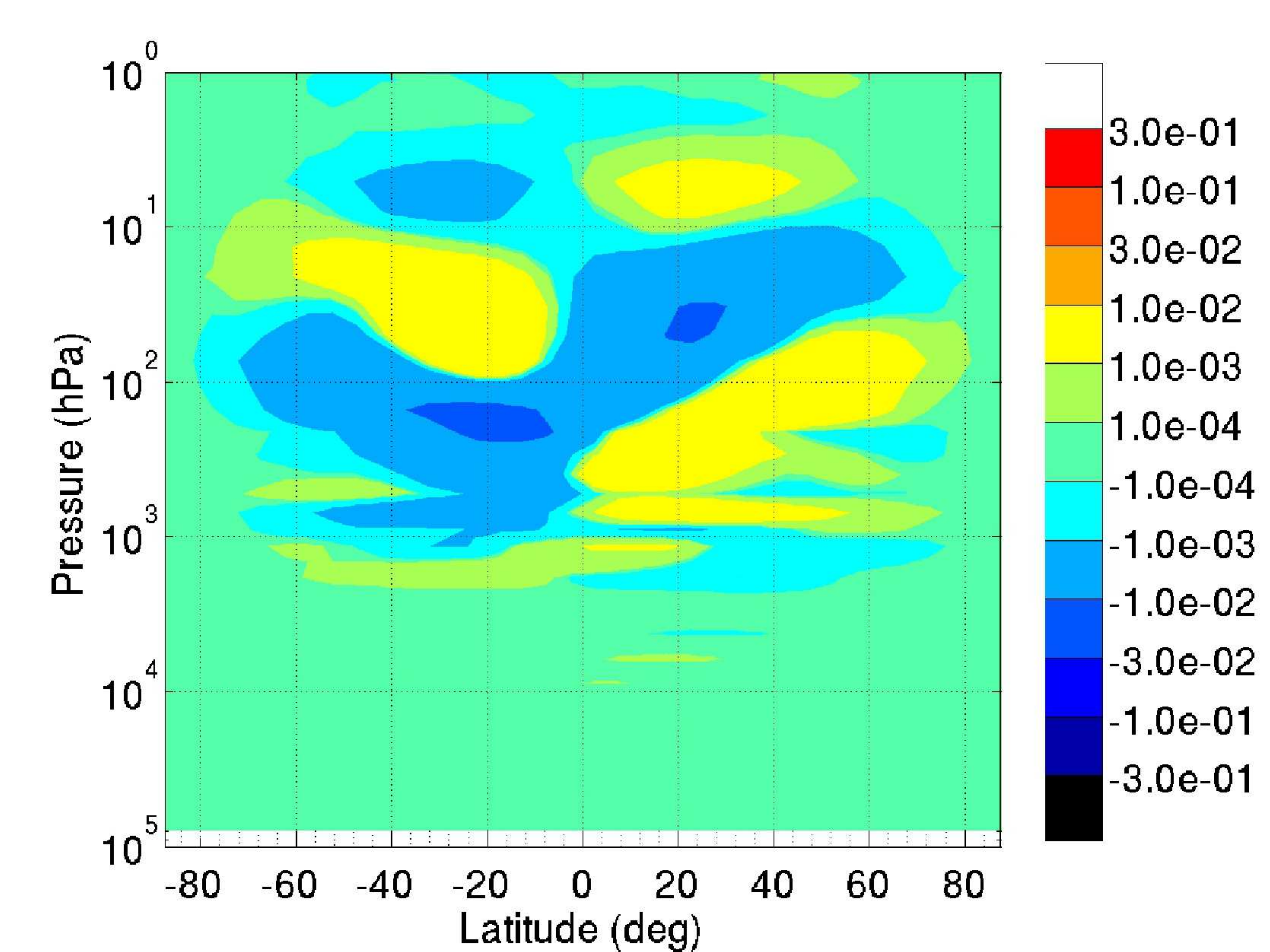}}
\subfigure[Transient wave.]{\label{fig:ux3-ht-sinc}\includegraphics[width=0.49\textwidth]{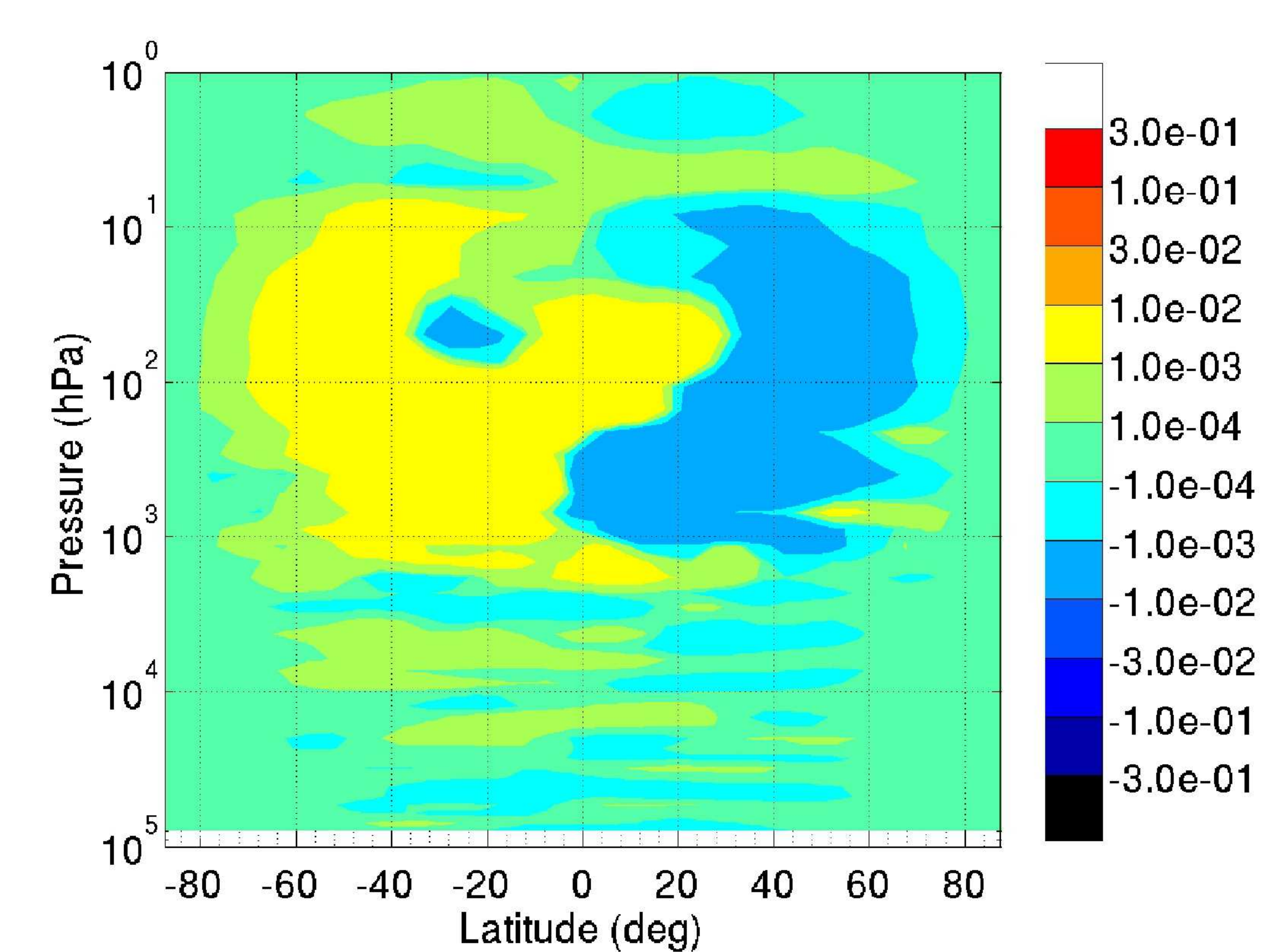}}
\caption[Meridional transport of angular momentum by waves in the reference results, calculated using a solar-fixed longitude.]{Meridional transport of angular momentum by waves in the reference results, calculated using a solar-fixed longitude. The data used to produce these results correspond to the last five days  of the $\sim$214 Venus day-long simulation. \textbf{(a)} and \textbf{(b)} are the horizontal transport by stationary ($[[\overline{v^{\star}}[\overline{M^{\star}}]$) and transient waves ($[\overline{v'M'}]$) respectively.  The units of the colour bars are in $10^{24}\times$(kg m$^3$ s$^{-2}$).}
\label{fig:ux-ht-sinc}
\end{figure}

\begin{figure}
\centering
\subfigure[Stationary waves.]{\label{fig:ux2-vt-sinc}\includegraphics[width=0.49\textwidth]{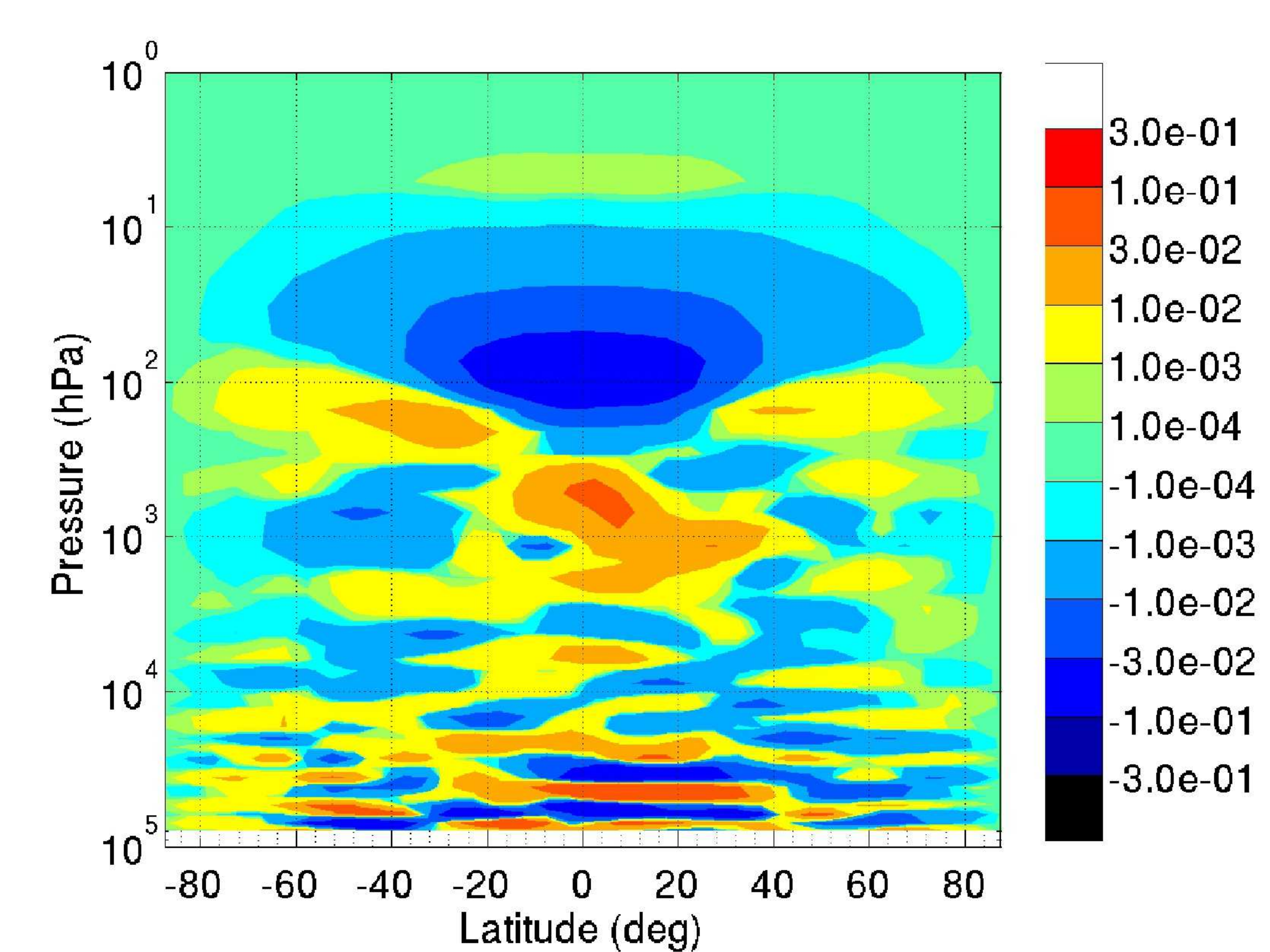}}
\subfigure[Transient waves.]{\label{fig:ux3-vt-sinc}\includegraphics[width=0.49\textwidth]{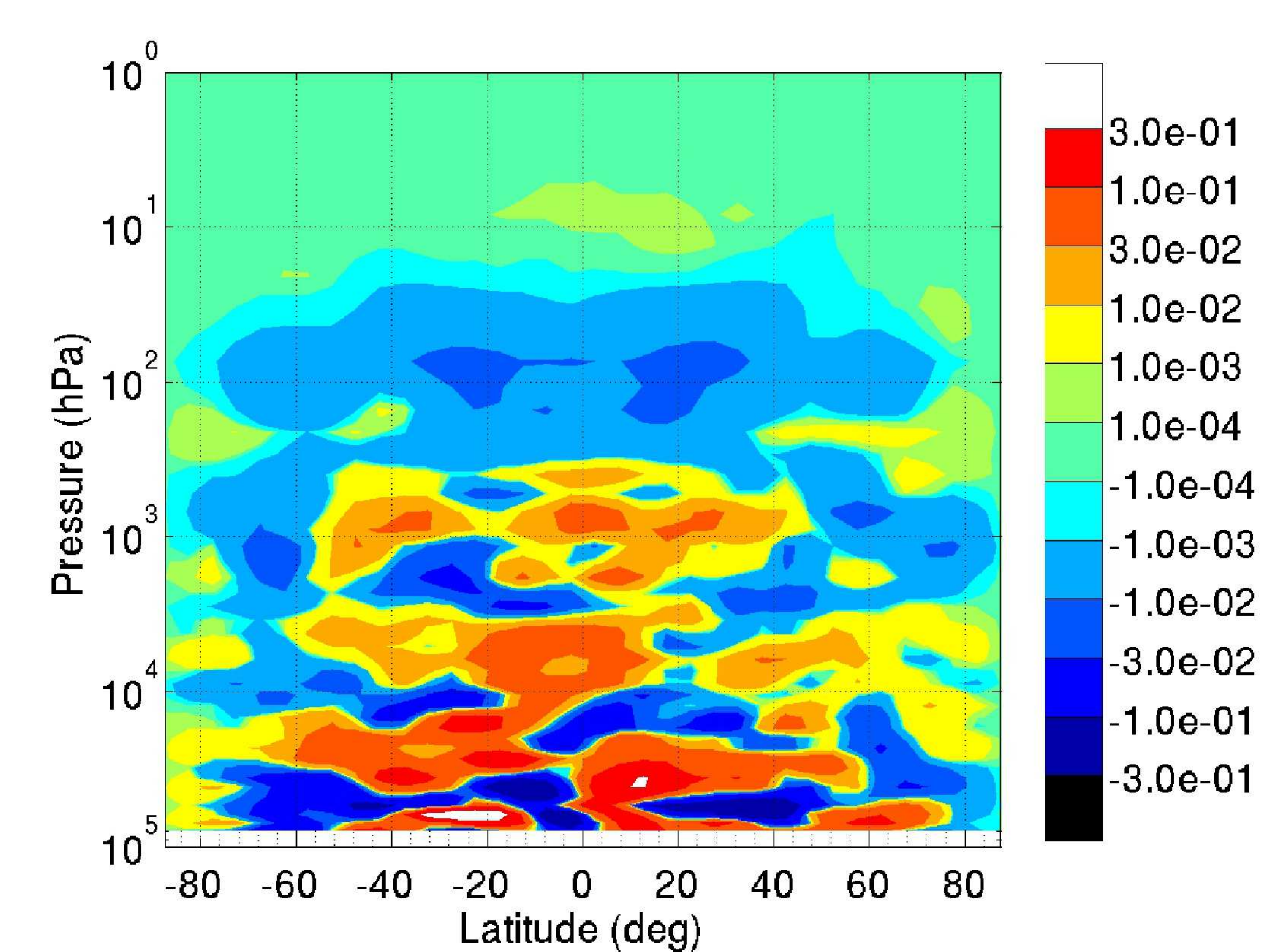}}
\caption[Vertical transport of angular momentum by waves in the reference results, calculated using a solar-fixed longitude.]{Vertical transport of angular momentum by waves in the reference results, calculated using a solar-fixed longitude. The data used to produce these results correspond to the last five days of the $\sim$214 Venus day-long simulation. \textbf{(a)} and \textbf{(b)} are the vertical transport by stationary ($[\overline{\omega^{\star}}\overline{M^{\star}}]$) and transient waves ($[\overline{\omega'M'}]$) respectively.  The units of the colour bars are in $10^{24}\times$(kg m$^3$ s$^{-2}$).}
\label{fig:ux-vt-sinc}
\end{figure}

In Figs. \ref{fig:ux-ht-sinc} and \ref{fig:ux-vt-sinc}, the data were analysed using a solar fixed longitude, which allows us to separate the transport due to the thermal tides from other transient waves. The thermal tides are stationary waves in this new reference frame and are forced waves, which are excited due to the diurnal cycle of the solar heating.  The amplitude of these waves becomes weaker below the main cloud deck where the radiative timescale becomes larger than one Venus day.

In Fig. \ref{fig:ux-ht-sinc}, it is clear that the thermal tides have an important contribution in transporting prograde momentum equatorward above and poleward below 100 hPa. The transient waves have a subsidiary role in the transport of angular momentum compared with the stationary waves.  The existence of these transient waves is clarified in the next section, where we use a wave analysis to explore the results.

The thermally-excited tides have an important role in the vertical transport of momentum as well, see Fig. \ref{fig:ux-vt-sinc}. Fig. \ref{fig:ux-vt-sinc}(a) shows the thermal tides transporting axial angular momentum in the downward direction from roughly 10 hPa (above the cloud region) into the cloud region. These waves accelerate the atmosphere in the region where they are excited and decelerate it in the regions where they are dissipated by radiative damping. This acceleration decreases the latitudinal wind variation between the two mid-latitude jets typically found at roughly $\pm 50^\circ$ latitude. The efficiency of the vertical transport of momentum by the mean circulation is weakened by the thermal tides. The transient waves also tranport axial angular momentum in the downward direction from the same region as the thermal tides. This transport is related to the phase tilt in the vertical of the free equatorial Rossby wave that peaks at mid-latitudes. We will discuss this mechanism in next section.

\subsubsection{Eliassen-Palm flux diagnostic}
\begin{figure}
\centering
\subfigure[$-\overline{\boldsymbol{u}}_{\star}\cdot\nabla\overline{m}$.]{\includegraphics[width=0.45\textwidth]{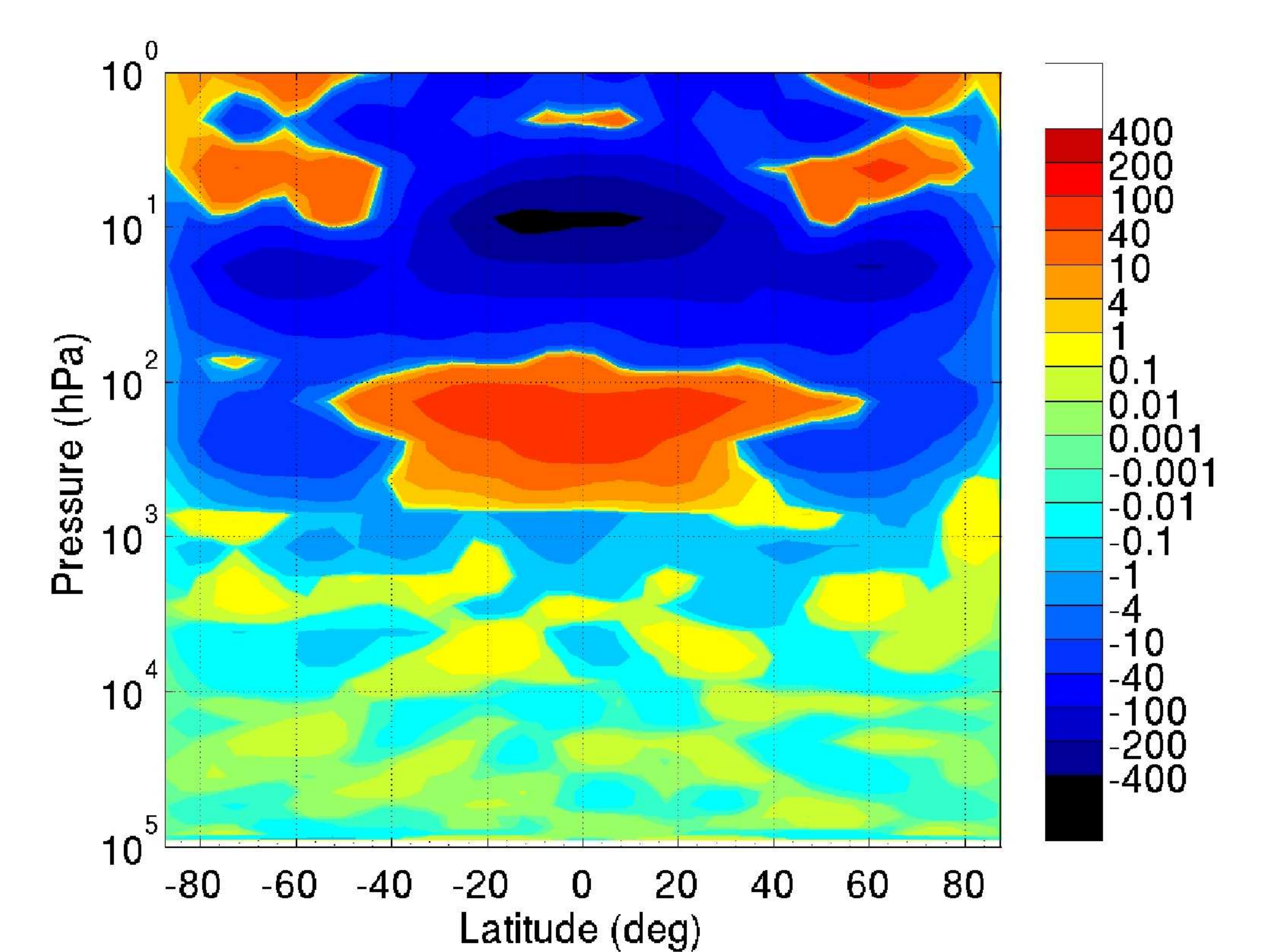}}
\subfigure[$-\nabla\cdot \boldsymbol{E}$.]{\includegraphics[width=0.45\textwidth]{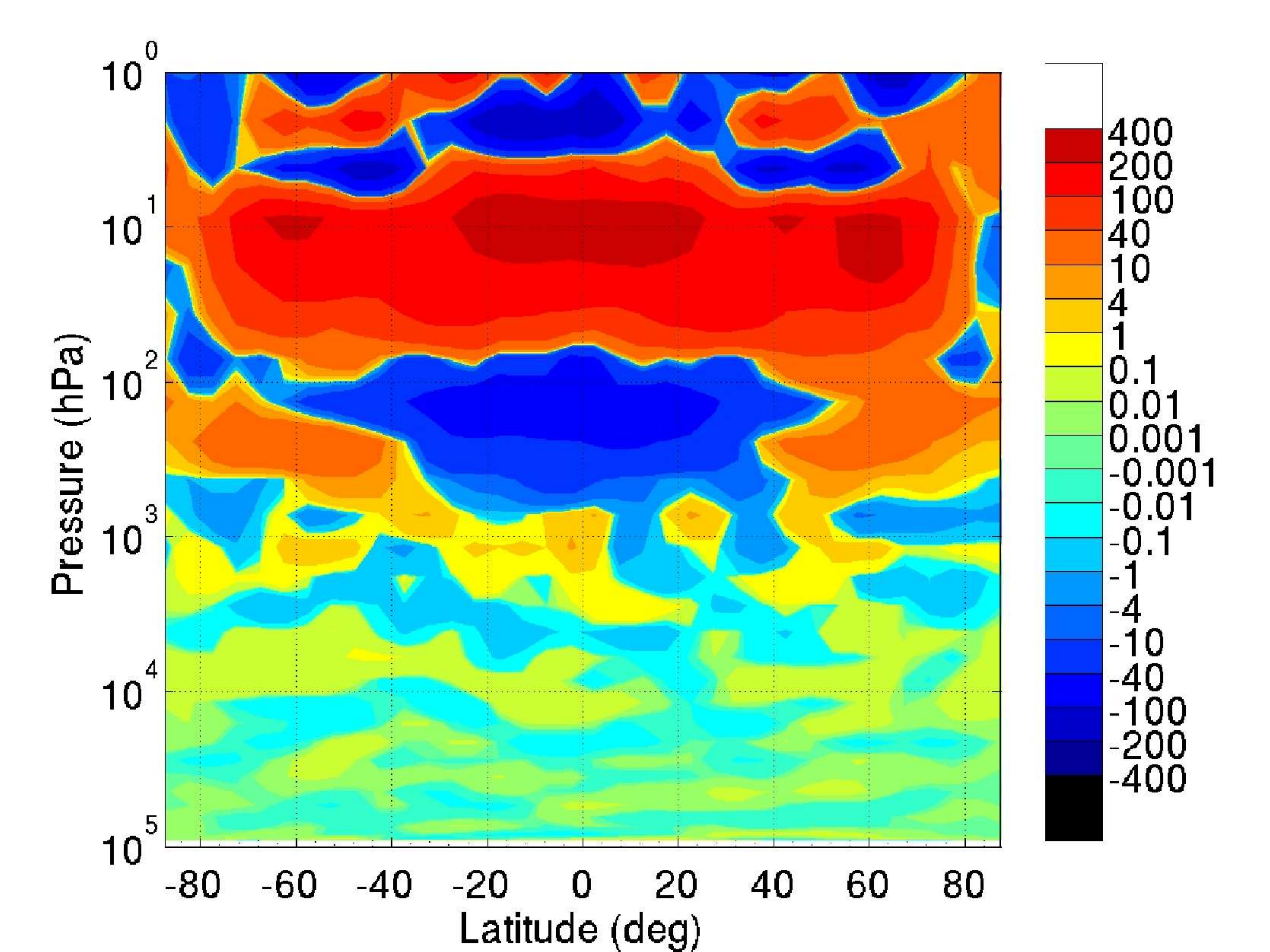}}
\caption{Eliassen-Palm flux diagnostics: (a) acceleration due to mean circulation; (b)  acceleration due to resolved waves. The units are in m$^2$s$^{-2}$.}
\label{fig:eli_palm}
\end{figure}

The sections above show how the axial angular momentum is being transported in the atmosphere due to the presence of mean circulation and waves, but they do not show clearly how much each component (mean circulation and waves) is contributing to the acceleration of the mean flow. Here, we use the Eliassen-Palm flux diagnostics in pressure coordinates (\citealt{1983Andrews}; \citealt{1986Read}) to analyse the acceleration of the mean axial angular momentum $m$ mainly in the upper cloud region, given by
\begin{equation}
\label{Eq:EliassenFlux}
\overline{m}_t+\overline{\boldsymbol{u}}_\star\cdot\nabla\overline{m}=-\nabla\cdot \boldsymbol{E} + \mathcal R.
\end{equation}
In Eq. \ref{Eq:EliassenFlux}, $\overline{\boldsymbol{u}}_\star$ is the residual mean velocity defined by:
\begin{align} 
\overline{\boldsymbol{u}}_\star = (\overline{v}_\star, \overline{w}_\star),\\
\overline{\boldsymbol{v}}_\star =\overline{v} -  \{(\overline{v'\theta'})/\overline{\theta}_p\}_p,\\
\overline{\boldsymbol{w}}_\star =\overline{w}  + \{(\overline{v'\theta'})(\cos\phi)/\overline{\theta}_p)\}_\phi/(A\cos\phi),
\end{align}
where $A$ is the planet mean radius. The variable $\boldsymbol{E}$ in Eq. \ref{Eq:EliassenFlux} is the Eliassen-Palm flux and $\cal R$ the residual which represents e.g., the dissipation. The Eliassen-Palm flux is defined here as:
\begin{equation}
\begin{split}
\boldsymbol{E} = A\cos\phi\Big[&(\overline{u'v'})-(\overline{u}_p(\overline{v'\theta'})/\overline{\theta}_p),\\
&(\overline{w'u'})+\Big((\overline{u}\cos\phi)_\phi/(A\cos\phi)-2\Omega\sin\phi)(\overline{v'\theta'})/\overline{\theta}_p\Big].
\end{split}
\end{equation}
The term $-\overline{\boldsymbol{u}}_\star\cdot\nabla\overline{m}$ in Eq. \ref{Eq:EliassenFlux} is associated with the acceleration of the mean angular momentum due to mean meridional circulation and $-\nabla\cdot \boldsymbol{E}$ due to the resolved waves. Our analysis focus on the balance between these two terms since the magnitude of $\mathcal R$ is much smaller than $-\overline{\boldsymbol{u}}_\star\cdot\nabla\overline{m}$ and  $-\nabla\cdot \boldsymbol{E}$. Fig. \ref{fig:eli_palm} shows two maps of $-\overline{\boldsymbol{u}}_\star\cdot\nabla\overline{m}$ and $-\nabla\cdot \boldsymbol{E}$ averaged over the last five Venus days of the simulation. Fig.  \ref{fig:eli_palm}(a) shows that the mean circulation is decelerating the prograde zonal winds in the upper cloud region where the strong jet is present. However, it is accelerating the prograde winds at low latitudes in the region 100-1000 hPa.  Fig.  \ref{fig:eli_palm}(b) shows that the waves are essential to the strong equatorial jet at low latitudes in the upper cloud region. In the next section, we filter  $-\nabla\cdot \boldsymbol{E}$ to study in more detail the contribution of the different types of waves. 

\subsection{Wave Analysis}
\begin{figure}
\centering
\subfigure[]{\label{fig:ux-peri1}\includegraphics[width=0.45\textwidth]{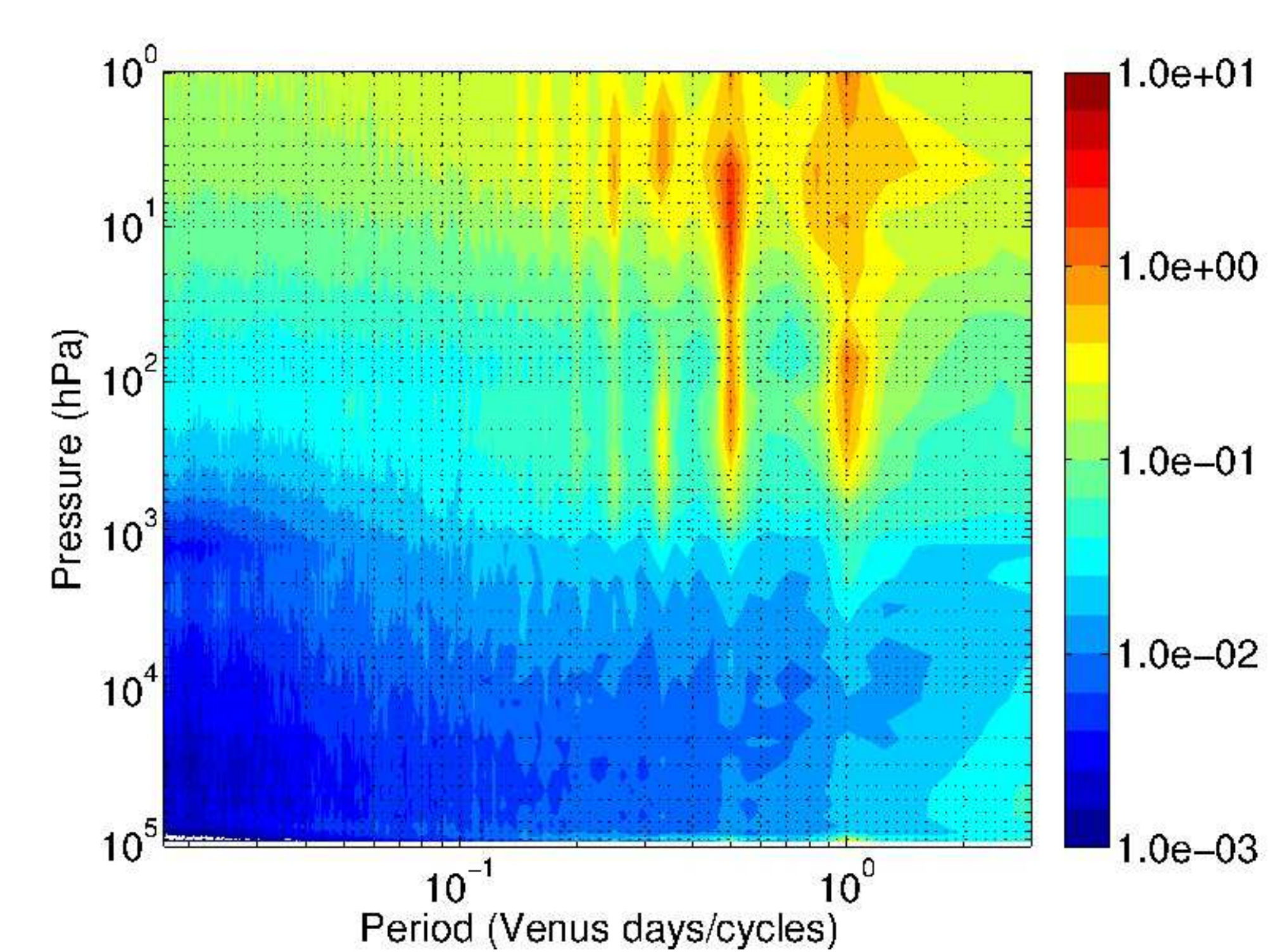}}
\subfigure[]{\label{fig:ux-peri2}\includegraphics[width=0.45\textwidth]{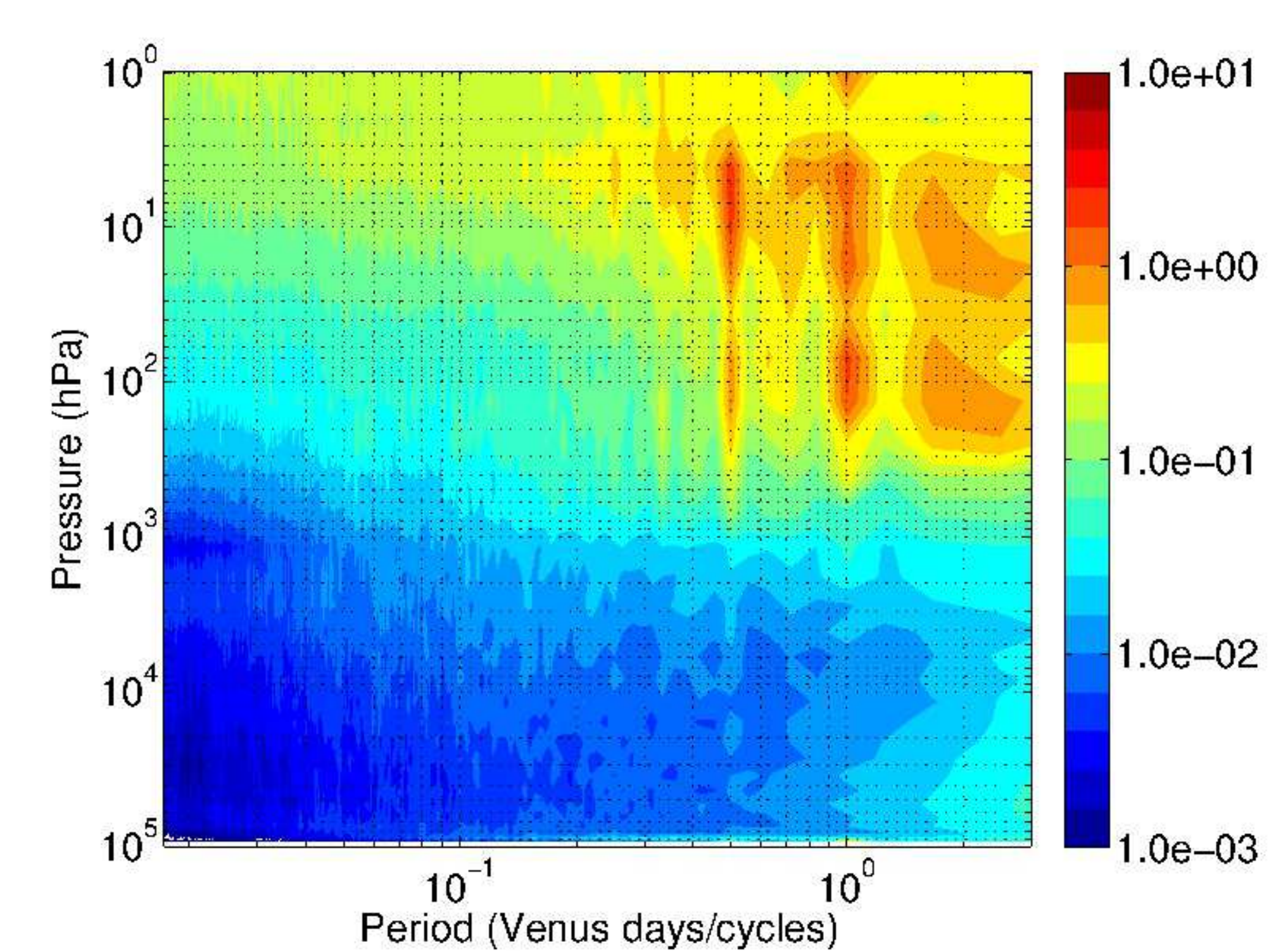}}
\subfigure[]{\label{fig:ux-peri3}\includegraphics[width=0.45\textwidth]{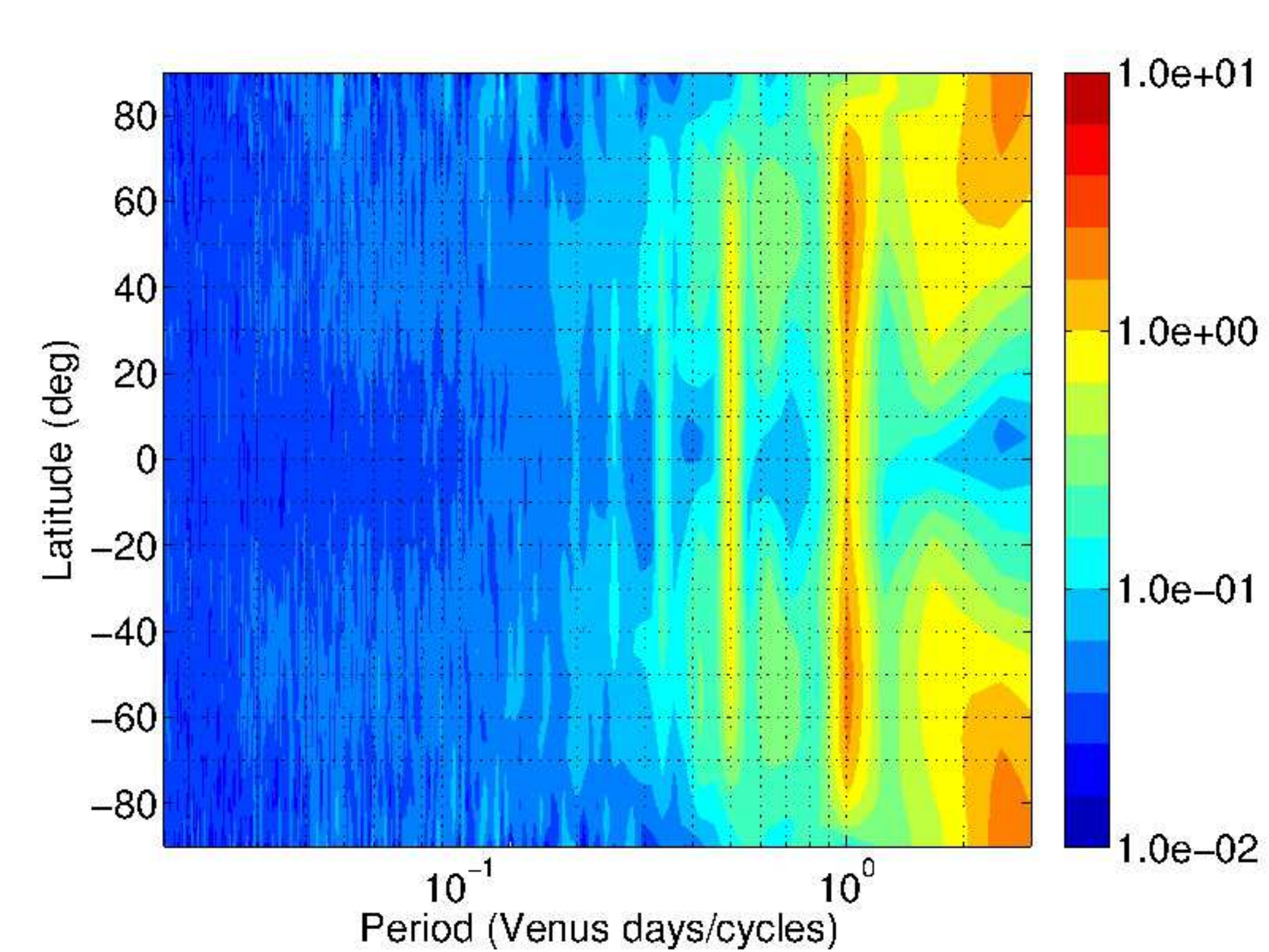}}
\caption[Amplitude spectra for the temperatures of the reference results.]{Amplitude spectra for the temperatures (K) of the reference results. The values of the spectra were zonally averaged and correspond to the last five Venus days of the long simulation. \textbf{(a)} and \textbf{(b)} show spectra contoured as a function of pressure and wave period (in Venus days) for the equator and 45$^{\circ}$N respectively; \textbf{(c)} shows spectra contoured as a function of latitude and wave period at a pressure level of 100 hPa.}
\label{fig:ux-peri}
\end{figure}

\begin{figure}
\centering
\subfigure[OPUS-Vr]{\label{fig:ux-hov31}\includegraphics[width=0.45\textwidth]{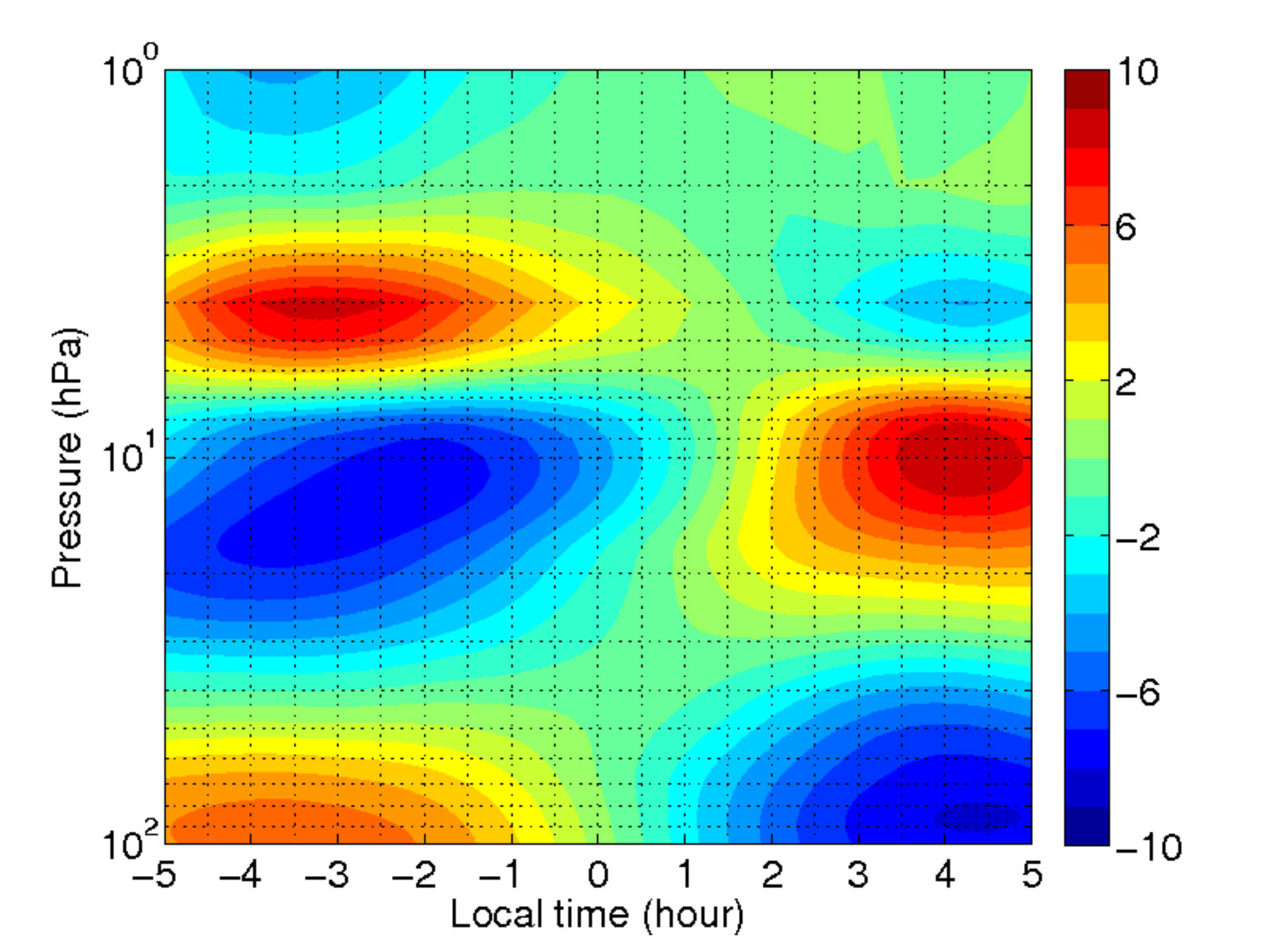}}
\subfigure[Venus Express]{\label{fig:ux-hov32}\includegraphics[width=0.45\textwidth]{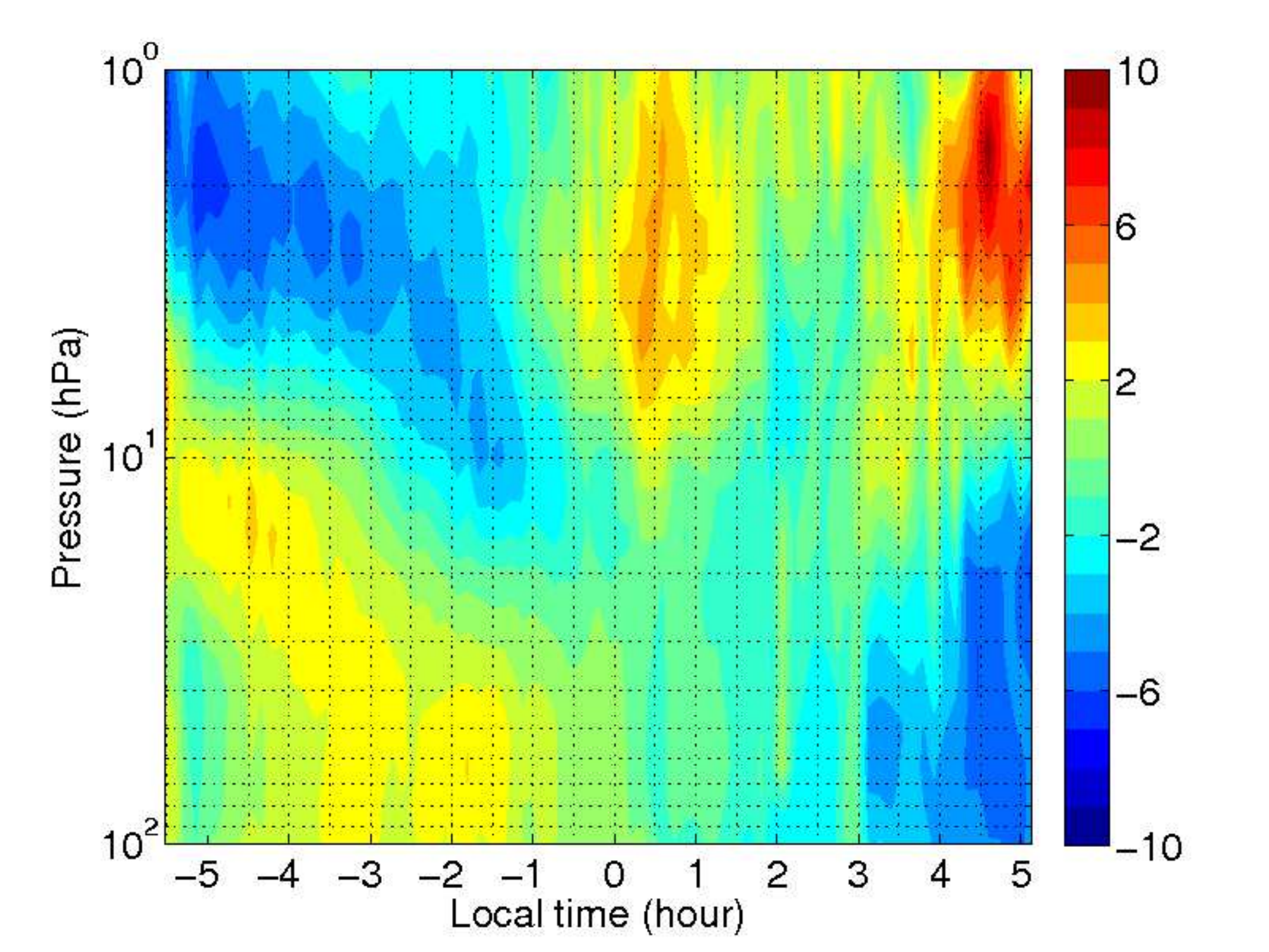}}
\caption[Temperature variability comparison between OPUS-Vr and Venus Express results.]{ Temperature variability (K) comparison between OPUS-Vr \textbf{(a)} and Venus Express results \textbf{(b)}. The results shown from the GCM results were averaged over five Venus days at 45$^\circ$S. The observational result used was based on data from \cite{2010Grassi} at 45$^\circ$S.}
\label{fig:ux-hov3}
\end{figure}

The temperature data analysed here from the reference simulation have a temporal resolution of one Earth day (with more than 100 points for each Venus day) and correspond to the last five Venus days of the $\sim$214 Venus day-long simulation. In Fig. \ref{fig:ux-peri}, transient travelling waves (relative to the underlying planet) are identified and characterized in terms of their period. These atmospheric waves are responsible for the momentum transport in the atmosphere, seen before in Figs. \ref{fig:ux-ht}(c) and \ref{fig:ux-vt}(c), and are an essential part of the mechanism to produce the atmospheric circulation obtained. In Fig. \ref{fig:ux-peri}, three amplitude spectra of the temperature field are shown: two as a function of pressure for two latitudinal locations (equator and mid-latitude) and one as a function of latitude at a pressure level of 100 hPa (in the upper cloud region). The signal associated with the thermal tides is very clear in all spectra. The thermal tides are forced waves formed due to the periodic solar heating (which travels with the sun's position) and can be decomposed into different harmonics. The more dominant harmonics are the diurnal and semidiurnal tides (with periods of 1 and 0.5 Venus days), and their amplitudes are larger in the upper cloud region and then above the 30 hPa pressure level at all latitudes. The largest amplitudes found for these waves at the regions mentioned are $\sim$3 K for the diurnal tide and $\sim$3.5 K for the semi-diurnal tide. For altitudes above 1 hPa (not shown in the maps) the diurnal tides become more intense with amplitudes reaching 11 K. In the maps that show the wave amplitude as a function of altitude, there is clear evidence of one convection cell located respectively at the cloud base (1000 hPa) preventing and/or weakening the wave propagation. In Fig. \ref{fig:ux-hov3} we show the comparison between the temperature variability, $T'$, of values from the OPUS-Vr and those based on a temperature field from Venus Express data  at 45$^\circ$S.  The variable $T'$ is defined by the difference between the temperature field and its average over the time period of data analysed. The figure shows clearly the waves excited by the dominant semi-diurnal tide, which is expected to have the strongest tidal response of the different harmonics due to having the longest wavelength (\citealt{1987Baker}). It is the clear tilt of the wave front with altitude that is responsible for the important vertical wave momentum transport. The values from the OPUS-Vr simulation were averaged over five Venus days in the sun-synchronized frame, and are similar to what has been obtained from the observations (\citealt{2010Grassi}) in terms of wave amplitude, location and the tilt of the phase front with altitude. The observational temperature map shown corresponds to time averaged values from data acquired in the period June 2006 to July 2008.

\begin{figure}
\centering
\subfigure[]{\label{fig:ux-hov1-eq}\includegraphics[width=0.45\textwidth]{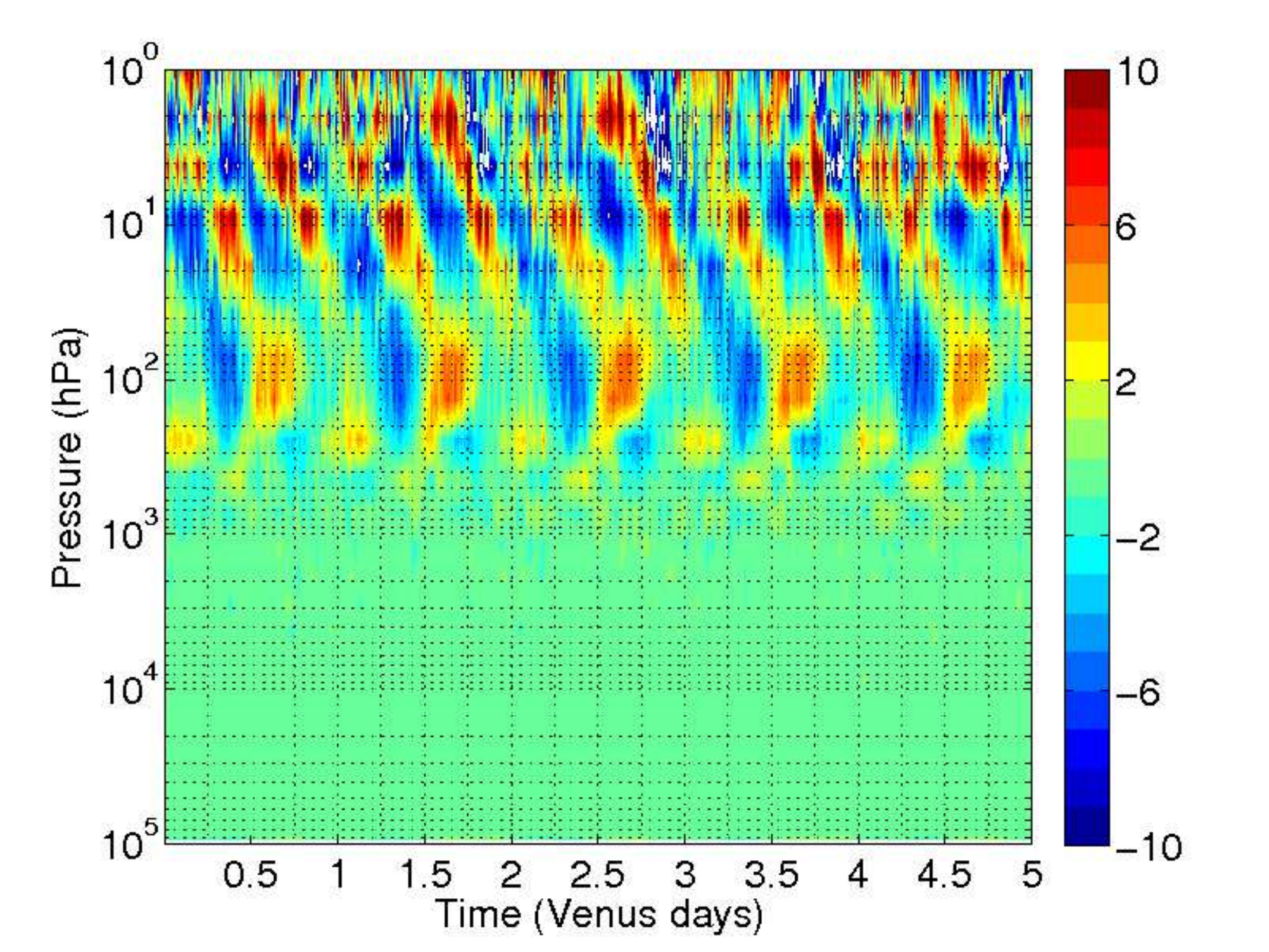}}
\subfigure[]{\label{fig:ux-hov2-eq}\includegraphics[width=0.45\textwidth]{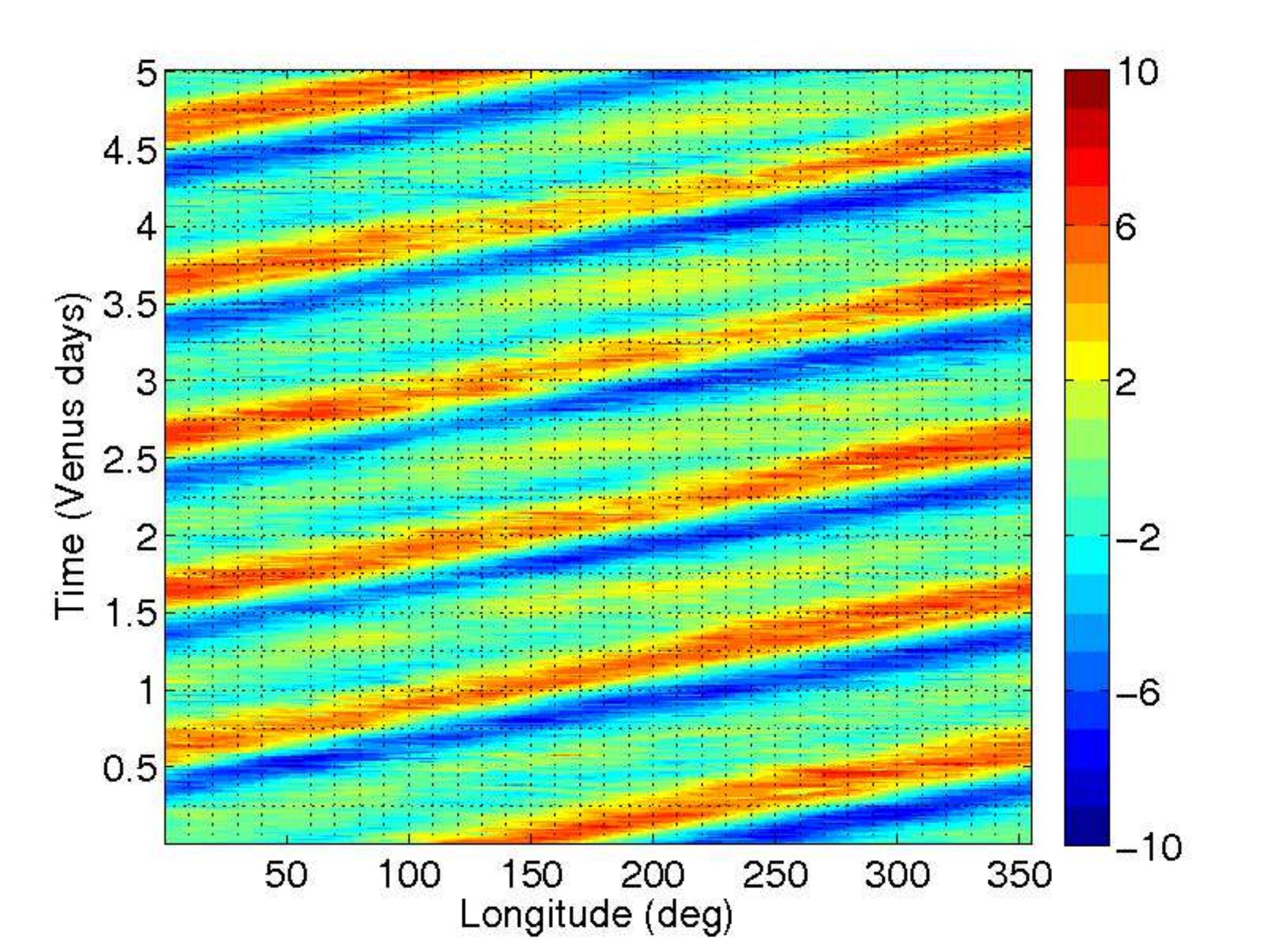}}
\caption[Hovm\"oller plots of temperature variability for reference results at the equator.]{ Hovm\"oller plots of temperature variability (K) for reference results at the equator. The temperature variability in \textbf{(a)} is measured at a fixed longitude of 0$^{\circ}$ and in \textbf{(b)} is measured at a pressure level of 100 hPa. These results correspond to the last five Venus days of the $\sim$214 Venus day-long simulation.}
\label{fig:ux-hov1}
\end{figure}

The spectrum as a function of latitude from Fig. \ref{fig:ux-peri}(c), shows that the mid-latitudes are regions of stronger wave activity. The waves with small amplitude are located predominantly in mid-latitude regions, which indicate that they are being produced in association with the unstable mid-latitude jets. These waves weakens the mid-latitude jets and strengthens the zonal winds in the equatorial region, however, in our work they have a much weaker impact than the meridional transport by the mean circulation or other transient waves, e.g. thermal tides  (discussed in the momentum transport section \ref{sec:MT}). From Fig. \ref{fig:ux-peri}, \ref{fig:ux-peri2} and \ref{fig:ux-peri22}, we found waves  with a period around 0.7 Venus days and with characteristics of free equatorial Rossby waves (e.g., \citealt{2012Arnold}). This wave has a minimum amplitude at the equator, and a further analysis shows the geopotential height of the disturbances to be symmetric about the equator. We also found that the free Rossby waves have an unstable resonance with the equatorial struture of the diurnal tide, which is similar to the mechanism found in \cite{2012Arnold}. This interaction results in residual amplitudes in the meridional wind map at low latitudes between periods of half and one Venus days, and also in a horizontal tilt of the phase front, which contributes to the angular momentum transport from mid latitudes towards low latitudes (see the discussion at the end of this section about the filtered Eliassen-Palm fluxs).

Figs. \ref{fig:ux-hov1}(a) and \ref{fig:ux-hov1}(b) show Hovm\"oller plots of $T'$, to demonstrate the propagation of waves identified. The Hovm\"oller plot is a useful tool to detect wave-like features in the atmosphere, and the axes are space versus time. In Fig. \ref{fig:ux-hov1}(a) and \ref{fig:ux-hov1}(b), altitude-time and longitude-time Hovm\"oller plots are shown for an equatorial latitude. As we would expect from the spectra, the dominant waves are excited by the first two harmonic components of the solar tides, propagating in the eastward (retrograde) direction. A downward component of the solar tides extends to a pressure level of roughly 500 hPa.

Two high latitude waves with quasi-bidiurnal periods are also found in the temperature and wind fields (Figs. \ref{fig:ux-peri}, \ref{fig:ux-peri2} and \ref{fig:ux-peri22}. The amplitude of these waves is roughly 1.5 K, and is related to the two low-frequency waves found before in the zonal wind field (Fig. \ref{fig:ux-var}).  A wave with a quasi-bidiurnal period, was also found by \cite{2010Lebonnois} but not explored. These simulated waves have a meridional component, as it is possible to see in Figs. \ref{fig:ux-peri2} and \ref{fig:ux-peri22}. These characteristics, together with the observed opposite phases in the waves' zonal winds for different hemispheres shown in Fig. \ref{fig:ux-var}, suggests that these low-frequency waves constitute an anti-symmetric equatorial wave mode, such as the mixed Rossby-gravity (or Yanai) wave (e.g., \citealt{andrewsbook}). These waves have a retrograde propagation in relation to the mean flow, just as mixed Rossby-gravity waves do (\citealt{andrewsbook}). In Fig. \ref{fig:ux_amp_rossb} the maximum amplitudes of the mixed Rossby-gravity waves analysed in the temperature field are shown. As expected, the larger magnitudes are localized at higher latitudes and a minimum is obtained at the equator. For these results, a Fourier bandpass filter was applied, retaining just the waves with periods shorter than 2.5 and longer than 1.5 Venus days. Other characteristics of these mixed Rossby-gravity waves can be found in  the Hovm\"oller maps  shown in Fig. \ref{fig:ux_mrg_hov}. In these results the model data were again filtered for periods shorter than 2.5 and longer than 1.5 Venus days to focus on the long-wave properties. The results show that the temperature and zonal velocity are anti-symmetric about the equator while the meridional component is symmetric. A more detailed analysis of the mixed Rossby-gravity wave properties would require an exploration of its dispersion relation.  The mixed Rossby-gravity waves are excited in the upper cloud region, and propagate upwards. It is in a region above the 1 hPa pressure level that they are mostly absorbed by the atmosphere via radiative damping (together with the inertia-gravity waves mentioned before). Other forms of dissipation such as diffusion, are taken into account numerically in the model, but this is much less relevant. These features of the mixed Rossby-gravity waves are reflected in the zonal wind field presented before on Fig. \ref{fig:ux-var}, where eddy zonal winds are in anti-phase between each hemisphere and have significant variations with a bi-diurnal period (1.86 and 2.14 Venus days). As it was pointed out before, these variations in the wind field were found in Venus Express data in \cite{2013Khatuntsev}, but claimed to be a product of the natural peridiocities of the instrument.

\begin{figure}
\centering
\subfigure[]{\label{fig:ux-peri1-u1}\includegraphics[width=0.45\textwidth]{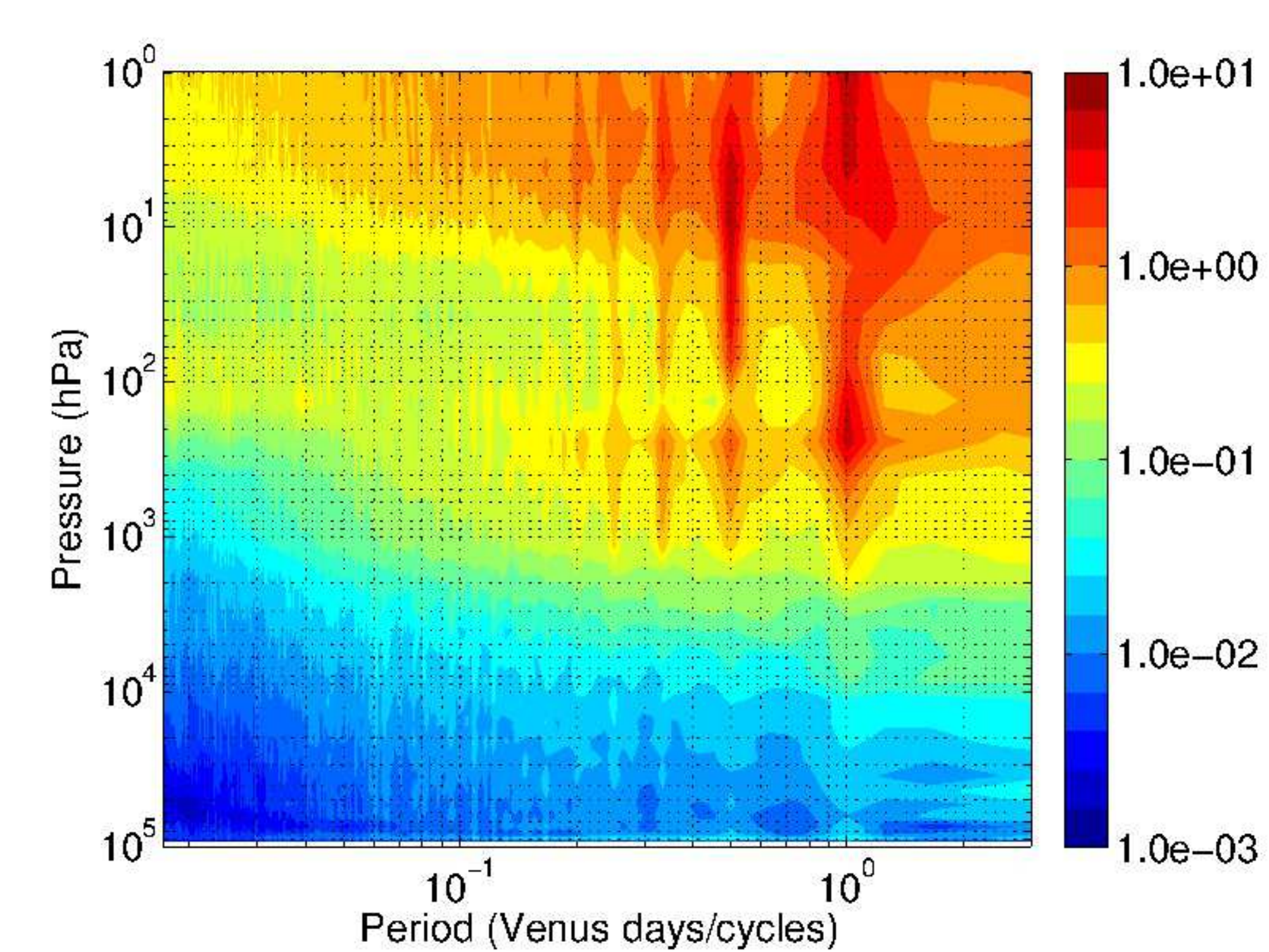}}
\subfigure[]{\label{fig:ux-peri1-v1}\includegraphics[width=0.45\textwidth]{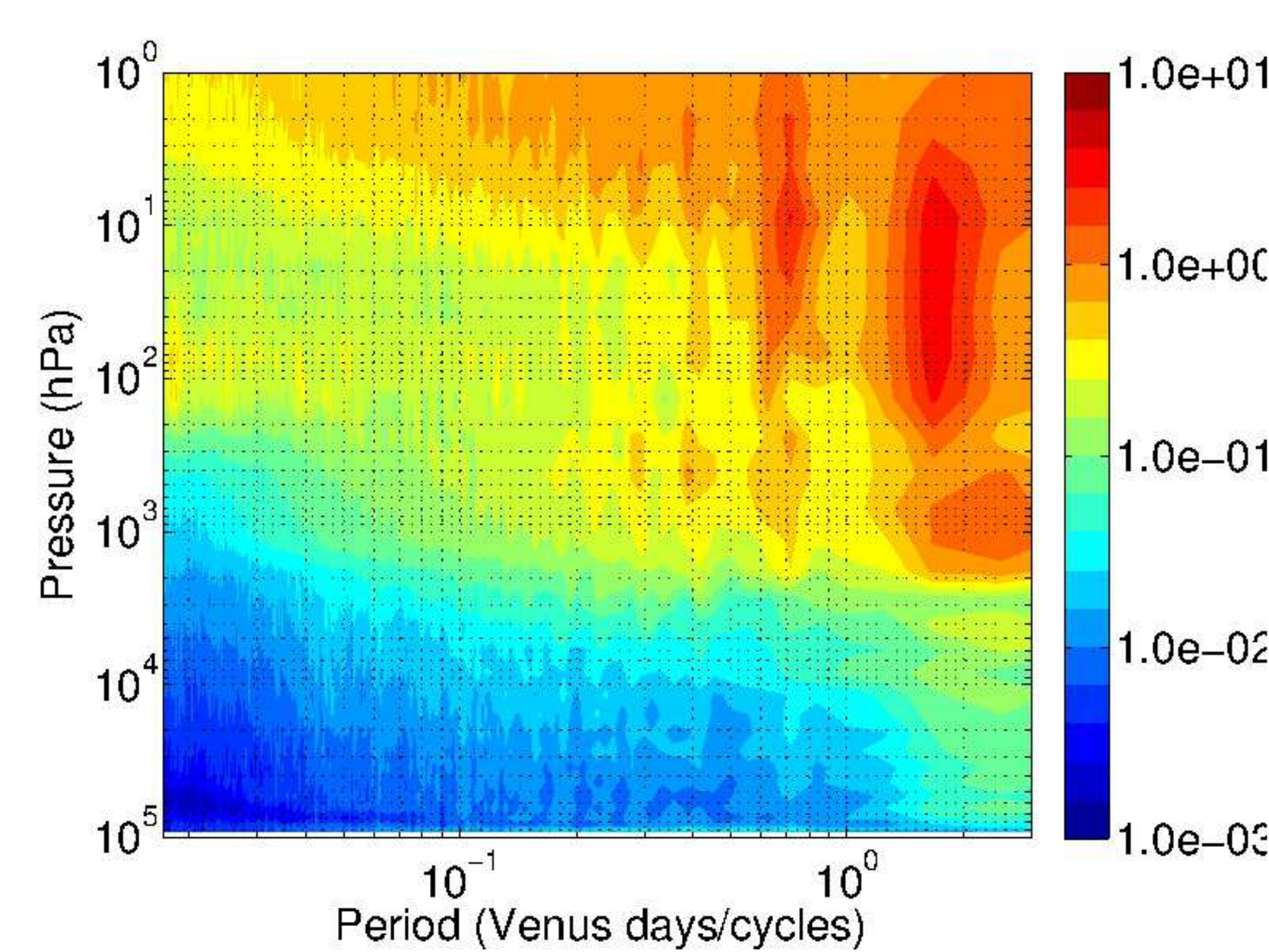}}
\caption[Zonally averaged amplitude spectra for the horizontal wind components of the reference results in the equatorial region.]{Zonally averaged amplitude spectra for the horizontal wind components (m$/$s) of the reference results in the equatorial region. \textbf{(a)} is a result retrieved from the zonal wind field and  \textbf{(b)} from the meridional wind component.}
\label{fig:ux-peri2}
\end{figure}

\begin{figure}
\centering
\subfigure[]{\label{fig:ux-peri1-u2}\includegraphics[width=0.45\textwidth]{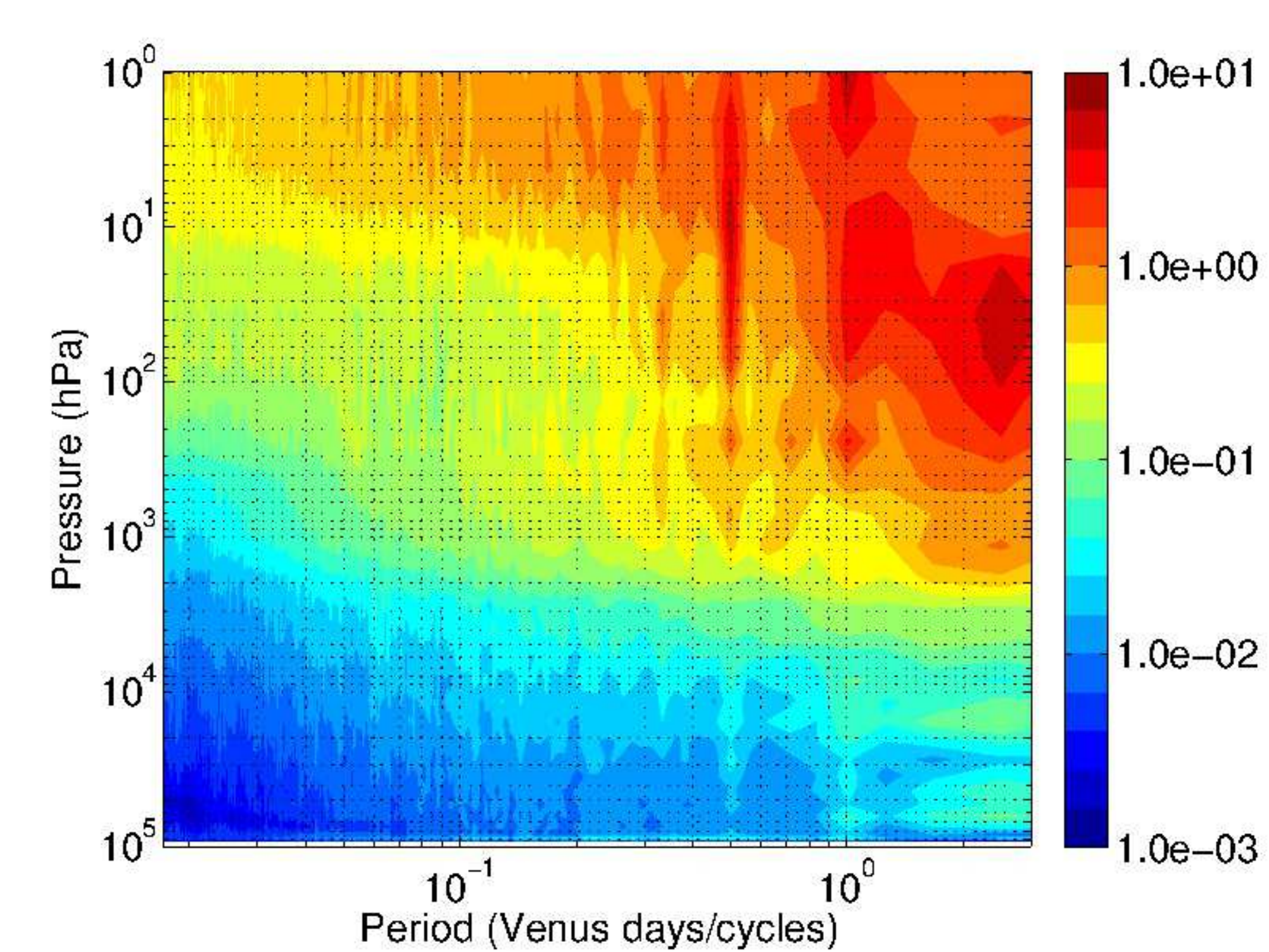}}
\subfigure[]{\label{fig:ux-peri1-v2}\includegraphics[width=0.45\textwidth]{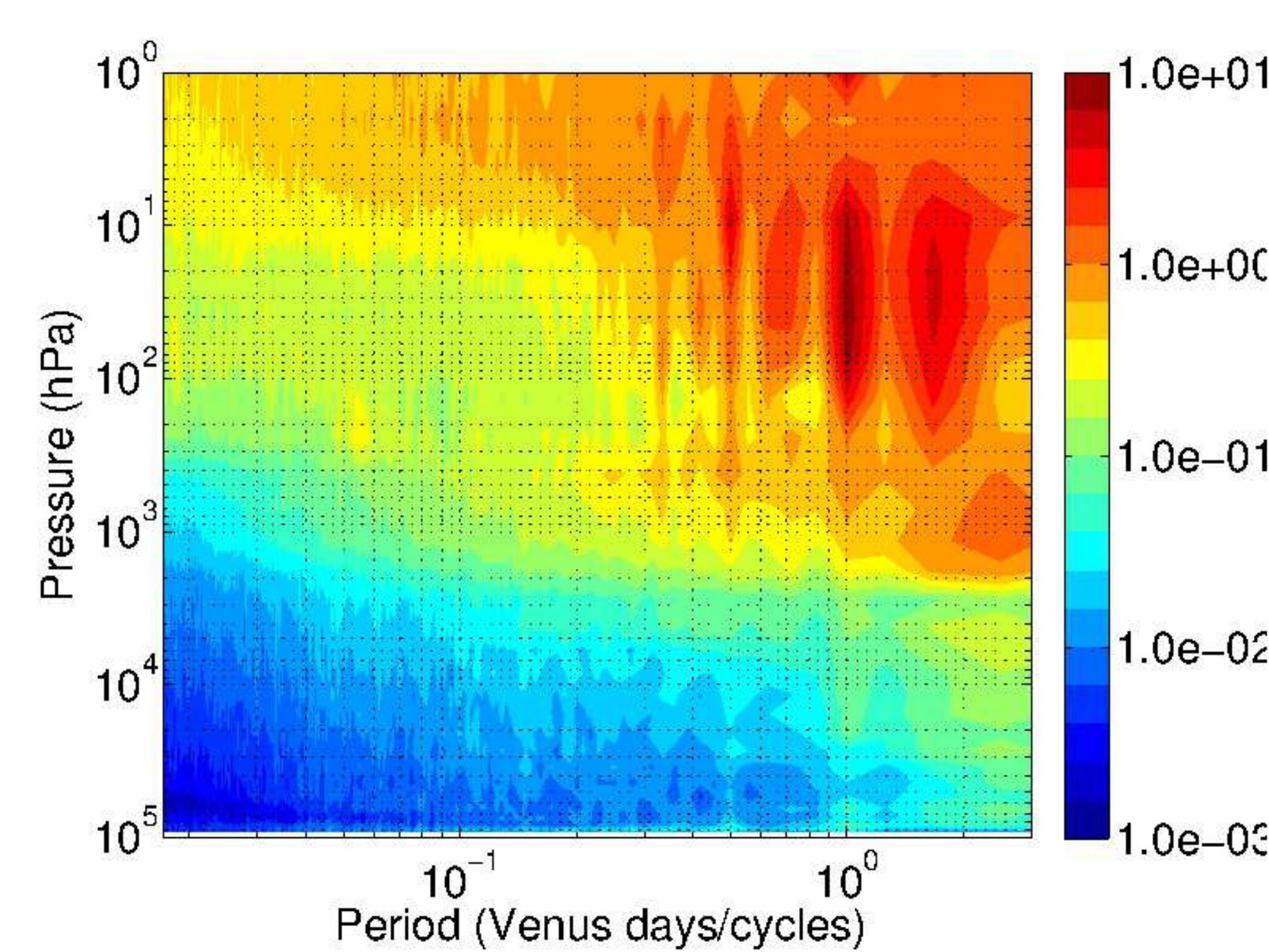}}
\caption[Amplitude spectra for the horizontal wind components of the reference results at mid-latitudes.]{Amplitude spectra for the horizontal wind components of the reference results at mid-latitudes. \textbf{(a)} is a result retrieved from the zonal wind field and \textbf{(b)} from the meridional wind component.}
\label{fig:ux-peri22}
\end{figure}

\begin{figure}
\centering
\includegraphics[width=0.6\textwidth]{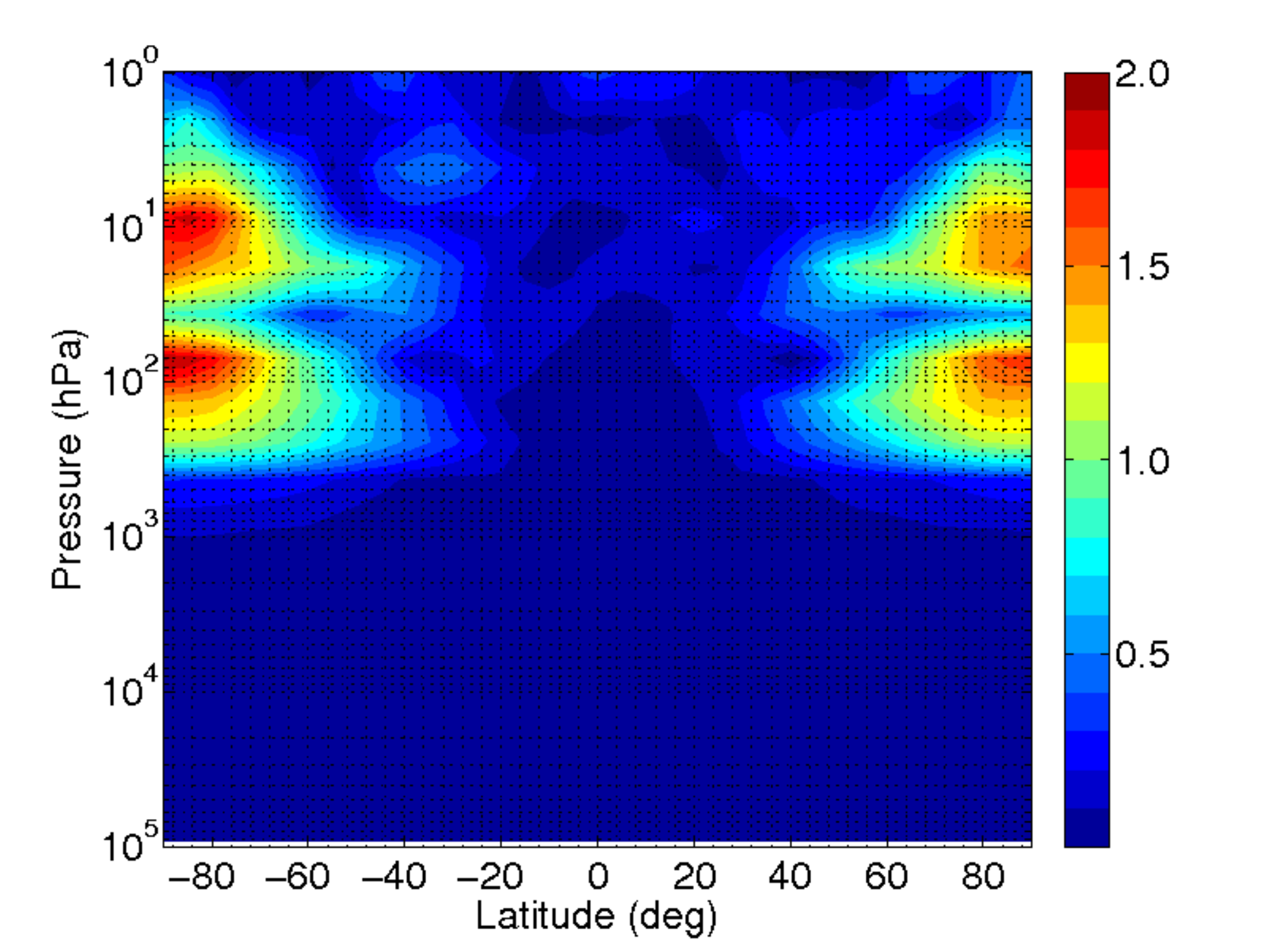}
\caption[Low-frequency waves maximum amplitudes.]{Low-frequency waves maximum amplitudes in the temperature field. A Fourier bandpass filtering was applied for periods shorter than 2.5 and longer than 1.5 Venus days.}
\label{fig:ux_amp_rossb}
\end{figure}

\begin{figure}
\centering
\subfigure[]{\label{fig:u}\includegraphics[width=0.32\textwidth]{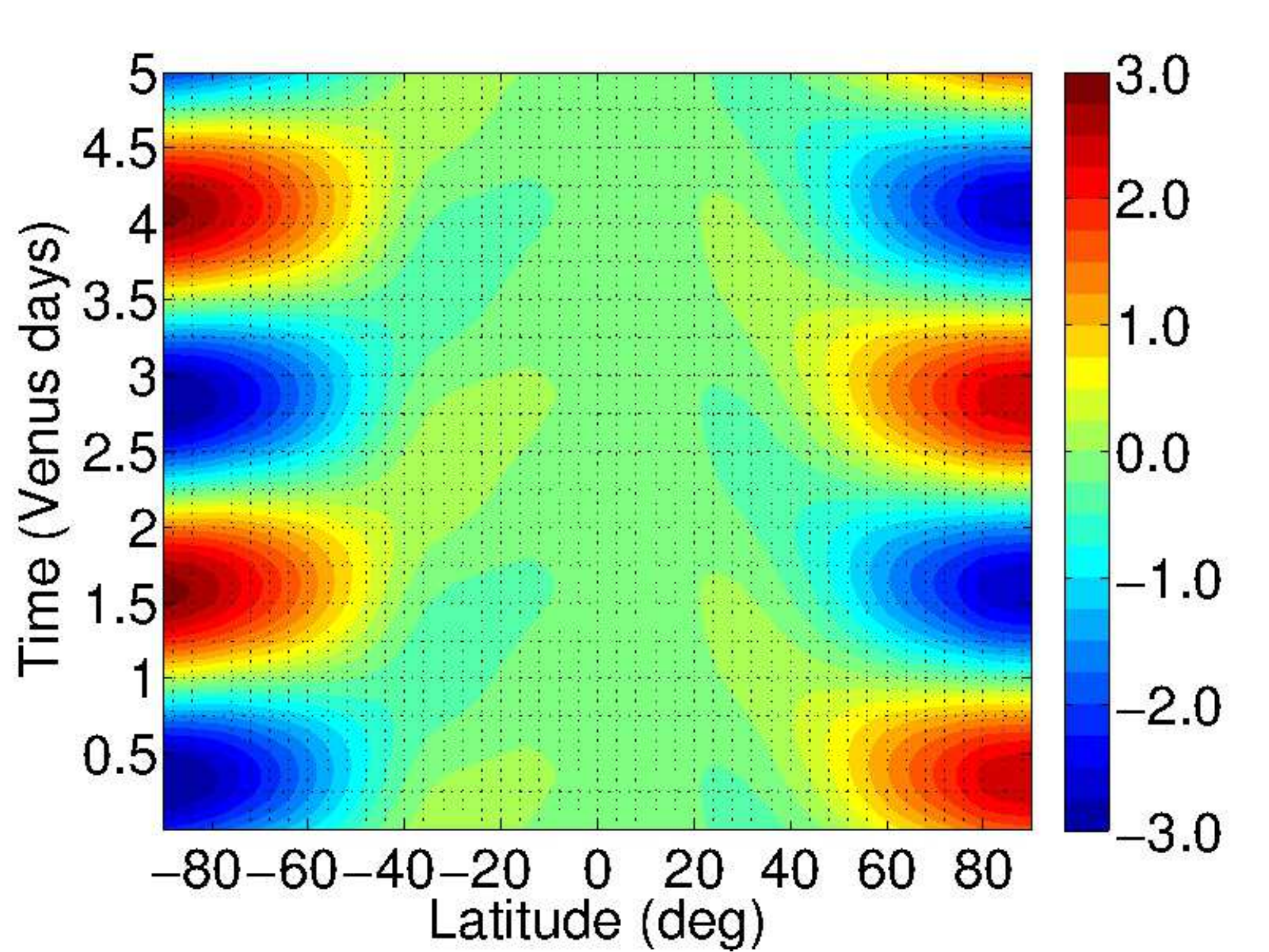}}
\subfigure[]{\label{fig:u}\includegraphics[width=0.32\textwidth]{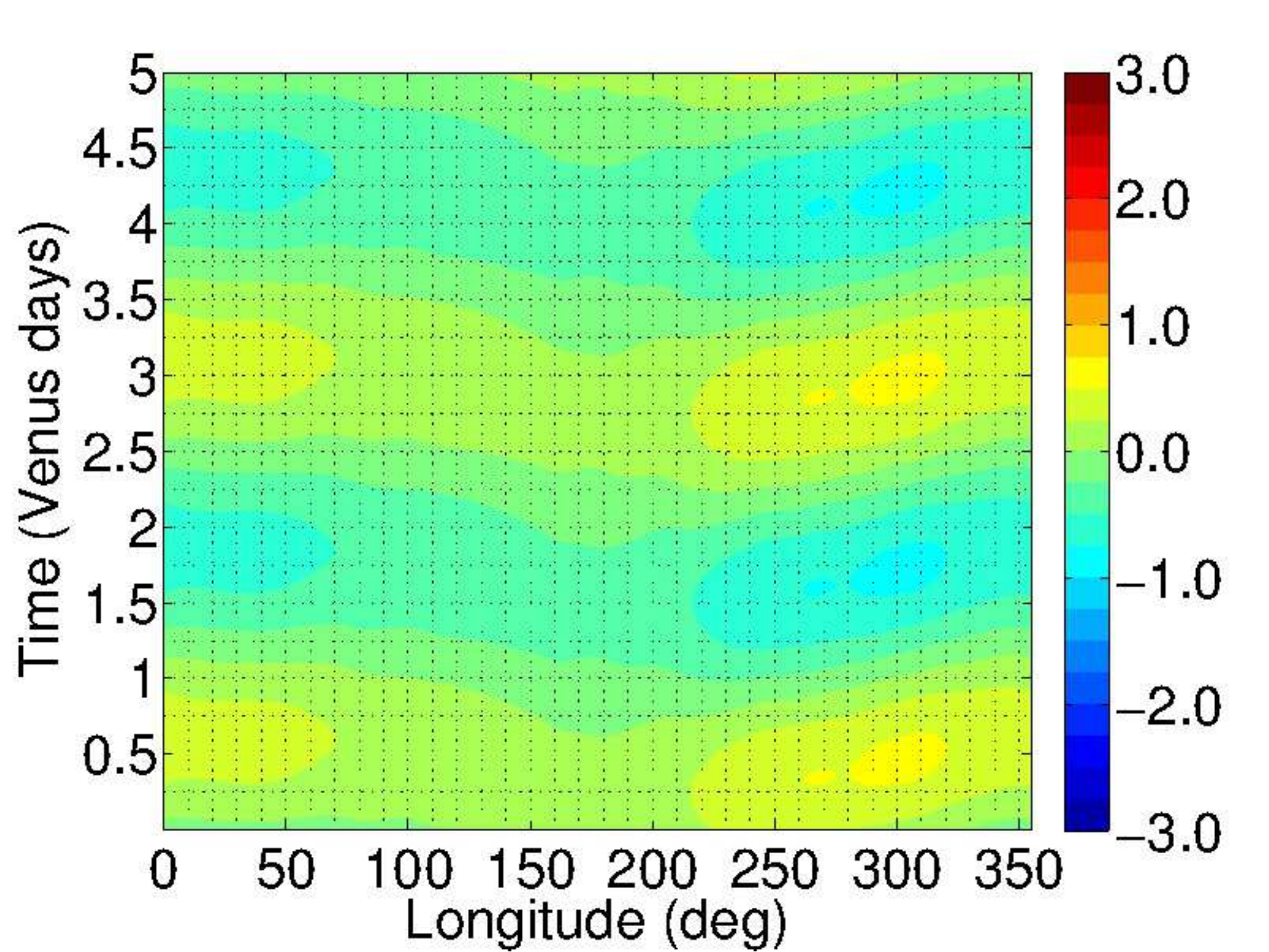}}
\subfigure[]{\label{fig:u}\includegraphics[width=0.32\textwidth]{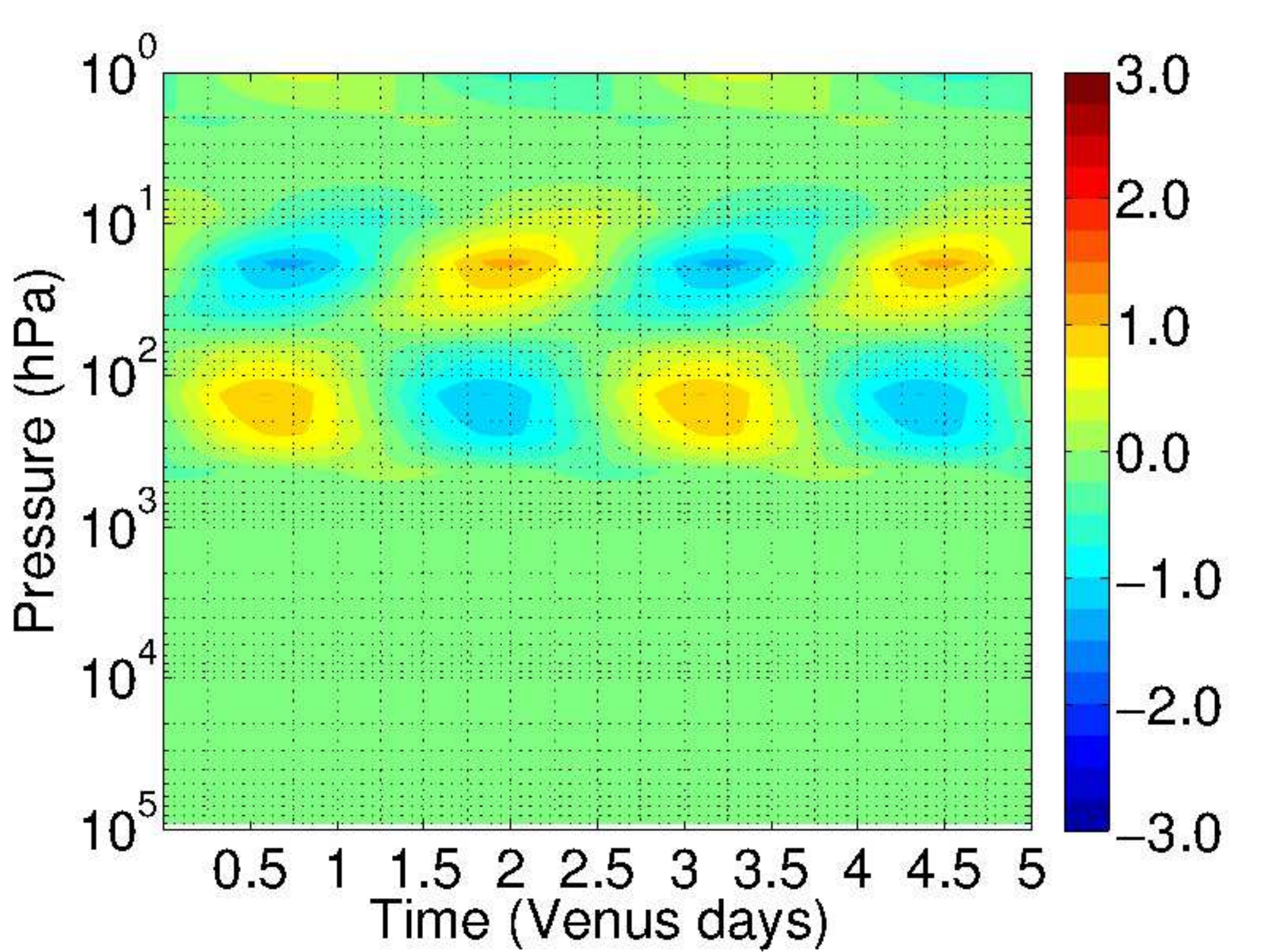}}
\\
\subfigure[]{\label{fig:u}\includegraphics[width=0.32\textwidth]{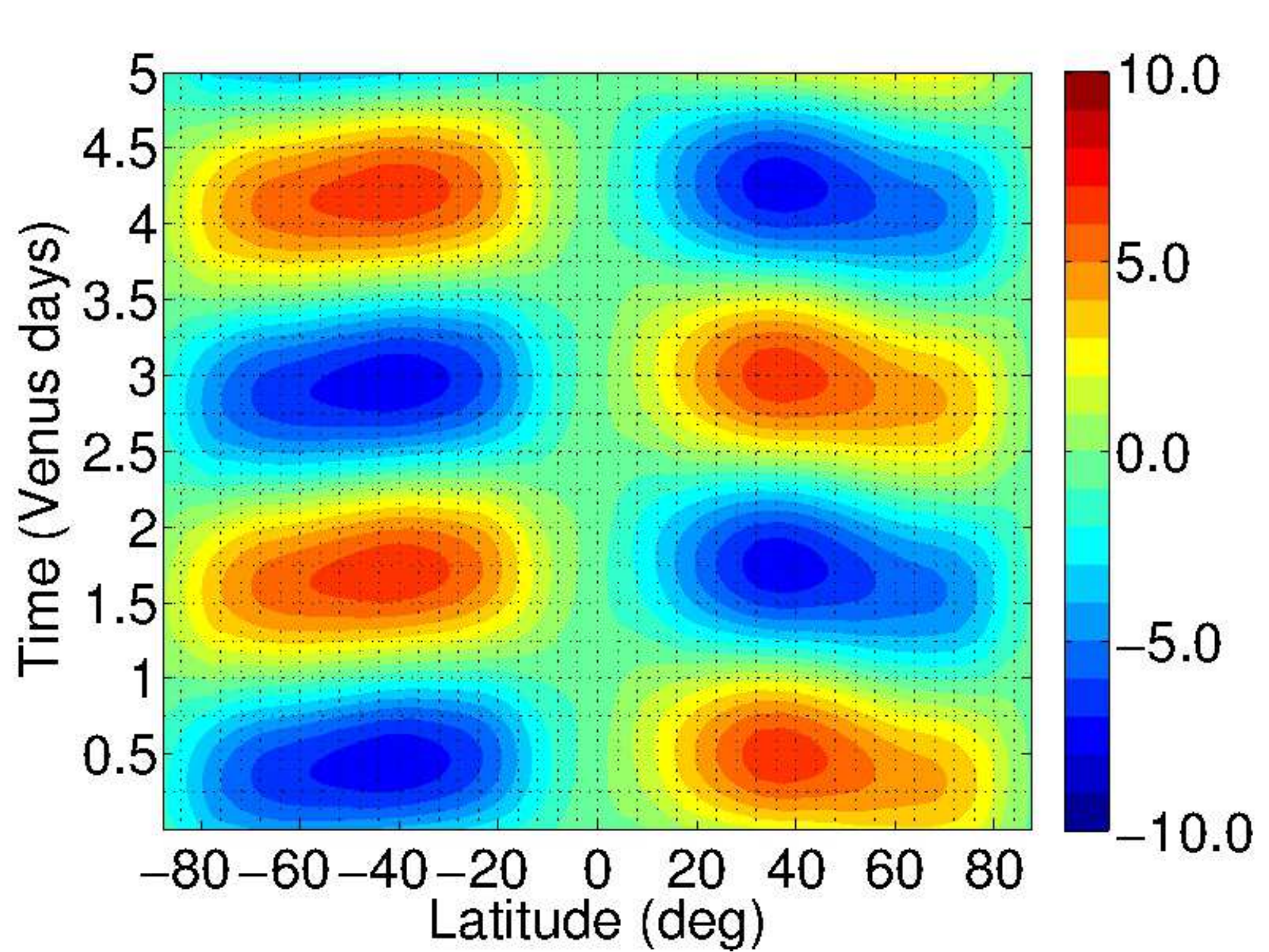}}
\subfigure[]{\label{fig:u}\includegraphics[width=0.32\textwidth]{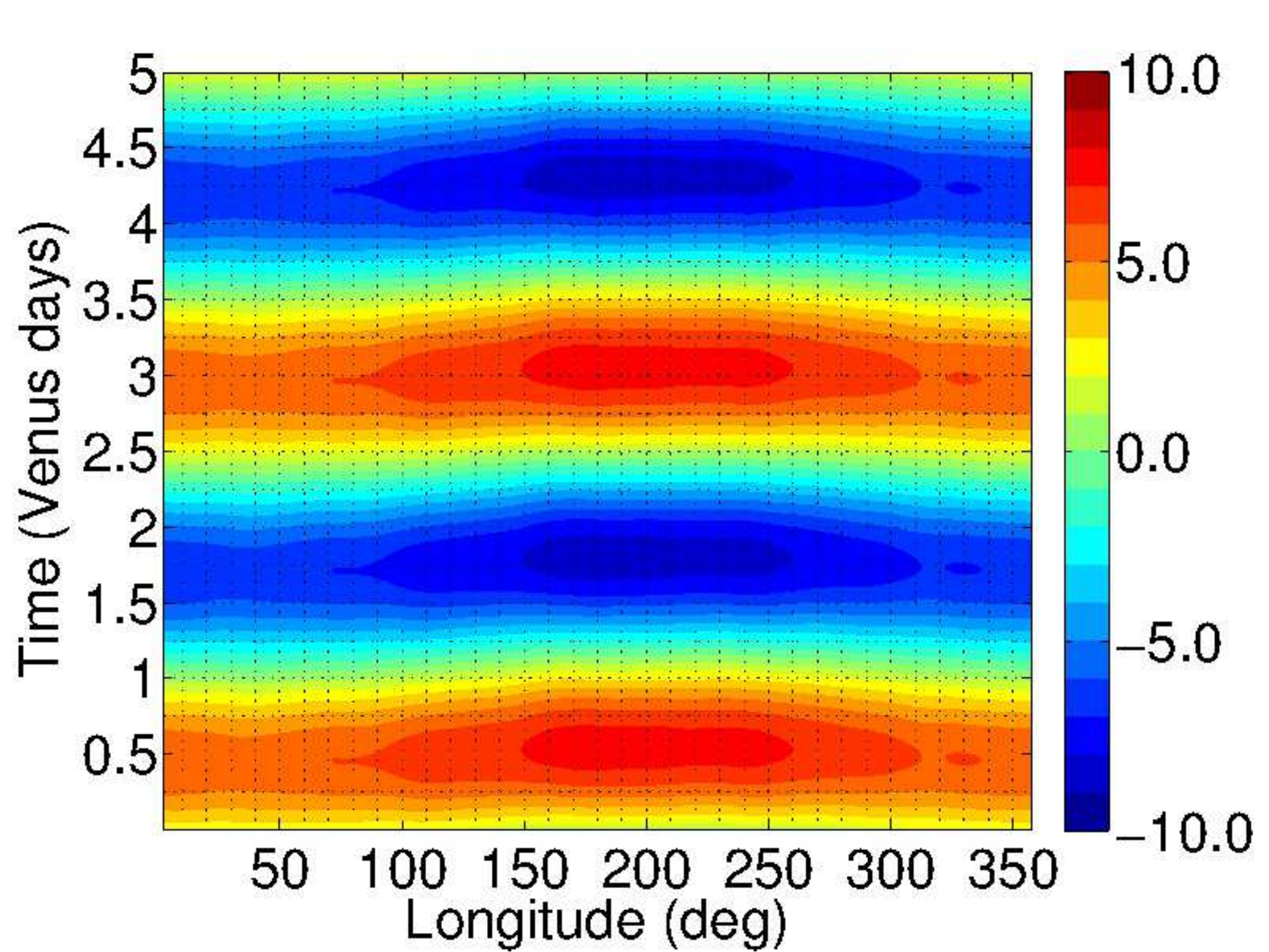}}
\subfigure[]{\label{fig:u}\includegraphics[width=0.32\textwidth]{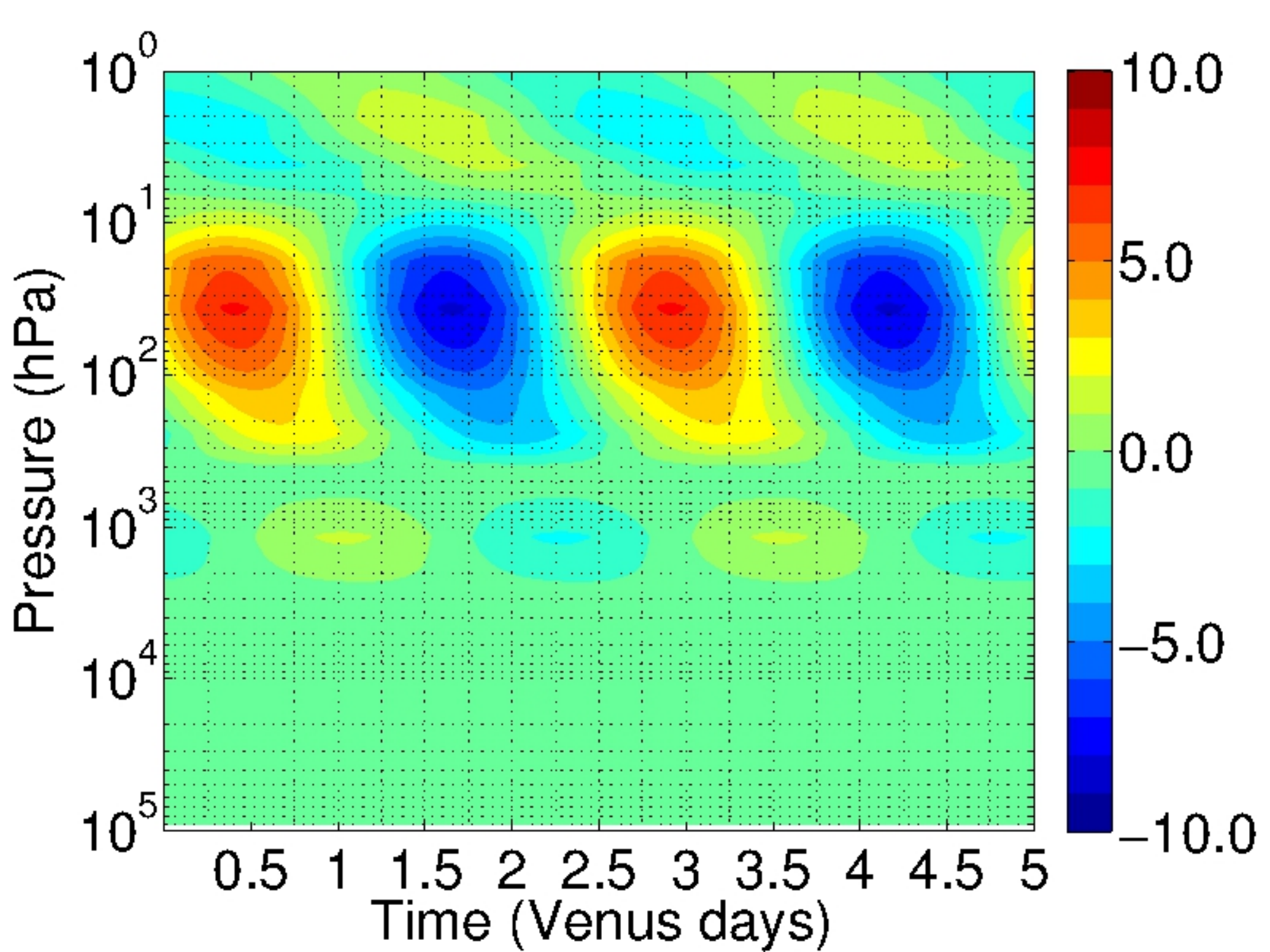}}
\\
\subfigure[]{\label{fig:u}\includegraphics[width=0.32\textwidth]{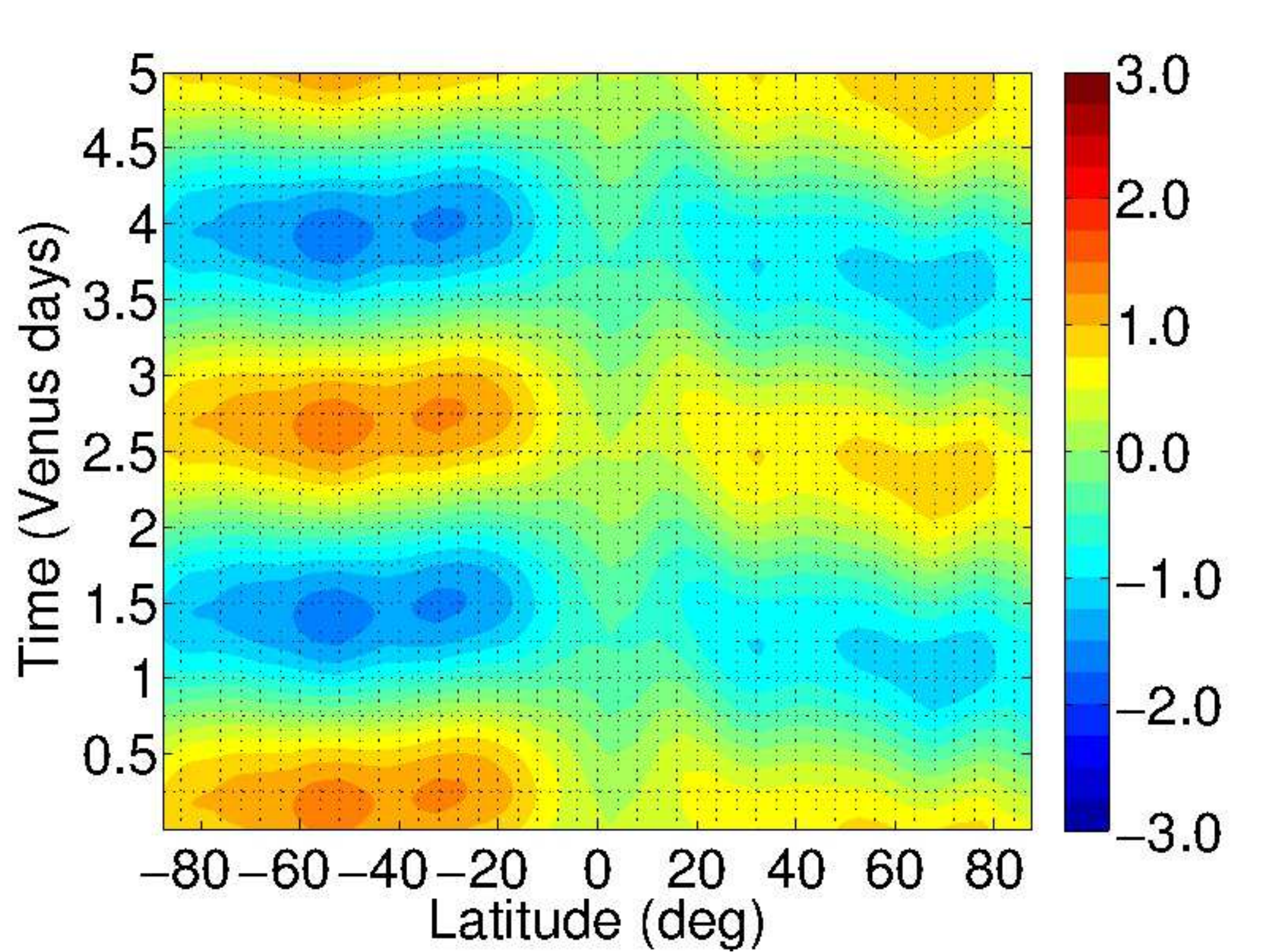}}
\subfigure[]{\label{fig:u}\includegraphics[width=0.32\textwidth]{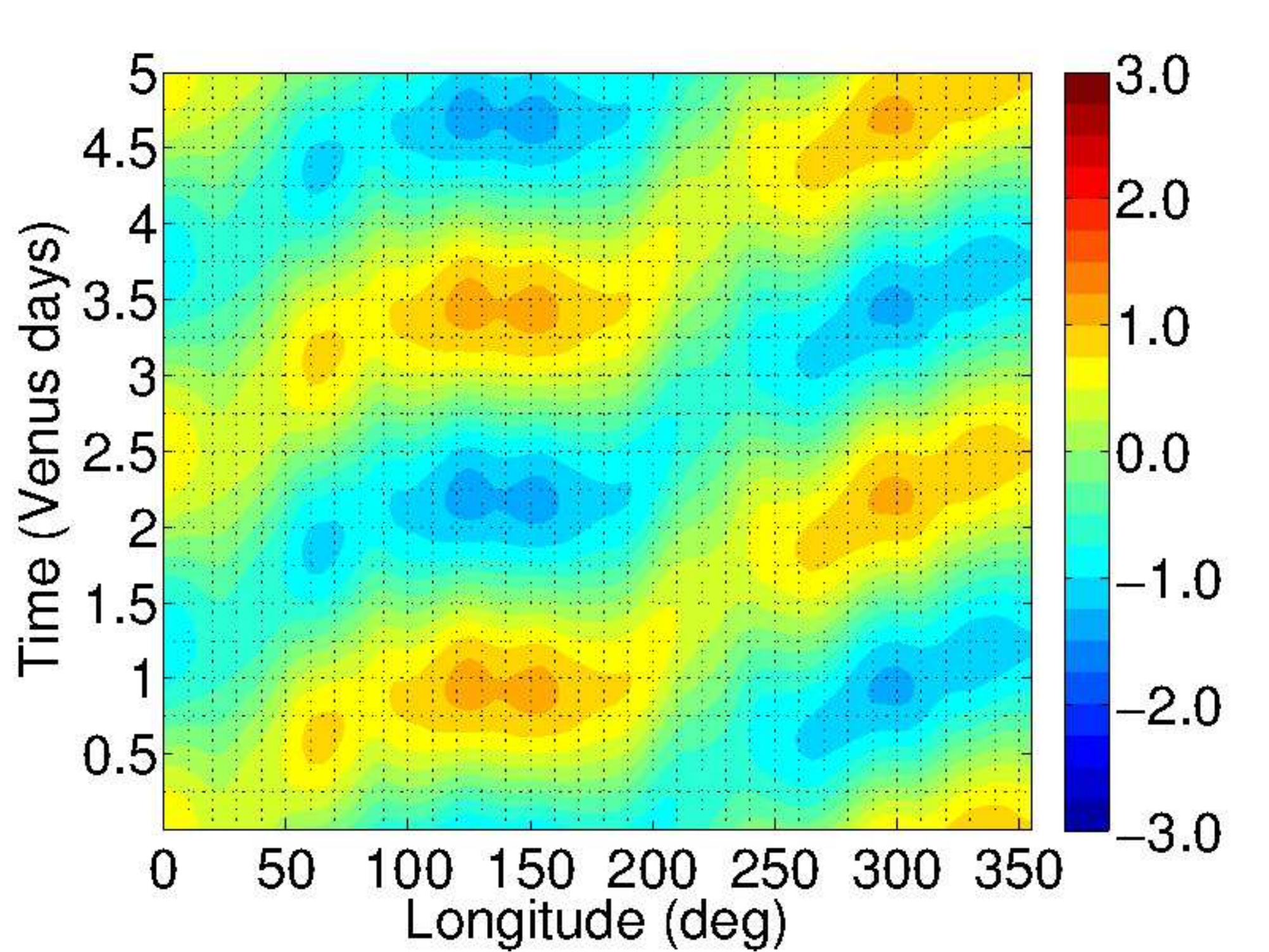}}
\subfigure[]{\label{fig:u}\includegraphics[width=0.32\textwidth]{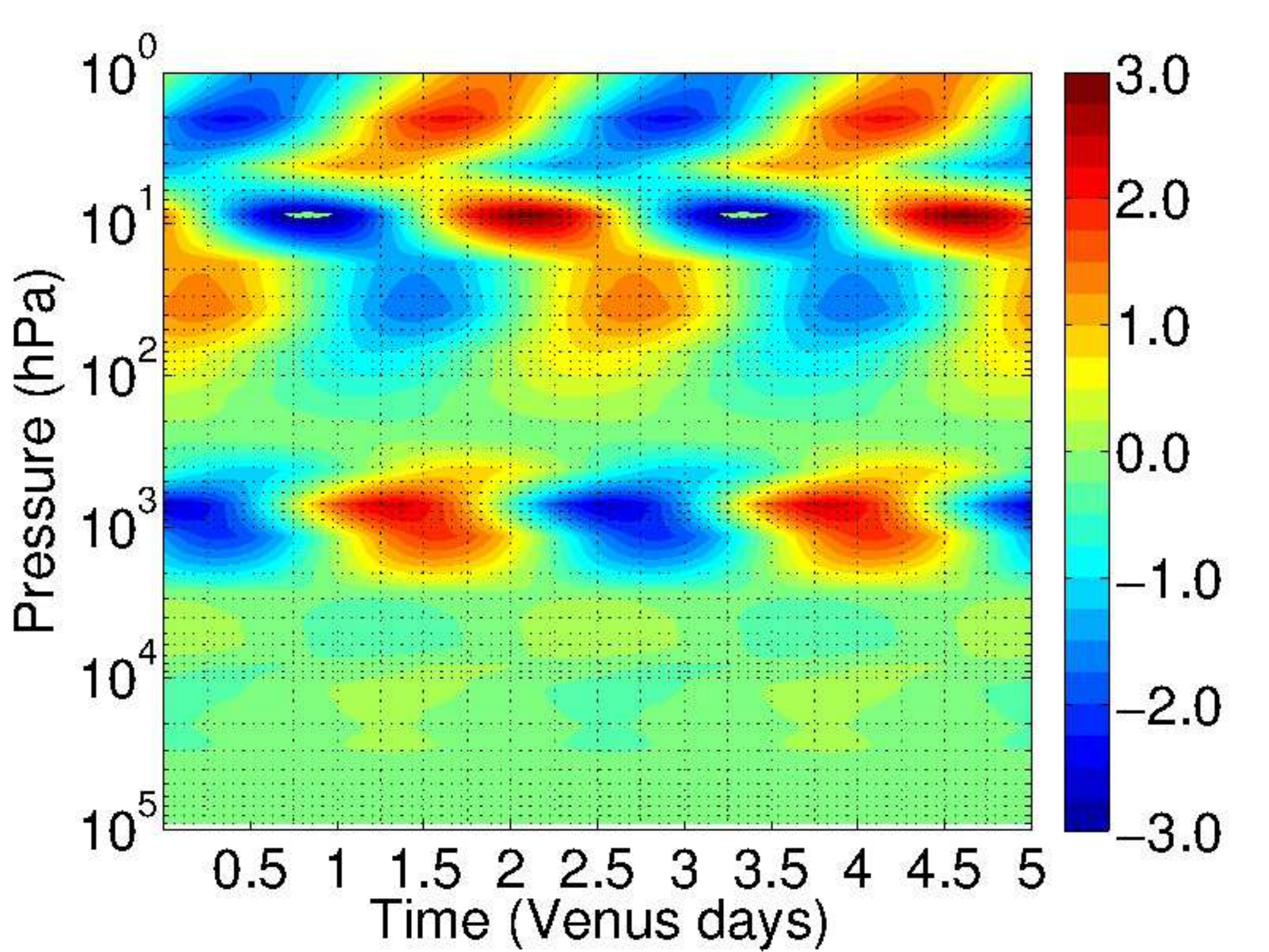}}
\caption[Hovm\"oller plots of the low frequency waves.]{Hovm\"oller plots of the low frequency waves. The data shown were filtered for periods longer than 1.5 and shorter than 2.5  Venus days. The three rows correspond to the analysis of three different variables: \textbf{(a)}, \textbf{(b)} and \textbf{(c)} temperature (K); \textbf{(d)}, \textbf{(e)} and \textbf{(f)} zonal wind (m s$^{-1}$); \textbf{(g)}, \textbf{(h)} and \textbf{(i)} meridional wind (m s$^{-1}$). The three columns are time series as a function of latitude (fixed in longitude $0^{\circ}$ and pressure 70 hPa), longitude (fixed in latitude 45N and pressure 70 hPa) and pressure (fixed in longitude $0^{\circ}$ and latitude 45N).}
\label{fig:ux_mrg_hov}
\end{figure}

\begin{figure}
\centering
\subfigure[]{\includegraphics[width=0.4\textwidth]{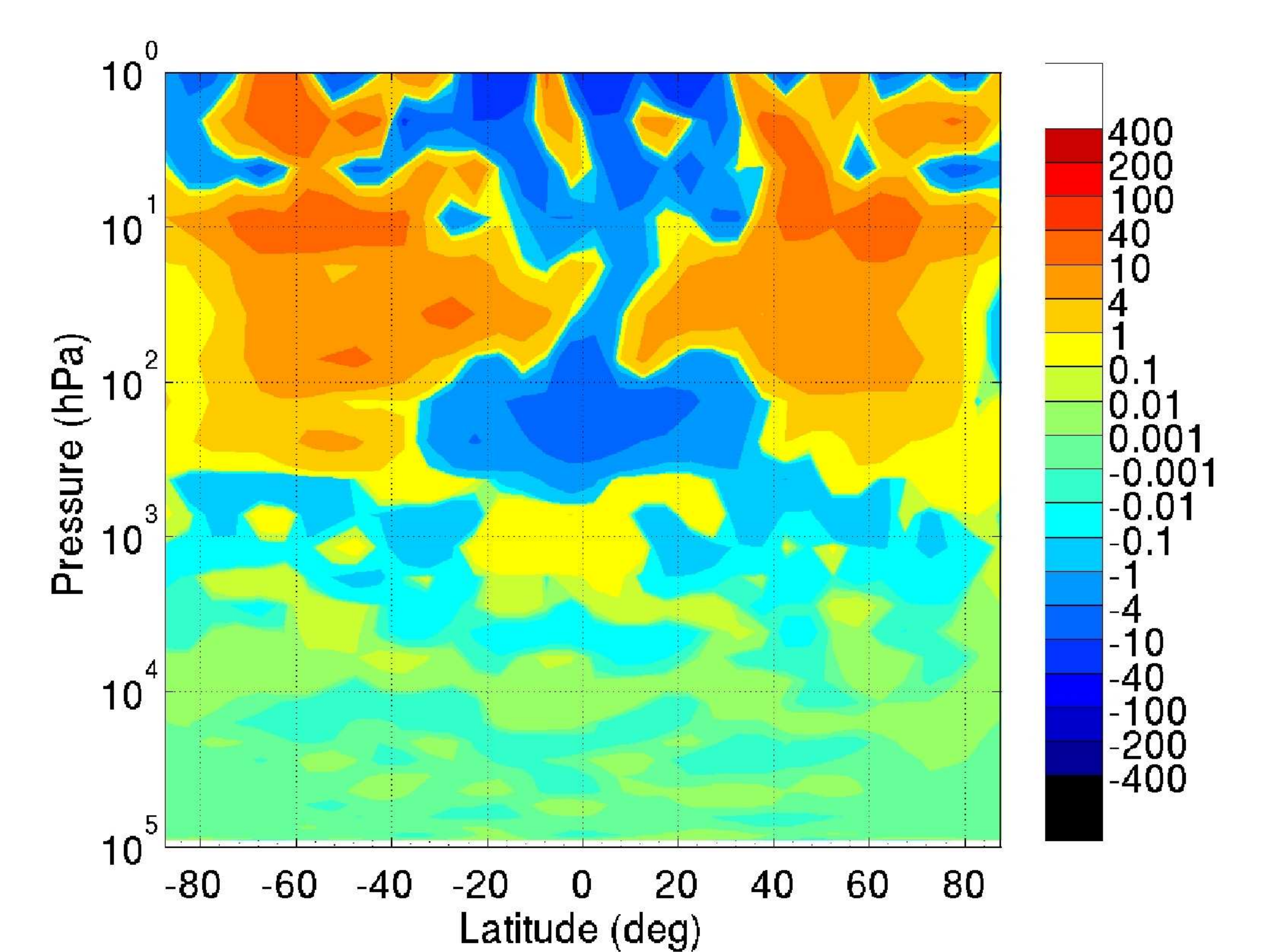}}
\subfigure[]{\includegraphics[width=0.4\textwidth]{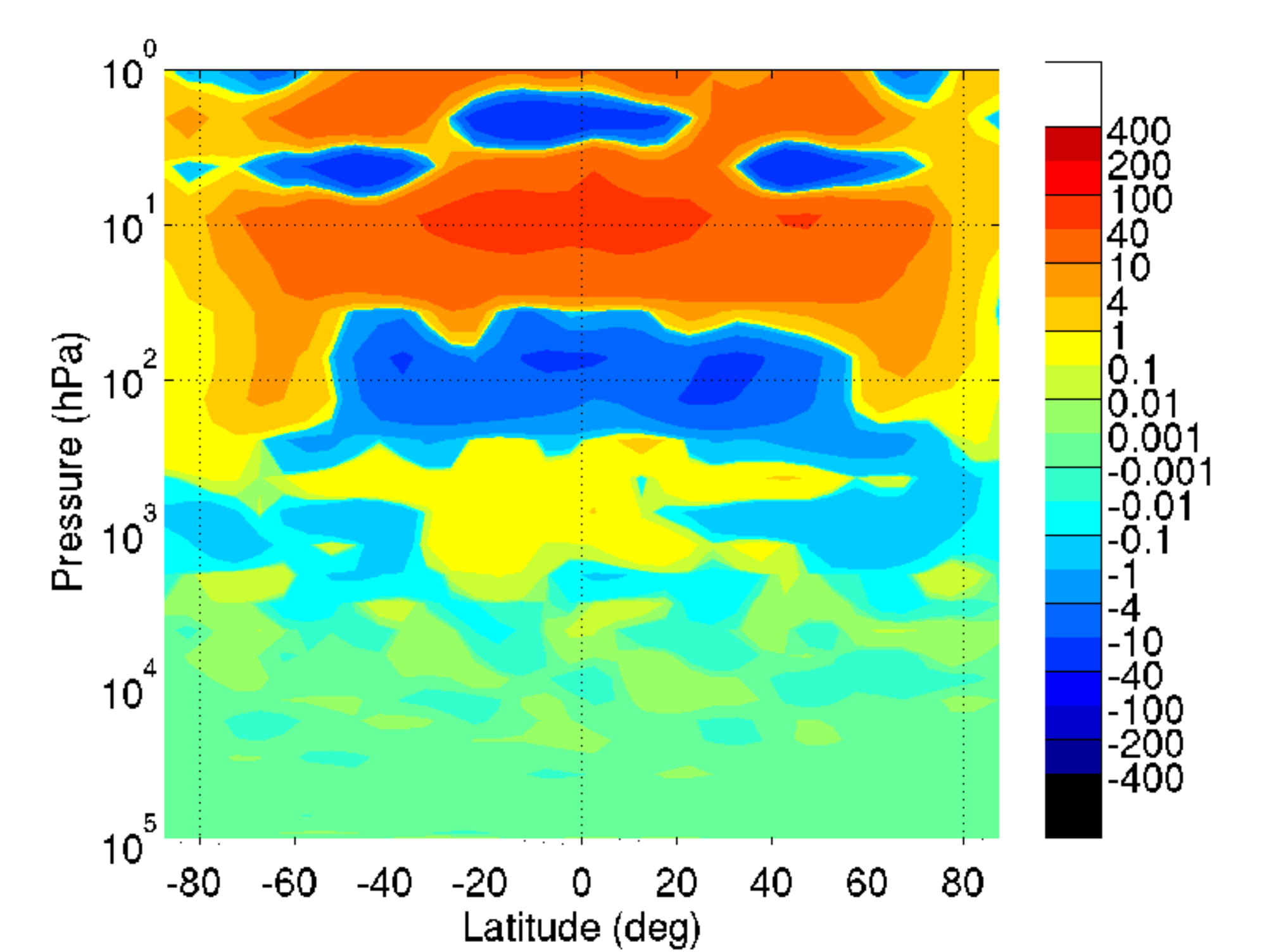}}
\\
\subfigure[]{\includegraphics[width=0.4\textwidth]{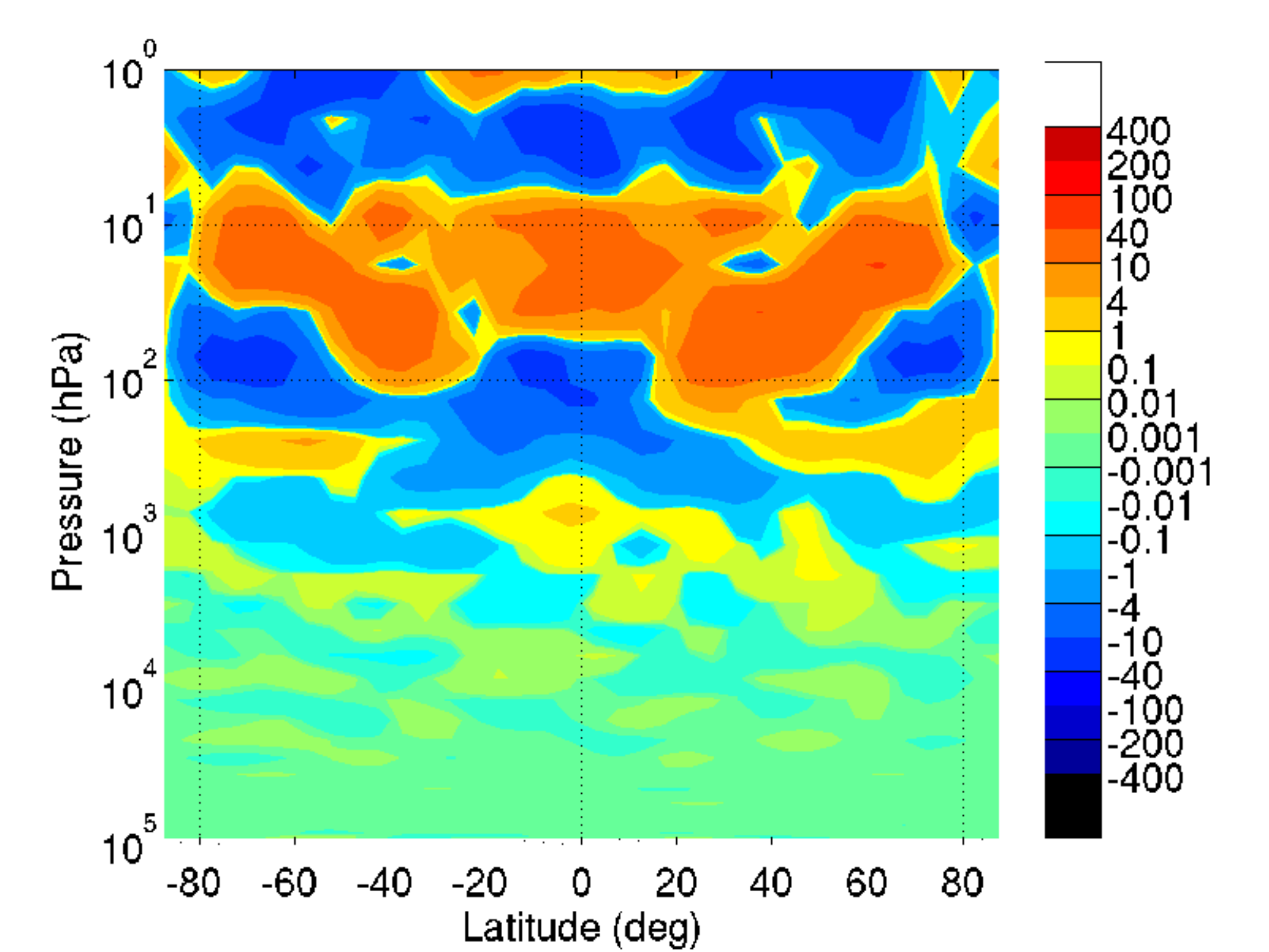}}
\subfigure[]{\includegraphics[width=0.4\textwidth]{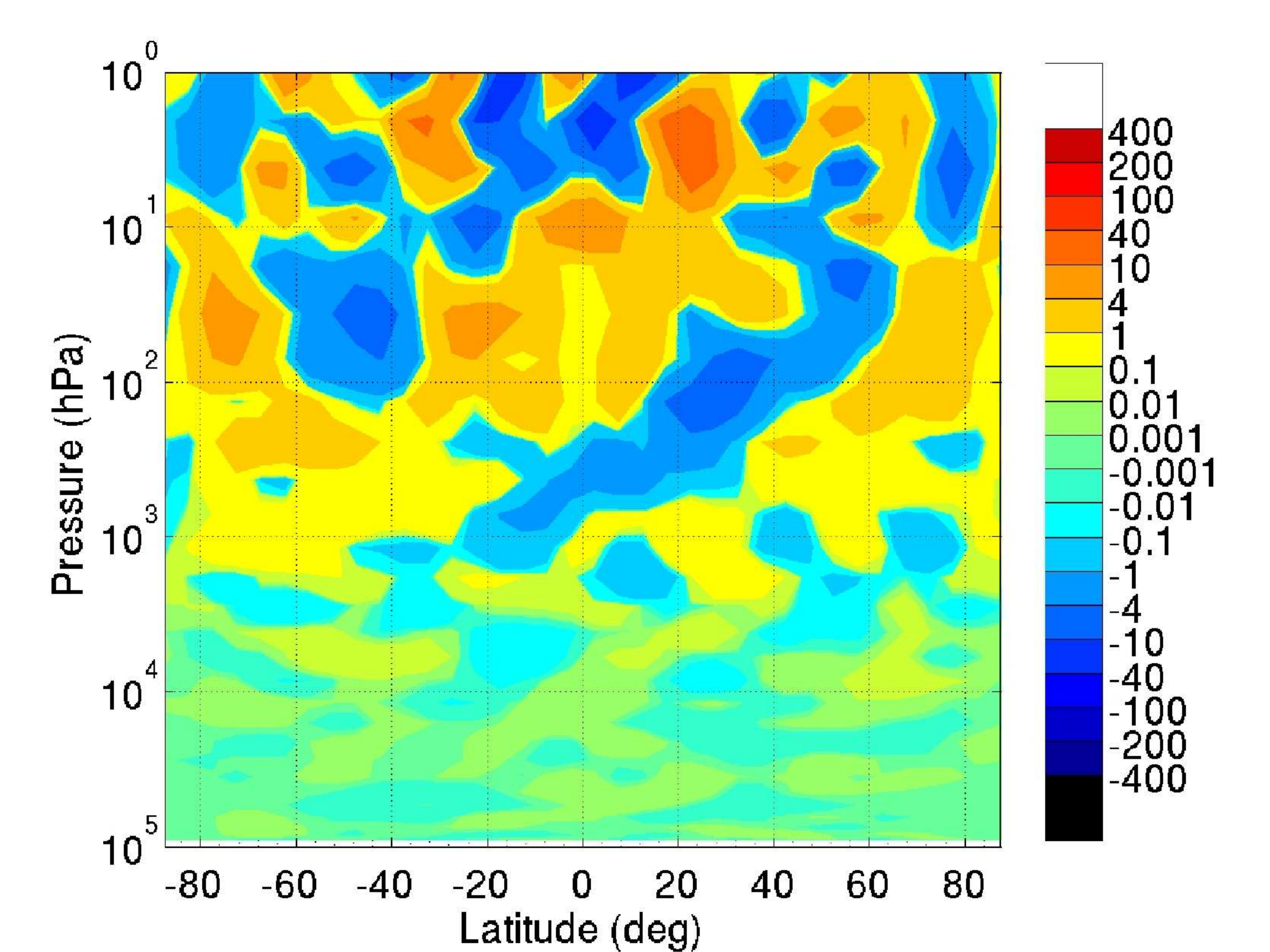}}
\caption{Contribution for the term  $-\nabla\cdot \boldsymbol{E}$ from waves with different periods:\textbf{(a)} $<$ 0.4 Venus days; \textbf{(b)} 0.4-0.6 Venus days; \textbf{(c)} 0.6-0.9 Venus days; \textbf{(d)}$>$ 0.9 Venus days}
\label{fig:fil_eli}
\end{figure}

In Fig. \ref{fig:fil_eli}, we filtered the model results to explore the contribution of the different type of waves to the Eliassen-Palm flux. In Fig. \ref{fig:fil_eli}(a) we show the term $-\nabla\cdot \boldsymbol{E}$ for waves with period shorter than 0.4  Venus days. The acceleration in the upper cloud region is related to the presence of the tides' harmonics with order higher than two. These forced waves, such as the semidiurnal tide, have a tilt of the wave phase front with altitude that exerts a pressure torque in the excited region and accelerates the flow in the prograde direction. The tilt is associated with the vertical shear and the motion of the Sun's position relative to the mean flow. In the region where these waves are dissipated they decelerate the flow, which is visible from the regions above and below the excited region. At low latitudes in the upper cloud region, the deceleration can be associated with instabilities produced by the convection region, which may be exerting a drag on the flow. This map, for short period waves, did not show any signs of barotropic waves transporting angular momentum from the mid-latitude jet region towards low latitudes, which would be associated with regions of deceleration at mid-latitudes and pressure level around 70 mBar, and regions of acceleration at the same pressure level but low latitudes. In Fig. \ref{fig:fil_eli}(b), we show the impact of the semi-diurnal tide. This wave type has the largest contribution to acceleration of the flow in the equatorial jet region, which as we saw before is associated with the vertical transport of  axial angular momentum from the upper atmosphere towards the upper cloud region. In Fig. \ref{fig:fil_eli}(c), we analyze the contribution of the waves with period between 0.6 and 0.9 Venus days. The acceleration shown is associated with the free equatorial Rossby waves found above. These waves have an important contribution due to their spatial structure. Rossby waves tend to accelerate the flow at latitudes where they are formed. However, due to an  interaction between the free Rossby wave and the diurnal tide, prograde angular momentum is also transported equatorwards (a process similar to the one found in \citealt{2012Arnold}). This mechanism is detected due to the deceleration at roughly 40$^o$ latitude in each hemiphere at the jet altitude, which will balance an aceleration at low latitudes. For waves with periods longer than 0.9 Venus days, which for example includes the potential contribution of the diurnal tide and the low frequency MRG wave decribed above (see Fig. \ref{fig:fil_eli}d), we do not find any relevant contribution to the acceleration of the flow compared to the role of other wave types (or mean circulation).

\section{Surface Albedo}
\label{sec:SurfAlb}
In this section the same model is run with a surface albedo calculated from a fit to the observational data from \cite{1986Pieters} for wavelengths shorter than 1 $\mu$m, and for larger wavelengths the surface albedo was set to 0.15 (\citealt{2015Mendonca}). The simulation in this section was integrated for 215.5 Venus days from a rest atmosphere as in the reference simulation.

\begin{figure}
\centering
\includegraphics[width=0.5\textwidth]{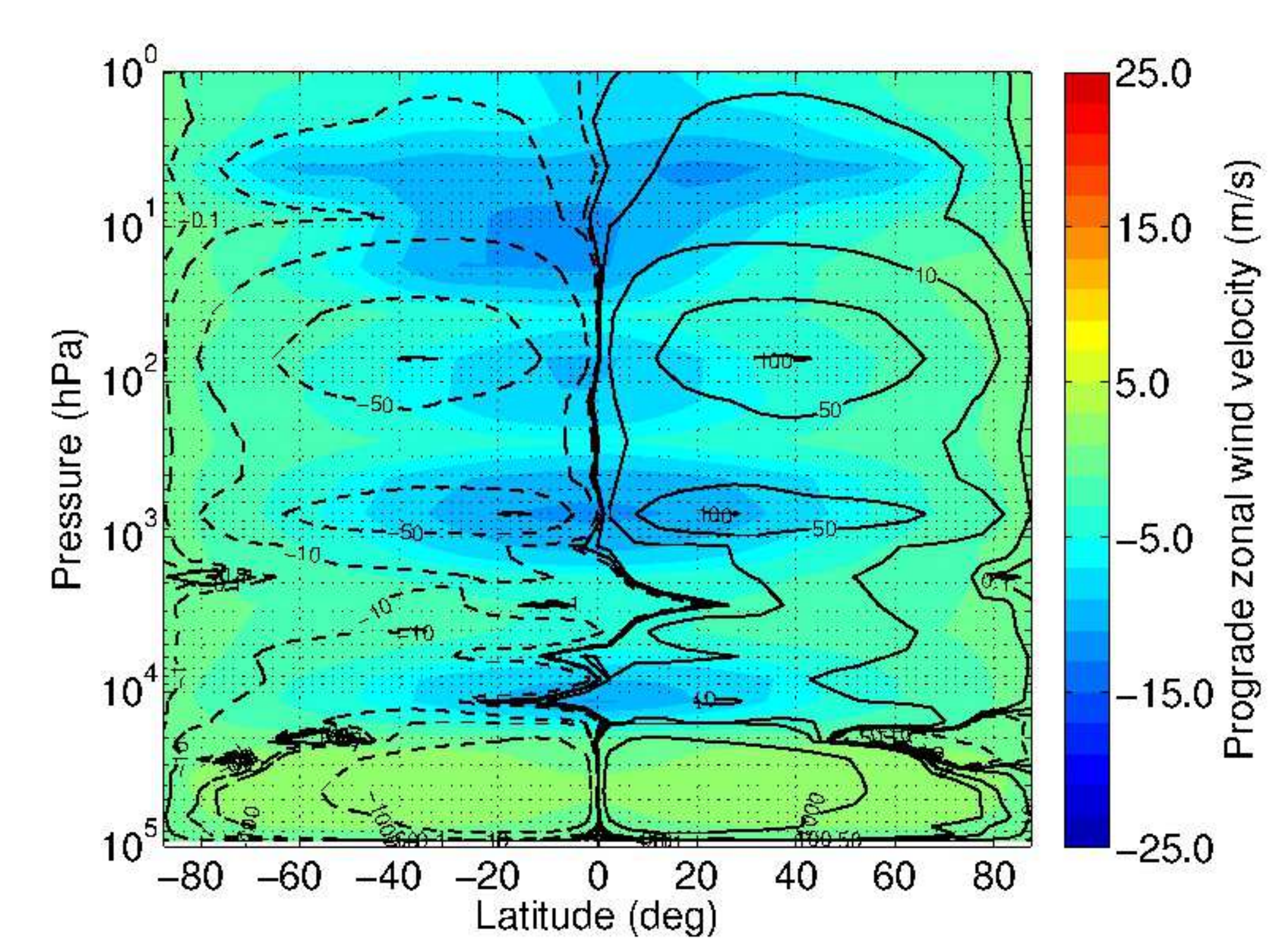}
\caption[Averaged zonal winds and mass stream function for the OPUS-Vr simulation testing the reference albedo.]{Averaged zonal winds and mass stream function (in units of $10^9$kg/s) from the OPUS-Vr simulation testing the observationally determined surface albedo. The dashed lines represent the anti-clockwise circulation and the solid lines the clockwise circulation. The values were zonal and time averaged for 5 Venus day.}
\label{fig:ut_u}
\end{figure}

\begin{figure}
\centering
\includegraphics[width=0.5\textwidth]{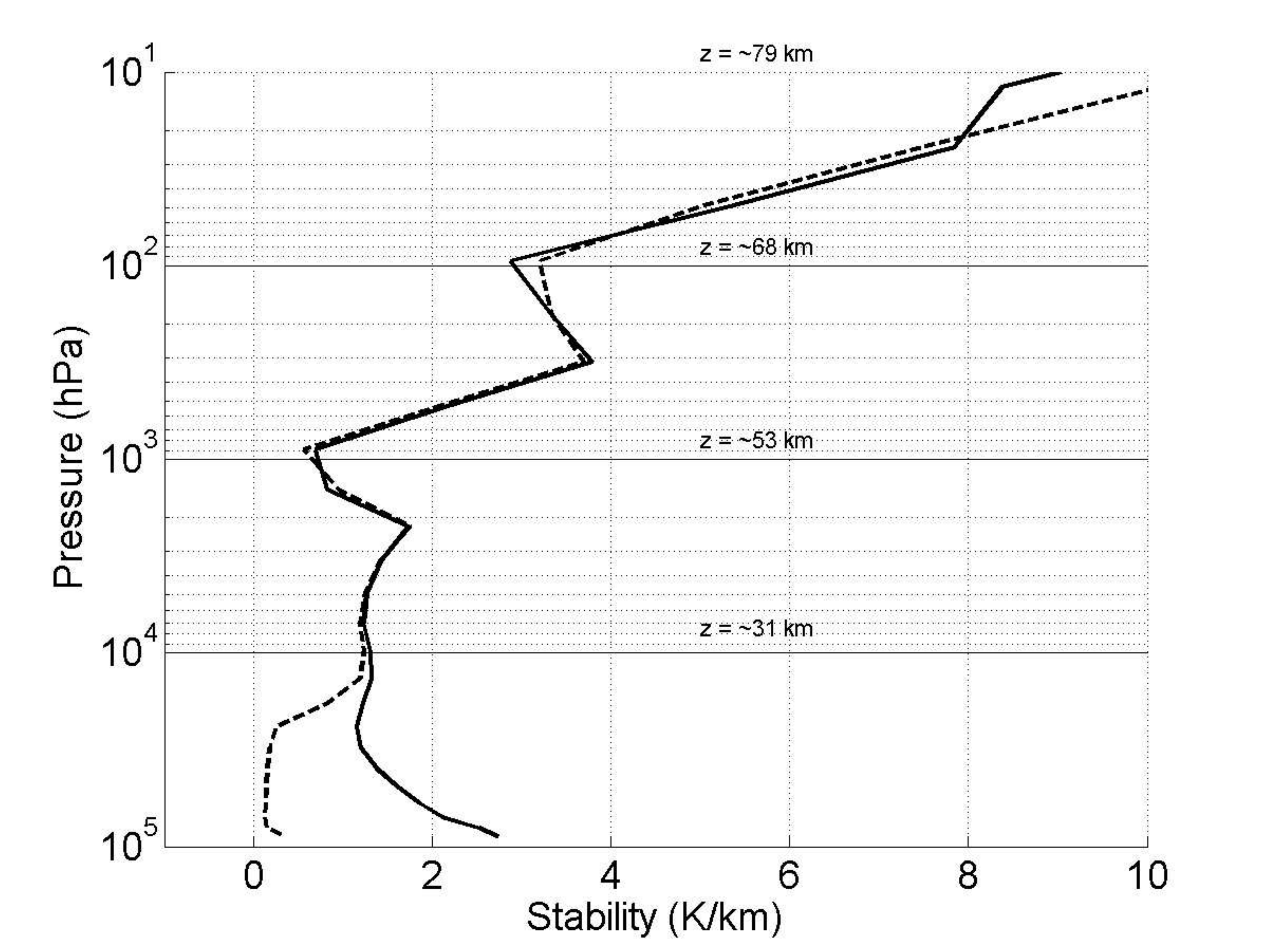}
\caption{Global mean profiles of atmospheric static stability ($d\theta/dz$, where $\theta$ is the potential temperature): solid line is the reference simulation; dashed line is the simulation with surface albedo calculated from a fit to the observational data from \cite{1986Pieters} for wavelengths shorter than 1 $\mu$m, and for larger wavelengths the surface albedo was set to 0.15 (\citealt{2015Mendonca}). The atmospheric static stability values were time averaged for five Venus days.}
\label{fig:sta_comp}
\end{figure}

Fig. \ref{fig:ut_u} shows zonal winds averaged in longitude and in time over the last 5 Venus days of the simulation between the surface and a pressure level of 1 hPa ($\sim$90 km). This map does not show any sign of strong prograde zonal winds comparable to the ones retrieved from observational data throughout the entire atmosphere (\citealt{1983Schubert}). The black lines in Fig. \ref{fig:ut_u} represent the zonally and time averaged mass stream function. In the deep atmosphere these lines show the clear presence of a large deep convection circulation pattern (see the stability profile in Fig. \ref{fig:sta_comp}). In this region of the atmosphere the momentum is transported upward in the equatorial region and transported poleward at the top of the deep convection cell, creating two peaks in the zonal wind field at 4.0$\times$10$^4$ hPa ($\sim$10 km) in each hemisphere. Above this region the mass stream function contours show the clear presence of two planet-wide thermally direct atmospheric cells situated within the clouds. These cells are the same as the ones described for the reference simulation. Despite the efficient latitudinal transport of momentum by the mean circulation at these pressures (within the upper clouds), the model did not produce any high latitude jets. The atmospheric circulation developed shows four retrograde equatorial maxima, despite the one at 100 hPa being very variable.

In this simulation disturbances in the deep convection patterns were identified as the source of pulses of gravity waves that travel upward. This very active convection region exerts a drag on the atmospheric circulation and disables the production of super-rotation. When the surface short-wave albedo is set to 0.95 just $\sim1\%$ of the total solar incoming energy is absorbed by the surface producing a statically stable atmosphere near the surface. In this case, the occurrence of vigorous convection near the surface is more unlikely. The shortwave albedo larger than $\sim$0.15 does not represent well the optical properties of the surface, which is thought to be mainly composed of basalt. However, this heuristic method of increasing the albedo appears to alleviate the problem of overestimating the amount of solar energy absorbed by the surface, and yet does not compromise the calculation of the radiative heating rates in the atmosphere above (\citealt{2013Mendonca}). This need to use large albedos can be interpreted as indicating an incomplete representation either of the radiative extinction in the extreme deep atmosphere, or of the possible existence of extra missing constituents of the atmosphere that interact with solar radiation, thereby modifying the static stability in this part of the atmosphere. In \cite{2013Mendonca} it is discussed that the need to adjust this albedo may be related with errors in the assumed radiative properties of the atmosphere and surface under the extreme conditions prevailing in the deep atmosphere and/or inadequacies associated with uncertainties in the distribution of clouds and aerosols in the deep atmosphere (e.g., \citealt{2004Grieger}). One possible source of such opacity could be the presence of suspended mineral dust, that could easily be lifted from the dry, dusty surface even by light winds in the very dense atmosphere close to the ground. There is little direct observational evidence for such dust lifting, although this should be explored further in both models and future observations. In \cite{2013Mendonca} it was estimated that the most easily lifted particle has a diameter of 106 $\mu$m and a saltation threshold of 3.2 cm/s.


\section{Discussion on the dynamics in the lower atmosphere}
\label{sec:dics}
The reference simulation obtained a global super-rotation index ($S$) of roughly 0.2, which is more than one order of magnitude lower than the value estimated from observational data and derived in section \ref{sec:intro}: $S = 7.7_{-3.6}^{+4.2}$. The variable $S$ is a mass weighted quantity, which means that the largest contributions come from the deeper atmosphere. The main cause for the poor representation of the atmospheric circulation in this region is still unknown. This problem is the next main challenge for all Venus circulation models. This work needs to be done in association with future observational studies of the deeper atmosphere. The characterization of the dynamical regime of the lower atmosphere from observational data is still relatively crude, since it is based on roughly a dozen entry probes from Pioneer Venus and Venera space missions (\citealt{1980Schubert}). Each of these observations correspond to one specific instant in time and location on the planet. Despite covering different parts of the planet and local times, they are insufficient to describe the global mean state of the atmosphere. 

The problems encountered by the current Venus GCMs (e.g., our OPUS-Vr and the LMD Venus GCM presented in \citealt{2010Lebonnois}) in reproducing strong prograde winds in the lower atmosphere may be related to the poor representation of certain physical mechanisms in the atmosphere. In the lower atmosphere the vertical profiles of static stability are correlated with profiles of zonal wind shear at low latitudes, where large shears are found in very stable layers (\citealt{1980Schubert}). In the OPUS-Vr results this correlation was also found, but it is less clear than the ones obtained from observational data (Figs. \ref{ux-12}b and \ref{ux-comp-u2}a). In the regions of high static stability, upward convection is less likely to occur and, as a consequence, vertical mixing is inhibited. The presence of vertical convective mixing can reduce the magnitude of atmospheric super-rotation, since the vertical mixing is likely to transport angular momentum down-gradient near the equator (in general, the zonal wind magnitudes increase from less than 10 m/s at 10 km to roughly 100 m/s at cloud heights). Following Hide's theorem (\citealt{1986Read}), the formation of an equatorial super-rotation (at any altitude) requires an up-gradient transport of angular momentum, which could be achieved by phenomena associated with upward-propagating waves.  In \cite{1980Schubert} a relatively large horizontal temperature contrast in the lower atmosphere (below the cloud base) is found in observational probe measurements. This thermal structure is an indication of significant eddy activity and a weak Hadley circulation (its existence is uncertain). The mechanism (or a combination of different mechanisms) involved in the formation of the strong zonal winds must be working under these constrained atmospheric conditions.

The excess of axial angular momentum in the atmosphere has to be associated with the surface-atmosphere mechanical interaction, where the surface pumps angular momentum into the atmosphere during the ``spin-up phase''. After the excess of momentum is deposited, it is likely to be redistributed through the atmosphere by the mean circulation and by eddies as we showed in this work. Another source of energy capable of influencing the atmospheric dynamics is the release of latent heat from condensation of H$_2$SO$_4$ cloud droplets. However, this is not very relevant for atmospheric circulation due to the small mass loading of the clouds (\citealt{1978Rossow}).

Despite just a small fraction of the incoming solar radiation reaching the deepest regions of the atmosphere, \cite{1987Hou} and \cite{2007Ikeda} suggested that this radiation can drive/initiate a mechanism to explain the strong prograde winds in the lower atmosphere. These authors proposed that such strong winds in Venus'  lower atmosphere (below the cloud deck), could be produced by critical level absorption of gravity waves from small-scale convection excited near the surface or in a deep atmospheric region with low static stability (e.g., between 11$\times 10^3$ hPa and 33$\times 10^3$ hPa). However, the fully self-consistent physical representation of this phenomenon needs simulations with high spatial resolution or the use of a non-orographic wave drag parameterisation. For the latter case, schemes suitable for GCM studies are already available such as the one developed by \cite{2005McLandress}. In \cite{2007Ikeda} and \cite{2016Lebonnois} the representation of wave drag due to breaking of non-orographic gravity waves in the atmosphere below the cloud region increased, in general, the zonal winds in the lower atmosphere. However, there are several uncertainties associated with this type of parameterisation that need to be explored in more detail. Some of those uncertainties are associated with a lack of adequate knowledge of the source spectra in the atmosphere and the estimation of the free parameters needed in the numerical scheme due to the typical simplifications used.

By using the LMD Venus GCM, which includes a radiative transfer formulation, \cite{2010Lebonnoisb} showed that simulations initiated with different atmospheric initial states (different magnitudes of global super-rotation) reach equilibrium towards different distributions of zonal winds in terms of amplitude and location. The mechanisms that maintain the strong winds remain largely the same and are similar to the ones shown in our reference simulation, although the magnitude of angular momentum transported in the atmosphere is different (e.g., by mean circulation and transient waves). This possibility of multiple atmospheric equilibrium states was also explored with OPUS-Vr. Some test simulations were made by starting the model with a very strong initial global super-rotation, which implied a more efficient heat transport in longitude at all altitudes. However, despite the model having reproduced realistic winds in comparison with the observed ones for a short period, the atmosphere  ended up relaxing towards the states shown in the reference simulation, which did not confirm \cite{2010Lebonnoisb}'s conclusion concerning the existence of multiple super-rotation equilibria.

Other factors that may have an important impact in the lower atmospheric circulation and that need to be better explored are the impact of a radiatively active dust cycle in the deep atmosphere, as suggested in \cite{2013Mendonca}. Regarding topographic effects, in general, mountain-induced internal gravity waves are usually considered to exert a stress on the atmosphere, dragging its motion towards a state of rest through the breaking of vertically propagating waves, the phase speed of which is stationary in the horizontal. \cite{1977Fels}, however, has also suggested that topographic scattering of downward propagating gravity waves may also be a possible atmospheric feature with an important impact on the tidal momentum flux in the Venus atmosphere. This phenomenon can actually accelerate the atmosphere, in contrast to mountain drag, and is suggested by \cite{1977Fels} to be an important possible mechanism for the angular momentum transport with a positive contribution to the equatorial super-rotation. However, this effect is still poorly explored in Venus GCMs. 

Another important topic in the Venus climate community that remains poorly understood is the impact of dissipative schemes, such as numerical diffusion, on the Venus atmospheric circulation simulated. A better comprehension of the kinetic energy spectra of the global Venus atmosphere can be the key to finally understand what mechanisms are producing the strong zonal winds in the lower atmosphere.

\section{Conclusions}
\label{sec:conclu}
Our new model is capable of producing an atmospheric circulation above the cloud base similar to the one observed. The mean circulation simulated is characterised in the cloud region by two planetary-scale Hadley cells in each hemisphere: one near the cloud base and another in the upper clouds. The existence of these two dynamical regions is mainly related to the significant absorption of solar radiation in the upper cloud region and the blocking at the cloud base of upwelling infrared radiation from the hot lower atmosphere. The atmospheric cells on Venus are in general larger than the ones found on the Earth's atmosphere, due to the weak coriolis acceleration on Venus (a slow rotating planet). These circulation patterns transport momentum poleward in the upper branches that drives the formation of mid-latitude jets but weakens the zonal winds at low latitudes.  The eddy-zonal flow interactions have a crucial role in replenishing the equatorial region with angular momentum. 

The model results showed that the nature of the mechanisms involved in the formation of atmospheric super-rotation above and below the cloud base are different.  In the upper cloud region, the radiative time-scale is smaller than a Venus day, and several harmonics of the thermal tides are produced. The components with the largest amplitudes comprise zonal wavenumbers one (diurnal tide) and two (semi-diurnal tide). In the upper cloud region, however, it is the wave number two component that has more impact on the atmospheric circulation. The Sun moves slowly in the retrograde direction in relation to the mean flow, which forces the tides to follow its position. This relative motion induces a positive acceleration of the flow in the region where the tides are excited (upper cloud region). In the region where the waves are absorbed (via radiative damping) by the atmosphere, predominantly above the cloud region, the forcing in the atmosphere acts in the reverse direction.  The presence of the thermal tides is very clear in the observational data, but the same cannot be said of the other transient waves.  The main difficulty in retrieving other eddy motions from observational data is the low resolution of the images available from any space mission to date at low latitudes. We also found that the flow in the upper cloud region is accelerated due to the wave structure of a free equatorial Rossby wave. In this case, the equatorward transport of momentum is achieved by a nonlinear interaction between the free Rossby wave and the equatorial structure of the diurnal tide. Also, this wave transport momentum verticaly due to the tilt of the phase front with altitude.

Below the cloud base the radiative time-scale becomes much larger than a solar day, and the influence of the thermal tides in the atmosphere becomes negligible. In general, the mean atmospheric circulation in this region is apparent from the simulations as large, deep, equator-to-pole cells, which extend from the surface to the cloud region. Nevertheless, in common with the LMD Venus GCM (\citealt{2010Lebonnois}), OPUS-Vr was not capable of reproducing strong zonal winds comparable with the observations in this region. More observational and theoretical work is needed to improve our understanding of what are the atmospheric mechanisms driving the circulation in the lower atmosphere.

The Venus atmospheric super-rotation is not a temporary state for its current atmospheric conditions, and supporting this idea are the results of long numerical simulations and the consistency between the observations over the last decades. However, numerical and observational studies show that the Venus atmospheric circulation is not steady. As an example, the variability of the zonal wind distribution at jet altitudes is significantly affected by long term oscillations (tens of Venus days) and by the approximately bidiurnal planetary mixed Rossby-gravity waves, which is seen in both observations and simulations. These low frequency waves (also found in the LMD Venus GCM, \citealt{2010Lebonnois}) may be the same ones as found in observations by \cite{2013Khatuntsev}.

With our new and flexible radiative transfer scheme, our model presents an excellent capability to explore different atmospheric conditions which is essential to study the Venus atmospheric dynamics. Our model shows great promise for future research in this area, in particular due to its unique ability to explore radiation parameters in the lower atmosphere, where the observational and modelling uncertainties are very high.

\section*{Acknowledgments}
The authors thank the Met Office for the use of the HadAM3 dynamical core in this work. J.M.M. thanks the Center for Space and Habitability (CSH) and the Space Research and Planetary Sciences Division (WP) of the University of Bern for financial, secretarial and logistical support. P.L.R. acknowledges support from the UK Science and Technology Research Council. 

\section*{References}

\bibliography{mybibfile}

\end{document}